\documentclass[prb,twocolumn,superscriptaddress]{revtex4-2}

\usepackage{graphicx,color}
\usepackage{amsmath}
\usepackage{amssymb}
\usepackage{natbib}
\usepackage{tabularx}
\usepackage{multirow}
\usepackage[colorlinks=true,linkcolor=blue,citecolor=blue,urlcolor=blue]{hyperref}
\usepackage{ulem}
\usepackage{placeins}

\newcommand{\rt}{$R_2T_2$O$_7$}

\newcommand{\hoti}{Ho$_2$Ti$_2$O$_7$}
\newcommand{\dyti}{Dy$_2$Ti$_2$O$_7$}

\newcommand{\ndzr}{Nd$_2$Zr$_2$O$_7$}
\newcommand{\ndhf}{Nd$_2$Hf$_2$O$_7$}
\newcommand{\ndzrx}{Nd$_2$(Zr$_{1-x}$Ti$_x$)$_2$O$_7$}
\newcommand{\ndtiA}{Nd$_2$Zr$_{1.95}$Ti$_{0.05}$O$_7$}
\newcommand{\ndtiB}{Nd$_2$Zr$_{1.8}$Ti$_{0.2}$O$_7$} 
\newcommand{\ndzry}{(Nd$_{1-y}$La$_y$)$_2$Zr$_2$O$_7$}
\newcommand{\ndlaA}{Nd$_{1.8}$La$_{0.2}$Zr$_2$O$_7$}
\newcommand{\ndlaB}{Nd$_{1.6}$La$_{0.4}$Zr$_2$O$_7$}
\newcommand{\ndlaC}{Nd$_{1.2}$La$_{0.8}$Zr$_2$O$_7$}
\newcommand{\ndmox}{Nd$_2$(TiZrHfScNb)$_2$O$_7$}
\newcommand{\nd}{Nd$^{3+}$}
\newcommand{\ndscnb}{Nd$_2$ScNbO$_7$}
\newcommand{\ndgasb}{Nd$_2$GaSbO$_7$}
\newcommand{\ndsn}{Nd$_2$Sn$_2$O$_7$}

\begin{document}

\author{M\'elanie L\'eger}
\affiliation{Institut N\'eel, CNRS and Universit\'e Grenoble Alpes, 38000 Grenoble, France}
\affiliation{Laboratoire L\'eon Brillouin, CEA, CNRS, Universit\'e Paris-Saclay, CE-Saclay, F-91191 Gif-sur-Yvette, France}

\author{Florianne Vayer}
\affiliation{Institut de Chimie Mol\'eculaire et des Mat\'eriaux d'Orsay (UMR CNRS 8182), Universit\'e Paris-Saclay, F-91405 Orsay, France}

\author{Monica Ciomaga Hatnean}
\altaffiliation[]{Current address: Laboratory for Multiscale Materials eXperiments, Paul Scherrer Institut, 5232 Villigen PSI, Switzerland}
\affiliation{Department of Physics, University of Warwick, Coventry, CV4 7AL, United Kingdom}

\author{Fran\c{c}oise Damay}
\affiliation{Laboratoire L\'eon Brillouin, CEA, CNRS, Universit\'e Paris-Saclay, CE-Saclay, F-91191 Gif-sur-Yvette, France}

\author{Claudia Decorse}
\affiliation{Institut de Chimie Mol\'eculaire et des Mat\'eriaux d'Orsay (UMR CNRS 8182), Universit\'e Paris-Saclay, F-91405 Orsay, France}

\author{David Berardan}
\affiliation{Institut de Chimie Mol\'eculaire et des Mat\'eriaux d'Orsay (UMR CNRS 8182), Universit\'e Paris-Saclay, F-91405 Orsay, France}

\author{Bj\"orn F\r{a}k}
\affiliation{Institut Laue Langevin, 6, rue Jules Horowitz, BP 156 F-38042 Grenoble, France}

\author{Jean-Marc Zanotti}
\affiliation{Laboratoire L\'eon Brillouin, CEA, CNRS, Universit\'e Paris-Saclay, CE-Saclay, F-91191 Gif-sur-Yvette, France}

\author{Quentin Berrod}
\affiliation{Univ. Grenoble Alpes, CNRS, CEA, Grenoble INP, IRIG, SyMMES, 38000 Grenoble, France}

\author{Jacques Ollivier}
\affiliation{Institut Laue Langevin, 6, rue Jules Horowitz, BP 156 F-38042 Grenoble, France}

\author{Jan P. Embs}
\affiliation{Laboratory for Neutron Scattering and Imaging, Paul Scherrer Institut, 5232 Villigen PSI, Switzerland}

\author{Tom Fennell}
\affiliation{Laboratory for Neutron Scattering and Imaging, Paul Scherrer Institut, 5232 Villigen PSI, Switzerland}

\author{Denis Sheptyakov}
\affiliation{Laboratory for Neutron Scattering and Imaging, Paul Scherrer Institut, 5232 Villigen PSI, Switzerland}

\author{Sylvain Petit}
\email[]{sylvain.petit@cea.fr}
\affiliation{Laboratoire L\'eon Brillouin, CEA, CNRS, Universit\'e Paris-Saclay, CE-Saclay, F-91191 Gif-sur-Yvette, France}

\author{Elsa Lhotel}
\email[]{elsa.lhotel@neel.cnrs.fr}
\affiliation{Institut N\'eel, CNRS and Universit\'e Grenoble Alpes, 38000 Grenoble, France}

\title{Impact of disorder in Nd-based pyrochlore magnets}

\begin{abstract}
We study the stability of the antiferromagnetic all-in--all-out state observed in dipolar-octupolar pyrochlores that have neodymium as the magnetic species. Different types of disorder are considered, either affecting the immediate environment of the \nd\, ion, or substituting it with a non-magnetic ion. Starting from the well studied \ndzr\ compound, Ti substitution on the Zr site and dilution on the Nd magnetic site with La substitution are investigated. The recently discovered entropy stabilized compound NdMox, which exhibits a high degree of disorder on the non magnetic site is also studied. Using a range of experimental techniques, especially very low-temperature magnetization and neutron scattering, we show that the all-in--all-out state is very robust and withstands substitutional disorder up to large rates. From these measurements, we estimate the Hamiltonian parameters and discuss their evolution in the framework of the phase diagram of dipolar-octupolar pyrochlore magnets.
\end{abstract}

\maketitle

\section{Introduction}
Geometrically frustrated magnetism is a forefront research topic within condensed matter physics, as testified by the wealth of exotic phenomena discovered over the past years \cite{Lacroix_2011, Moessner_2006}. 
To explore this physics, pyrochlore oxide compounds (\rt, $R$ is a rare-earth and $T$ is usually a transition metal), in which both $R$ and $T$ sites form a corner sharing tetrahedra lattice are a rich playground. Depending on the rare-earth element, different spin anisotropy (Ising, easy-plane, Heisenberg) and interactions (antiferromagnetic, ferromagnetic, long-range, anisotropic) come into play. This results in amazing physics, ranging from delicate balance between ferro- and antiferromagnetism, order by disorder mechanisms to Coulomb and quantum spin liquid phases \cite{Gardner_2010, Gingras_2014}.

\hoti\, and \dyti\, are for instance experimental realizations of the celebrated ``spin ice'' state \cite{Harris_1997}. They correspond to the case of ferromagnetic interactions between rare earth magnetic moments, constrained to lie along the local $\langle 111 \rangle$ axes (linking the center of each tetrahedron to its vertices) by a crystal field scheme featuring a strong Ising anisotropy. 

In contrast to spin ice, Nd-based magnets, \ndzr\ \cite{Ciomaga_2015, Lhotel_2015, Xu_2015}, \ndsn\, \cite{Bertin_2015} and \ndhf\, \cite{Anand_2015,Anand_2017} are Ising antiferromagnets, hence, at first glance, more conventional materials. The \nd\ magnetic moments indeed exhibit a strong local $\langle 111\rangle$ Ising anisotropy with an effective value of about 2.4-2.5 $\mu_{\rm B}$. Magnetization measurements together with powder neutron diffraction experiments further reveal that these compounds undergo an antiferromagnetic transition towards an all-in--all-out magnetic configuration, where the four spins point inwards to or outwards from the center of a tetrahedron. This situation corresponds in principle to a simple non-collinear antiferromagnetic state.
Surprisingly, experiments demonstrate a much more complex case: in \ndzr\ and \ndhf, the measured ordered magnetic moment is only about 1~$\mu_{\rm B}$, thus less than half the full moment. Furthermore, the Curie-Weiss temperature obtained from the magnetic susceptibility is positive, indicating ferromagnetic interactions that should stabilize a spin ice state, contrasting with the all-in--all-out ground state. In addition, inelastic neutron scattering measurements reveal distinctive excitations, encompassing a flat mode at finite energy, above which dispersive modes appear \cite{Petit_2016, Lhotel_2018, Xu_2019}. Both modes have remarkable structure factors, the flat mode being characterized by pinch points, suggesting dynamic spin ice correlations, while the dispersive mode is characterized by a half moon pattern stemming out of the pinch point positions. 

To solve this conundrum, Benton \cite{Benton_2016} showed that it is necessary to consider the dipolar-octupolar nature \cite{Huang_2014} of the \nd\ magnetic moment in the pyrochlore lattice symmetry. At low temperature, the relevant degree of freedom is an effective spin $1/2$ whose $z$ component can be interpreted as the magnetic moment, while its $x$ and $y$ components represent combinations of octupolar moments.
The observed ordered phase corresponds to an ordering of this pseudo-spin, which adopts an all-in--all-out structure, each spin pointing however in an inclined ``direction" in the $({\bf x}, {\bf z})$ local plane. It is therefore physically both a magnetic and octupolar structure, the ordered dipole moment being reduced compared to the ground state doublet value because of the octupolar contribution. It was further shown that this peculiar ordering competes with a ferromagnetic Coulomb phase, which gives rise to spin ice correlations above the ordering temperature \cite{Xu_2020, Leger_2021}, so that the system may be close to a $U(1)$ spin liquid state \cite{Huang_2014, Benton_2016, Benton_2020, Patri_2020}.

The aim of this work is to study the stability of this unconventional all-in--all-out state when disorder is introduced into the \ndzr\ compound. 
In standard magnets, disorder is known to progressively suppress the magnetic long range order. If too strong, exchange disorder for instance favors a spin glass ground state. When the magnetic species is substituted by a non magnetic element, the order survives up to a certain level called the percolation threshold. The latter can be quite large (and depends on the lattice), so the ordering can stand strong disorder. In frustrated magnets, disorder has a more dramatic effect. The reason is that the energy landscape of frustrated systems comprises a macroscopic number of states that are very close in energy, or even fully degenerate in some special cases, such as nearest neighbor spin ice for instance. Any minute perturbation is then likely to favour a particular ground state. Several studies on pyrochlore materials have documented this physics, as for instance in Tb$_{2+x}$Ti$_{2-x}$O$_{7+y}$ \cite{Taniguchi_2013}, Er$_{2}$Ti$_{2-x}$Sn$_x$O$_{7}$ \cite{Shirai_2017} and Yb$_{2+x}$Ti$_{2-x}$O$_{7+y}$ \cite{Arpino_2017}. In Tb$_{2+x}$Ti$_{2-x}$O$_{7+y}$, a long range quadrupolar order is stabilized beyond a minute threshold of $x=0.005$, to the detriment of a spin liquid phase. In Er$_{2}$Ti$_{2-x}$Sn$_x$O$_{7}$, the Sn/Ti substitution induces a change of the magnetic structure. Yb$_{2+x}$Ti$_{2-x}$O$_{7+y}$ lives just at the edge between ferromagnetic and antiferromagnetic phases \cite{Robert_2015, Jaubert_2015, Scheie_2020}; achieving a very high sample quality was necessary to eventually observe the ferromagnetic long range order. In \ndzr, it was shown that off-stoichiometry in the Nd/Zr ratio has an impact on the lattice parameter and magnetic properties, and is at the origin of the different ordering temperatures observed between polycrystalline and single crystal samples \cite{Zoghlin_2021}.

In the present study, different kinds of disorder are considered: on the one hand, substitution on the magnetic site, replacing neodymium by lanthanum, and on the other, substitution on the zirconium (non-magnetic) site. Two cases are envisaged: substitution by titanium, and another original route, which consists of an entropy stabilized mixing of five different cations at the zirconium site. In the latter case, the global pyrochlore structure is preserved but the disorder is maximum on the $T$ site, which is expected to strongly affect the magnetic properties. All those different substitutions contribute to modifying the environment and exchange interactions between \nd\ ions. 

We show that the all-in--all-out state is very robust and withstands disorder and / or dilution up to very large levels. This paper is organized as follows. The first section focuses on experimental techniques and the sample synthesis. The next sections describe the crystalline structures, crystal field schemes and magnetic ground states. We then investigate the low energy excitations, propose effective interactions parameters and briefly present the in field evolution of the structure. We finally discuss the influence of the different types of disorder, by comparing our results with the properties of previously reported Nd-based pyrochlore compounds.

\section{Experimental techniques}
\label{Experimentaltechniques}

\subsection{Sample synthesis}
Three types of samples were synthesized for this study. The \ndzrx\ samples are the same as studied in Ref. \onlinecite{Leger_2021b}, and were also grown as single crystals. The \ndzry\ samples were synthesized  in polycrystalline form using solid state reactions between starting powders Nd$_2$O$_3$, La$_2$O$_3$ and ZrO$_2$ in a stoichiometric mixture \cite{Ciomaga_2015b}. The powders were mixed and pressed in pellets to facilitate the chemical reaction. To eventually obtain \ndzry\ compounds, several heat treatments were necessary. Finally, three samples were obtained: Nd$_{1.8}$La$_{0.2}$Zr$_2$O$_7$ (10 \%-substituted sample), Nd$_{1.6}$La$_{0.4}$Zr$_2$O$_7$ (20 \%-substituted sample) and Nd$_{1.2}$La$_{0.8}$Zr$_2$O$_7$ (40 \%-substituted sample). They were then characterized by X-ray diffraction measurements to check the purity of the phase.

The NdMox (\ndmox) samples were synthesized, as described in Ref.~\onlinecite{Vayer_2022}.

\subsection{Bulk magnetic measurements}

The ``high" temperature magnetization measurements were performed on a Quantum Design MPMS VSM-SQUID magnetometer from 2 to 300 K. The powder samples were packed in cellophane with an approximate spherical shape. The data were corrected from demagnetizing effects with the demagnetization factor for a sphere.

Magnetization and AC susceptibility measurements were performed down to 80 mK using a superconducting quantum interference device (SQUID) magnetometer equipped with a $^3$He-$^4$He dilution refrigerator developed at the Institut N\'eel-CNRS Grenoble \cite{Paulsen01}. The powder samples were placed in small rectangular pouches made from thin copper sheets and covered with Apiezon grease to favor proper thermalization. The NdMox polycrystalline sample was pressed into a parallelepiped shape. 

The specific heat of the NdMox sample was measured with a Quantum Design PPMS equipped with a $^3$He system. To estimate the lattice contribution, a YMox (Y$_2[$(TiZrHfSn)$_{0.205}$Ge$_{0.18}]_2$O$_7$) sample was also measured.

\subsection{Powder neutron diffraction}

For the \ndzrx\ and \ndzry\ samples (except the 40 \% substituted-sample), powder neutron diffraction experiments were carried out at G4.1 (LLB-Orph\'ee, France) operated with $\lambda = 2.43$ \AA\ and equipped with a $^3$He-$^4$He dilution refrigerator. A dedicated vanadium sample holder was used, loaded with $^4$He pressure of about 10 bars to ensure thermalization. 

Neutron powder diffraction data for the NdMox and \ndlaC\ samples were acquired with the high resolution neutron powder diffractometer HRPT \cite{Fisher_2000} at Paul Scherrer Institute. The data for NdMox were collected at 300 K with neutron wavelengths $\lambda=1.494$ and 1.1545 \AA\ in the high intensity mode of the instrument and for \ndlaC\ with $\lambda=1.044$ and 2.450 \AA\ in the medium resolution mode of the instrument. The \ndlaC\ sample has been found to contain ~4\% wt. of a Nd(OH)$_3$ impurity, which was included into the refinements.

\subsection{Inelastic neutron scattering}

For the \ndzry\ samples with $y=10$ and 20 \%, inelastic neutron scattering measurements were performed on the time-of-flight spectrometer PANTHER (ILL, France) \cite{Fak_2022} to determine the crystal electric field (CEF) excitations \cite{doi_panther}. The spectrometer was operated with an incident energy $E_i=50$ meV, yielding a good compromise in terms of energy resolution and flux. The samples were placed in an orange cryostat capable of cooling down to 1.5 K, and data were taken up to 200 K with 50 K steps. The scattering from the empty sample holder was measured and subtracted from the raw data. 

Similar experiments were performed on the triple axis spectrometers 2T (LLB-Orph\'ee, France) for \ndzrx\ samples, and at EIGER (PSI, Switzerland) \cite{Stuhr_2017} for the NdMox and \ndlaC\ samples. 
We used a constant final wave vector $k_f=2.662$~\AA$^{-1}$, yielding an energy resolution at the elastic line of the order of 1 meV. Several $Q$-scans were measured with $1 < Q < 6$ \AA$^{-1}$ and the spectra were then assembled to produce intensity maps as a function of $Q$ and energy transfer $\omega$. 
A pyrolytic graphite (PG) filter was used to eliminate spurious signals due to harmonics in the incident beam being scattered from the elastic incoherent response of the sample and sample holder.  Depending on the strength of such scattering, the length of the filter may not be sufficient to completely eliminate this effect.  At 2T we also measured the aluminum can sample holder under the same conditions to eliminate background and spurious scattering as far as possible by subtraction.  When measuring \ndlaC\ at EIGER we used a long filter ($\sim 70$ mm), which also minimizes the spurious scattering. For the measurements of NdMox at EIGER we used the normal short filter ($\sim 36$ mm) and time constraints prevented the measurement of the empty holder.  We have therefore fitted the spurious peaks at 17.5 meV ($2k_i=3k_f$) and 43.5 meV ($k_i=2k_f$) using Gaussian profiles and subtracted these from the data. This treatment works well for the peak at 17.5~meV but is less effective for the peak at 43.5 meV.

To determine the low energy magnetic response, typically below 1 meV, we used the time of flight spectrometers SHARP (LLB CRG @ILL, France) \cite{doi_sharp} with an incident wavelength of 5.1 \AA\ for the \ndzry\ samples with $y=10$ and 20 \%, and FOCUS (PSI, Switzerland) with incident wavelengths of 5 and 5.75 \AA , for the NdMox sample. The samples were placed in a Cu sample holder and inserted into a dilution refrigerator, reaching temperatures below 100 mK. The \ndzry\ spectra present a Gaussian profile at the elastic position, due to incoherent and Bragg scattering. To extract the low temperature inelastic magnetic signal, data measured at 5 K were used as a reference. Nevertheless, because a low energy quasielastic signal is still visible at this temperature, amplified by the detailed balance factor, these data cannot be directly used for subtraction. The spectrum measured at 5 K was thus reduced to a series of energy scans, for each $Q$ position. Each scan was then fitted in a small energy interval $-\omega_1 < \omega< \omega_1,~\omega_1\approx$ 0.4 meV to a function consisting of a Lorentzian function of width $\gamma$ on top a Gaussian profile of width $\sigma$:
$$I_{\rm model}= I_o  e^{-4 \ln{2}\times  (\omega/\sigma)^2 }+ (1+1/(e^{\omega/T}-1))   \frac{A}{T} \frac{\omega}{(\omega^2+\gamma^2 )}$$
The Gaussian and Lorentzian contributions represent the elastic and quasielastic responses respectively. We then considered the quantity: $$S(Q,\omega)=I_{\rm exp} (Q,\omega,T)-I_o  e^{-4 \ln{2}\times (\omega/\sigma)^2}$$

The low energy excitations of the \ndzrx\ samples were probed on single crystals on the IN5 (ILL, France) time-of-flight spectrometer. Data for the $x=2.5$ \% single crystal sample \cite{doi_IN5_1} are reported in Ref. \onlinecite{Leger_2021}. The $x=10$~\% sample was measured in the $(100)-(010)$ scattering plane with incident wavelengths of 6, 8 and 10 \AA\ \cite{doi_IN5_2}.
 

\section{Structural properties}
\label{Structuralproperties}

\begin{table*}[th!]
\begin{tabular}{*{9}{c}}
\hline
\hline
\multirow{2}{*}{Sample} 	&Ti/La occupancy 	& \multirow{2}{*}{$x$(O2)} & Lattice parameter	&$T_{\rm CW}$	& $\mu _{\rm eff}$ &  $T_{\rm N}$	& $m_{\rm ord}$ 	& $\epsilon_0$ \\ 
					& [\%] 			&					& [\AA] 			& [mK] 		& [$\mu _{\rm B}$] &  [mK]  	 	& [$\mu _{\rm B}$] 	& [meV] \\ \hline
\ndzr					& - 	 			& 0.3366			 	& 10.69  			& 175  		& 2.45  			& 370 		& $1.15 \pm 0.05$ 	& 0.075 \\ 
\ndtiA				& 2.8 			& 0.3360				& 10.67			& 145  		& 2.46 			& 420 		& $1.36 \pm 0.04$ 	& 0.07 \\
\ndtiB   				& 11.2 			& 0.3359			 	& 10.64 			& 180  		& 2.49 			& 430 		& $1.33 \pm 0.03$ 	& 0.075 \\ 
\ndlaA\ 				& 9.6   			& 0.3346				& 10.69 			& 275 		& 2.61 			& 330 		& $1.27 \pm 0.06$ 	& 0.05 \\
\ndlaB\ 				& 20.5  			& 0.3344				& 10.70  			& 220 		& 2.75 			& 240 		& $1.25 \pm 0.06$ 	& 0.025 \\ 
\ndlaC				& 40	  			& 0.3330				& 10.73			& 170		& 2.61 			&  -			& -  				& - \\
NdMox				& -  	  			& 0.330				& 10.56			& 30 			& 2.60 			& 550 		& $1.22 \pm 0.09$ 	& 0.11 \\\hline 
\hline
\end{tabular}
\caption{\label{Table_charac} Structural (at room temperature) and magnetic parameters for the investigated polycrystalline samples. The values for the pure compound have been updated compared to Refs. \cite{Ciomaga_2015,Lhotel_2015} from new measurements on D1B and D2B (ILL). For all compounds, the space group is $Fd\bar{3}m$ (\#227), the atoms are: Nd(La) - $16d$ $(1/2,1/2,1/2)$ position, Zr(Ti,$M$) - $16c$ $(0,0,0)$ position, O1 - $8b$ $(3/8,3/8,3/8)$ position, and O2 - $48f$ $(x,1/8,1/8)$ position. The Curie-Weiss temperatures and effective moments have been obtained from Curie-Weiss fits between 4 and 50 K for all samples. Note that the obtained Curie-Weiss temperature is affected by the fitting range, as well as the demagnetization effects, which may explain the differences with some values reported in the literature. For \ndlaC, no ordering temperature could be observed down to 75 mK. For \ndzr, \ndtiA\ and \ndtiB, the gap values $\epsilon_0$ were obtained for single crystal samples.}
\end{table*}

\begin{figure}[b!]
\centering{\includegraphics[width=7.5cm]{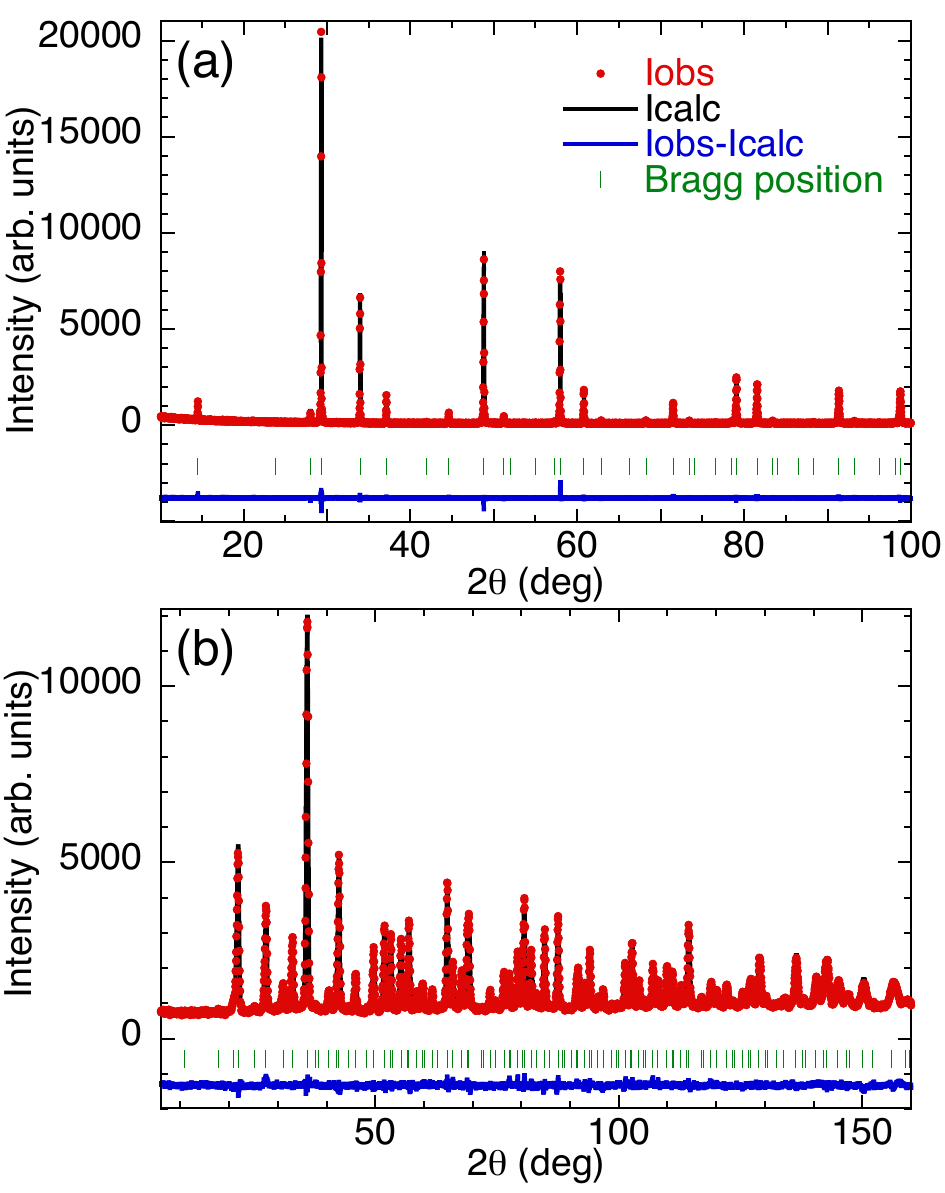}}
\caption{(a) X-ray and (b) neutron diffractograms (taken at HRPT) measured on NdMox (red points) at room temperature together with the refinement with a pyrochlore structure (black line). The difference between measurements and calculations is shown in blue. Green bars show the Bragg positions.}
\label{fig_struct_ndmox}
\end{figure}

As shown below, all the compounds discussed in this article were found to crystallize in the $Fd{\bar 3}m$ space group. Rietveld refinement using {\sc FullProf} \cite{Fullprof} was performed to refine the lattice parameter as well as the position of the $48f$ oxygen atom, which is the single positional parameter in the pyrochlore structure. For the substituted samples, the lanthanum occupancy could not be determined with a good accuracy as neodymium and lanthanum elements have very similar scattering lengths.

The \ndzrx\ samples structural properties are discussed in Ref. \cite{Leger_2021}. 
\ndzry\ structural properties at ambient temperature were determined from X-ray diffraction for $y=10$ and 20 \% and neutron scattering for $y=40$ \%. The percentages of lanthanum occupancy are consistent with the synthesis expectations. The NdMox structure was refined from X-ray and neutron diffraction measurements (see Figure \ref{fig_struct_ndmox}). Accounting for a possible anti-site disorder, which could occur between the Nd$^{3+}$ and the Sc$^{3+}$ ions, did not improve the refinement.

The obtained structural properties are summarized in Table \ref{Table_charac}. They show that the lattice parameter and the oxygen position are little affected by substitution in \ndzr\ compounds. The lattice parameter seems to slightly decrease with the introduction of Ti, and to increase with the introduction of La, which is consistent with Ref. \onlinecite{Harvey_2005}. In the case of NdMox, the lattice parameter is significantly reduced compared to the pure \ndzr\ compound. \\

\section{Crystal electric field}
\label{Crystalelectricfield}

\begin{figure*}[th!]
\includegraphics[width=14cm]{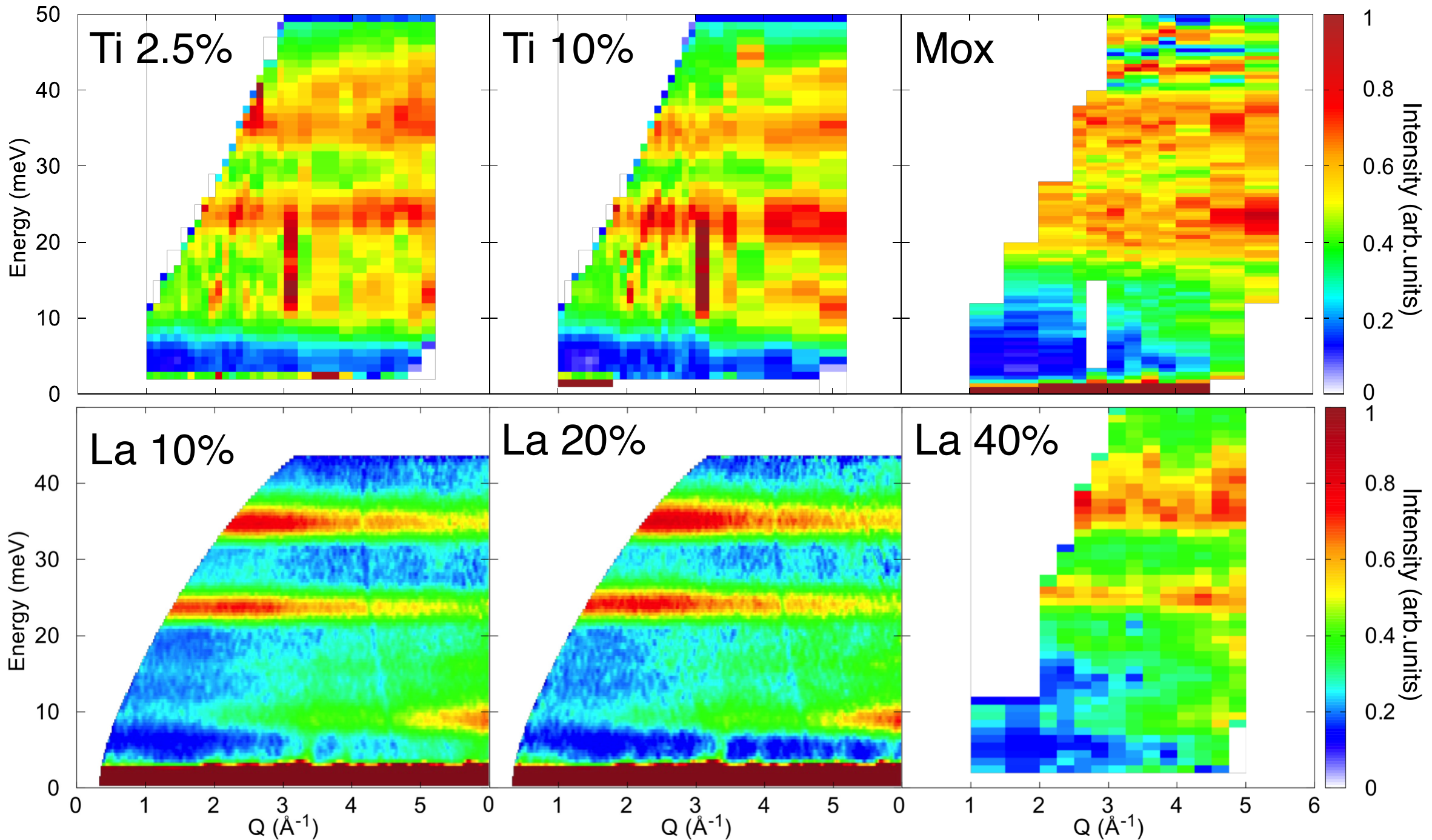}
\caption{\label{fig_CEF} Inelastic neutron scattering spectra at $T=3.8$ K in \ndzrx\ with $x=2.5$ and 10 \% measured on 2T, and at $T=1.5-1.6$ K for \ndzry\ with $y=10$ and 20 \% measured on PANTHER and  \ndmox\ and \ndzry\ with y=40 \% measured on EIGER. $Q$-cuts of these spectra are shown in Appendix \ref{appendix_cef}.}
\end{figure*}

The inelastic neutron scattering spectra recorded at low temperature for all samples are shown in Figure \ref{fig_CEF}. For the Ti and La-substituted compounds, the spectra are characterized by two lines, centered around 23-25 and 34-38 meV, the latter peak being broader. In \ndzry\ with $y=10$ and 20 \%, the instrument resolution allows us to distinguish a shoulder on the higher energy mode that can be due to the presence of two very close modes. The energy widths are almost identical (considering the instruments resolution), whatever the doping (see also Table \ref{Table_CEFmeas}). For the high entropy compound, the situation is very similar, with excitations centered at about the same energies. Their half width at half maximum, however, is much larger, about 6-8~meV. Phonon excitations, characterized by their larger intensity at high $Q$, are also visible in all samples around 10-15~meV. 

\begin{table}[h!]
\begin{tabular}{*{6}{c}}
\hline
\hline
Sample & $E_1$ & $\Delta_1$ & $E_{2}$ & $E_{3}$ & $\Delta_{2,3}$  \\
\hline
\ndzr\ \cite{Xu_2015}		& 23.4	& - 		& 34.4	 & 35.7 		        	& - \\
\ndtiA *				& 23.2   	& 3.9		& \multicolumn{2}{c}{35.9} 	& 5.9 \\ 
\ndtiB *			   	& 22.8 	& 4.8 	& \multicolumn{2}{c}{36.3} 	& 5.7 \\
\ndlaA $^{\dagger}$ 		& 23.8	& 1.9		& 34.5 	& 36.5 			& 1.5 \\			
\ndlaB $^{\dagger}$ 		& 24.2	& 1.9		& 35.0 	& 38.4 		         & 2.4 \\		
\ndlaC *                    		& 25.4	& 2		& \multicolumn{2}{c}{37}    	& 6  \\
NdMox *				& 21 		& 6		& \multicolumn{2}{c}{36} 		& 8	 \\ 
\hline
\hline
\end{tabular}
\caption{\label{Table_CEFmeas} Energy and half-width at half maximum of the CEF levels measured on the different Nd-based pyrochlore compounds, in meV. Compounds marked with a * were measured on triple axis spectrometers, where the energy resolution is expected to be about 3.5 and 5 meV at 23 and 36 meV respectively. Compounds marked with a $^\dagger$ were measured on PANTHER with a resolution of about 2 meV \cite{Fak_2022}.}
\end{table}

These results should be discussed in the perspective offered by the reference compound \ndzr. Its crystal field scheme has been studied in references \cite{Lhotel_2015, Ciomaga_2015, Xu_2015}, and has levels at about 23, 34, 35 and 104 meV. All the studied compounds thus have very similar schemes compared to the ``pure" compound (the 104 meV level could not be observed within the energy range of our measurements). While this result could have been expected for the La-substitution, it is more surprising for the other compounds, namely \ndzrx\ and NdMox, for which the local environment of the \nd\ ion is directly perturbed by the substitutions and the disorder on the $T$ site. 

To estimate how the small variations in the CEF level energies reflect in the crystal field coefficients, we have analyzed the data using the multiplet description of the CEF levels previously used in Refs. \onlinecite{Ciomaga_2015,Xu_2015}. In this case, the crystal field Hamiltonian is written \cite{Wybourne_1965}:
\begin{eqnarray}
V_{\rm CEF} &=& \sum_{k=0,2,4,6} B_{k,0} C_{k,0} \\
& &+\sum_{k=2,4,6}\sum_{m=1}^k B_{k,m} \left(C_{k,-m}+(-1)^k C_{k,m}\right)
\end{eqnarray}
The $C_{k,m}$ operators, usually called Wybourne operators, are defined as $C_{k,m} = \sqrt{\frac{4\pi}{2k+1}}Y_{k,m}$, where $Y_{k,m}$ are the spherical harmonic functions. The Wybourne coefficients $B_{k,m}$ are considered as empirical coefficients. The $B_{k,m}$ parameters which are not zero in the $D_{3d}$ symmetry were computed using the energy level values detailed in Table~\ref{Table_CEFmeas}. The obtained values are close to the ones obtained in the pure compound and differ only by about 5 \% (see Appendix \ref{appendix_cef}). 

In the \ndzrx\ and \ndzry\ samples, the levels are not significantly broadened, so one can assume that the \nd\ crystal field states are homogeneous within the sample. In the case of NdMox, disorder induces a larger distribution that could correspond to local variations up to about 35~\% of the $B_{k,m}$ values  (see Appendix \ref{appendix_cef}). This may affect locally the properties of the Nd$^{3+}$ magnetic ion, and in particular the value of the magnetic moment. Nevertheless, it is worth noting that the dipolar-octupolar character of the ground doublet, which confers to \ndzr\ its peculiar properties, remains present in the obtained wavefunctions. This is especially interesting since it makes possible the study of the disorder impact on the collective properties of the Nd-substituted compounds, without considering as a first approximation the local variations in the Nd$^{3+}$ single ion properties.

\section{Magnetic ground state}
\label{Section_GS}

\begin{figure}[h!]
\centering{\includegraphics[width=7.5cm]{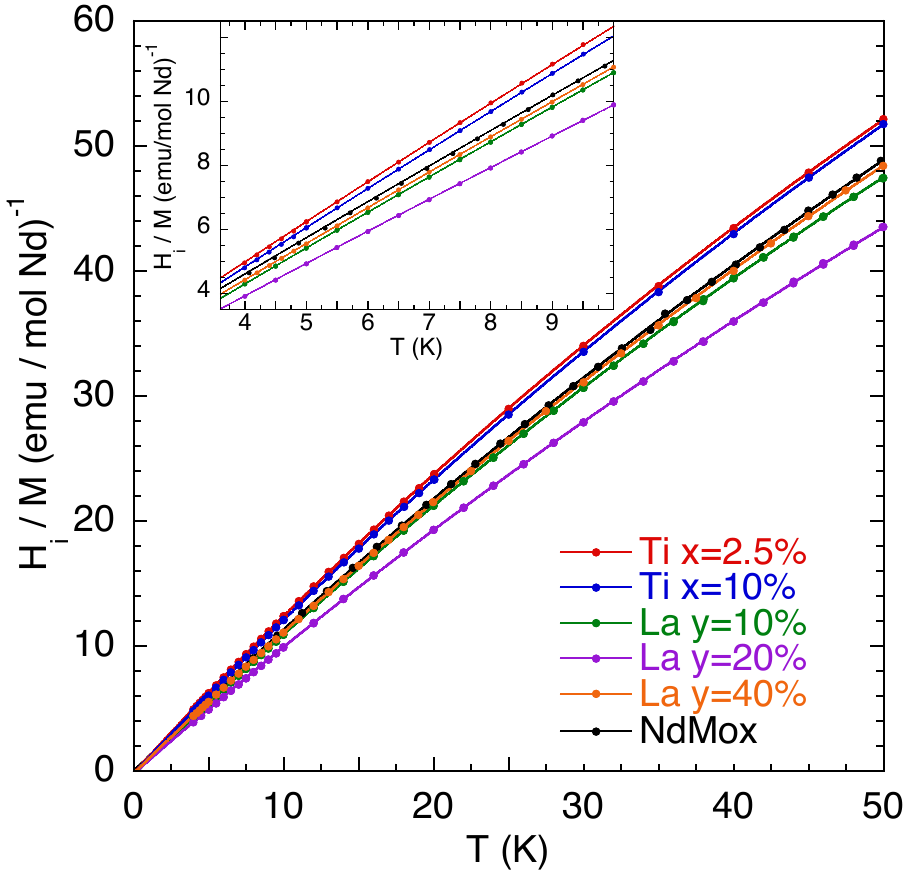}}
\caption{Inverse approximate susceptibility $H_i/M$ vs temperature $T$ between 4 and 50 K with an applied field $H\approx 1000$~Oe, where $H_i$ is the internal field obtained after demagnetization corrections. The inset shows a zoom in these curves between 4 and 10 K. Lines correspond to fits using a Curie-Weiss law with a Van Vleck component. The resulting Curie-Weiss temperatures and effective moments are summarized in Table~\ref{Table_charac}.}
\label{fig_CW}
\end{figure}

For all samples, magnetization measurements performed between 2 and 300 K show a continuous behavior, excluding the possible existence of magnetic transitions due to magnetic impurities. Estimations of the Curie-Weiss temperature $T_{\rm CW}$ and the effective moment $\mu_{\rm eff}$ for all samples were performed by fitting the inverse susceptibility $\chi^{-1}$ (obtained from magnetization data, assuming that $M$ is linear in $H$) as a function of temperature using a Curie-Weiss law $\chi=C/(T-T_{\rm CW})$, where $C=\mu_0 {\cal N}_{\rm A} \mu_{\rm eff}^2/(3k_{\rm B}$), as shown in Figure \ref{fig_CW} ($\mu_0$ is the vacuum permeability, ${\cal N}_{\rm A}$ the Avogadro constant, and $k_{\rm B}$ the Boltzmann constant). A Van Vleck contribution was also included in the fit and gives for all samples values between 0.0035 and 0.0039 emu/mol Nd, in agreement with the literature \cite{Ciomaga_2015, Xu_2015}. The obtained values are summarized in Table \ref{Table_charac}. All Curie-Weiss temperatures are positive, similarly to the pure compound, which suggests effective ferromagnetic interactions. Although the determination of $T_{\rm CW}$ is only approximate (the fit is very sensitive due to the small values, and $T_{\rm CW}$  is affected by demagnetization corrections), the following tendencies are observed: in the Ti-substituted samples, the Curie-Weiss temperatures are in the same range or smaller than in the pure sample.
Surprisingly, in the La-substituted samples, where the magnetic Nd$^{3+}$ ions are diluted, Curie-Weiss temperatures are larger than in the pure sample, reaching the pure sample value only for the 40 \% substituted sample. In NdMox, $T_{\rm CW}$ is strongly reduced and is close to zero.

It is also interesting to discuss the obtained effective moments. They are all larger than in the pure compound, reaching 2.75 $\mu_{\rm B}$ in the \ndzry\ compound with $y=20$ \%. In magnetization as a function of field measurements, the magnetization is almost saturated at 100~mK and 8 T for all samples (see Appendix \ref{comp_data}). The magnetization reaches values corresponding to half of the moment value, which is the expected result for polycrystalline pyrochlore compounds with local $\langle 111 \rangle$ anisotropy \cite{Bramwell_2000}. A finite slope is nevertheless observed at high field. 
While effective moment values between 2.5 and 2.6 $\mu_{\rm B}$ are consistent with the crystal electric field parameters deduced from neutron scattering, the high value of 2.75~$\mu_{\rm B}$ obtained for the 20~\% La-substituted compound does not seem compatible with the INS results. It may be due to an overestimation of the La substitution rate in the sample, which would naturally induce an overestimated average Nd magnetic moment.

\begin{figure}[h!]
\centering{\includegraphics[width=8cm]{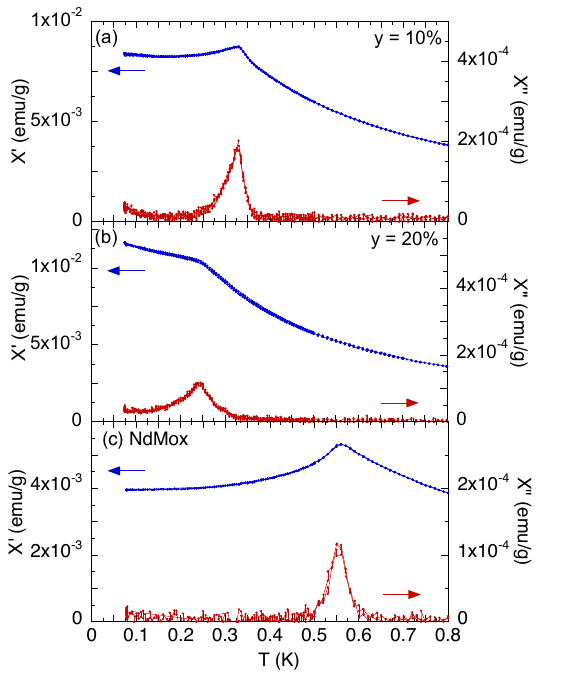}}
\caption{AC susceptibility, in-phase $\chi'$ (blue, left) and out-of-phase $\chi''$ (red, right), as a function of temperature measured with an applied field $H_{\rm ac}=1.8$ Oe and $f=0.57$ Hz for \ndzry\ with (a) $y=10$ \%, (b) $y=20$ \% and for (c) NdMox. }
\label{fig_Xac}
\end{figure}

The existence of a magnetic transition below 1 K was probed through AC susceptibility and magnetization measurements. An antiferromagnetic transition is observed in all samples except the most diluted \ndzry\ sample. This transition manifests as a maximum in $\chi'$ and $\chi''$, whose position does not depend on frequency (see Figure \ref{fig_Xac} and Appendix \ref{comp_data}), as well as a peak in magnetization. For NdMox, a peak in the specific heat (see Appendix \ref{comp_data}) confirms the existence of a transition (note that it was not possible to cool down to a low enough temperature to reach the transition temperature of the other samples with our specific heat set-up). 

The N\'eel temperatures of the La-substituted compounds are smaller than in the pure powder \ndzr\ (330 and 240 mK for $y=10$ and 20 \% respectively), which may be expected due to the dilution of the magnetic Nd$^{3+}$ ions. The transition is even suppressed in the measured range ($T>75$ mK) for the $y=40$~\% sample for which a paramagnetic behavior is observed down to the lowest temperature. Conversely, in other substituted samples, the N\'eel temperature $T_{\rm N}$ is larger (see Table \ref{Table_charac}), which indicates that disorder has reinforced the antiferromagnetic ordering. It is worth noting that the ratio $T_{\rm N}/ T_{\rm CW}$ is much larger in NdMox and to a lesser extent in \ndzrx\ than in \ndzr\ and \ndzry\ samples, pointing to different magnetic couplings as discussed below. In all samples, the susceptibility peaks are wider than in the pure sample, suggesting a small distribution of the transition temperatures. In addition, in the \ndzry\ samples, and especially in the $y=20$ \% one (see Figure \ref{fig_Xac}(b)), the real part of the susceptibility keeps increasing at low temperature, which may be due to the persistence of disordered Nd magnetic moments even below the N\'eel temperature. 

\begin{figure}[h!]
\includegraphics[width=8cm]{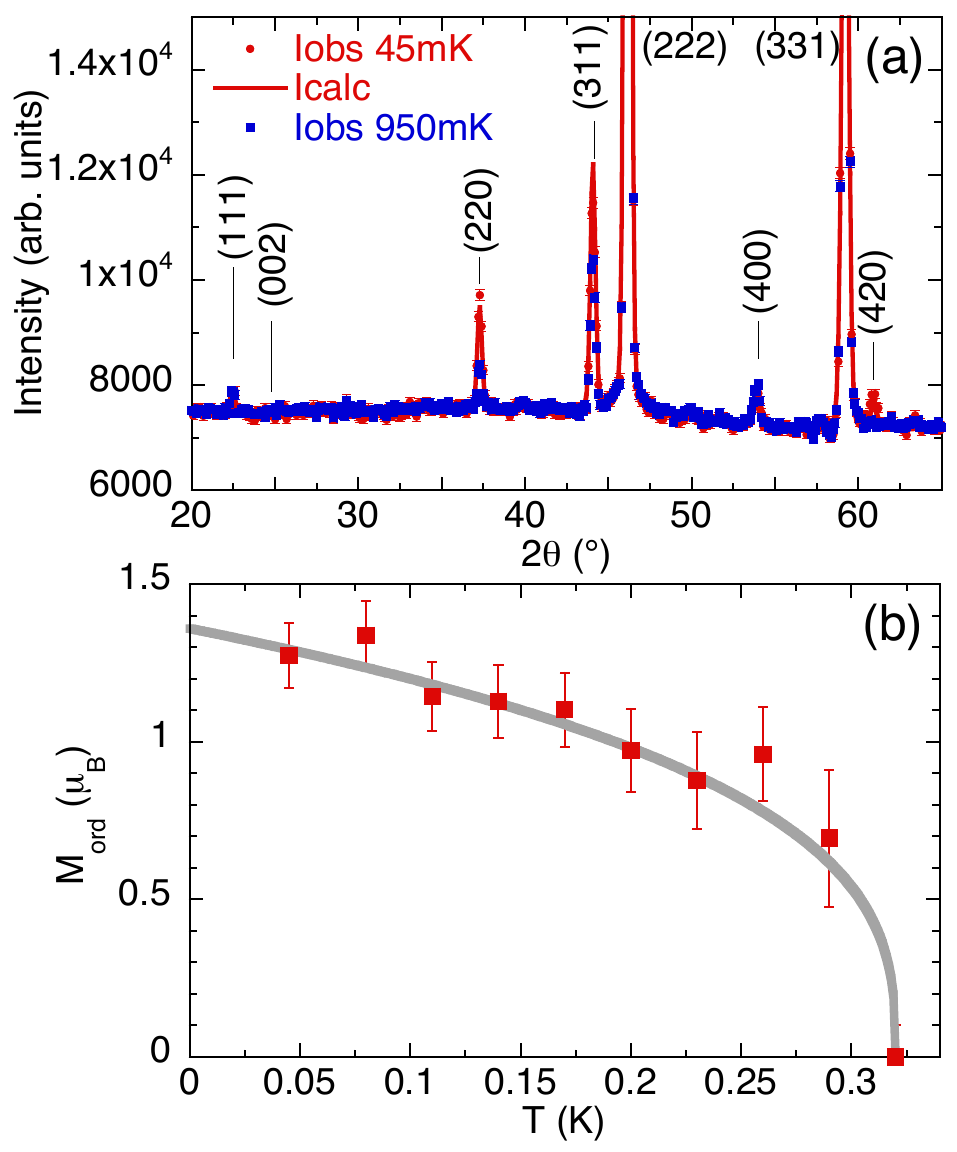}
\caption{\label{fig_diff_NdLa} (a) Neutron diffractogram measured on G4.1 for \ndzry\ with $y=10$ \% at 45 mK (red dots) and 950~mK (blue squares).  The red line is a refinement to an all-in--all-out magnetic structure. (b) Ordered magnetic moment as a function of temperature determined from the diffractogram. The line is a guide to the eye.}
\end{figure}

\begin{figure}[h!]
\includegraphics[width=8cm]{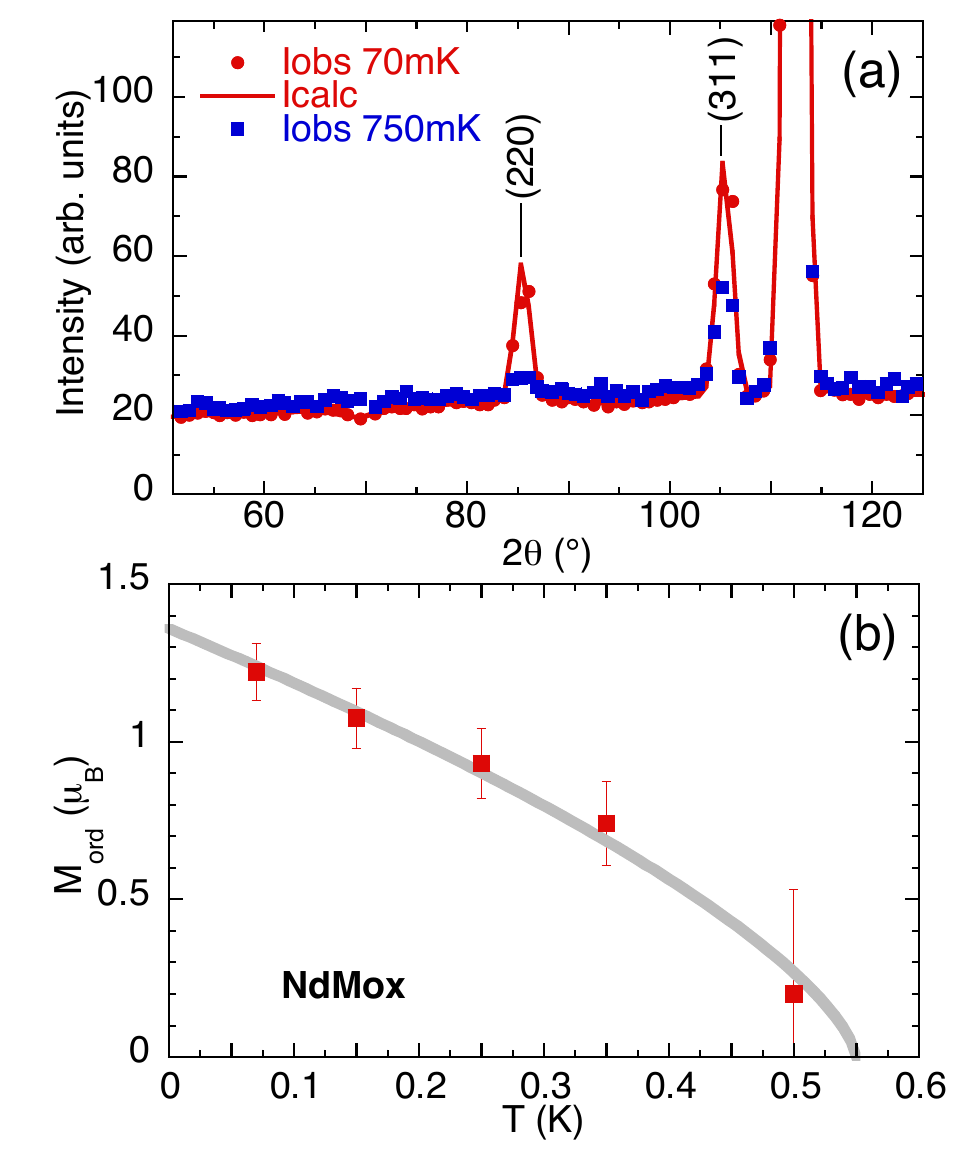}
\caption{\label{fig_diff_NdMox} (a) Neutron diffractogram measured for NdMox by integrating the inelastic data measured on FOCUS between -0.1 and 0.1 meV at 70 mK (red dots) and 750 mK (blue squares). The red line is a refinement to an all-in--all-out magnetic structure. (b) Ordered magnetic moment as a function of temperature determined from the diffractogram. The line is a guide to the eye.}
\end{figure}

The magnetic structure associated with this magnetic ordering was determined from neutron diffraction measurements using a {\sc Fullprof} refinement \cite{Fullprof}. Examples of diffractograms measured on \ndzry\ and NdMox are shown in Figures \ref{fig_diff_NdLa}(a) and \ref{fig_diff_NdMox}(a), where it can be seen that the magnetic structure has a ${\bf k}={\bf 0}$ propagation vector. (220) and (311) peaks have a strong magnetic intensity at low temperature, while the (002) peak has not, which is characteristic of the all-in--all-out structure \cite{Petit_2017} observed in the pure \ndzr\ compound \cite{Lhotel_2015}. The temperature dependence of the ordered magnetic moment $m_{\rm ord}$ could also be determined (see Figures \ref{fig_diff_NdLa}(b) and \ref{fig_diff_NdMox}(b)). At low temperature, the obtained ordered moment is always larger than in the pure compound (see Table \ref{Table_charac}) and reaches about 1.35 $\mu_{\rm B}$ for the Ti-substituted samples. It is nevertheless far from the full moment of the Nd$^{3+}$ ion. These results thus suggest that the ground state of all the substituted compounds is similar to the pure compound, i.e. an antiferromagnetic all-in--all-out state in which the ordered moment has a dipolar-octupolar character \cite{Benton_2016}.

\section{Low energy excitations}
\begin{figure}[h!]
\includegraphics[width=8.5cm]{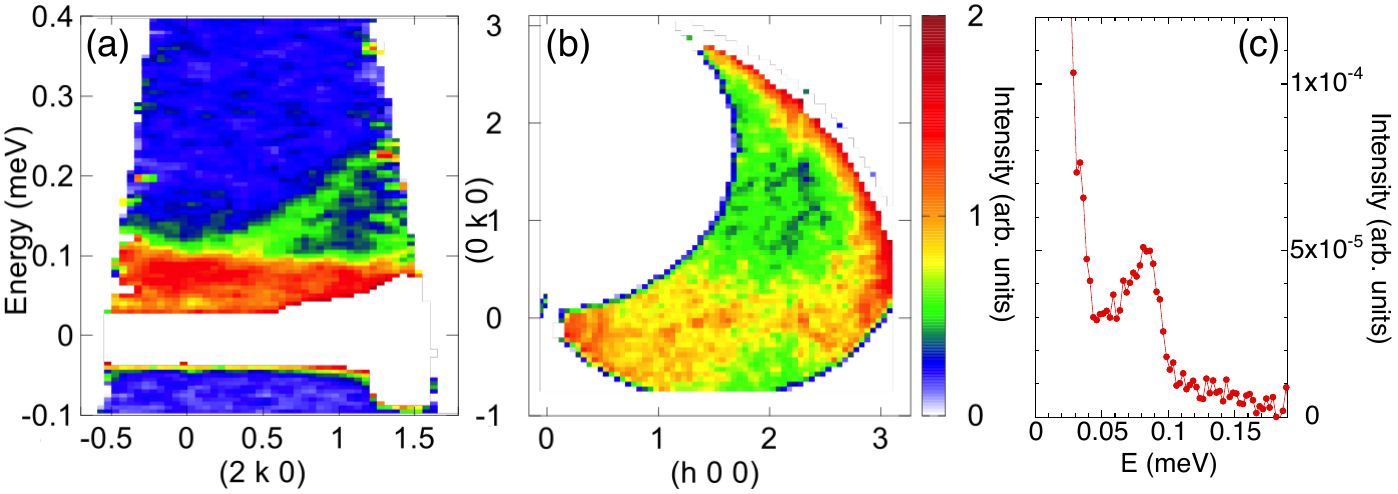}
\caption{\label{fig_IN5_NdZrTip10} Low energy excitations measured on \ndzrx\ with $x=10$ \% at 50 mK on IN5. (a) Spectra along $(2k0)$ showing the flat mode and the dispersive branches measured with $\lambda=6$ \AA. (b) Energy cut at $E=70 \pm 0.05\ \mu$eV showing the broadened pinch points measured with $\lambda=8$ \AA. (c) $Q$-cut measured with $\lambda=10$ \AA\ showing the persistence of a gap in the excitations. }
\end{figure}

A key feature of this dipolar-octupolar ordered state is the existence of low energy excitations with peculiar characteristics: flat mode with pinch points from which stem dispersive excitations. We have thus performed measurements to check whether these excitations are affected by disorder and substitutions. In the case of 2.5~\% titanium substituted single crystals, we have shown that the excitations remain almost intact \cite{Leger_2021}. In the $x=10$~\% sample, these excitations are still present at almost the same energy (the energy of the flat mode is about 70~$\mu$eV), but features are broadened, especially the pinch points (see Figure \ref{fig_IN5_NdZrTip10}).

\begin{figure}[h!]
\includegraphics[width=8cm]{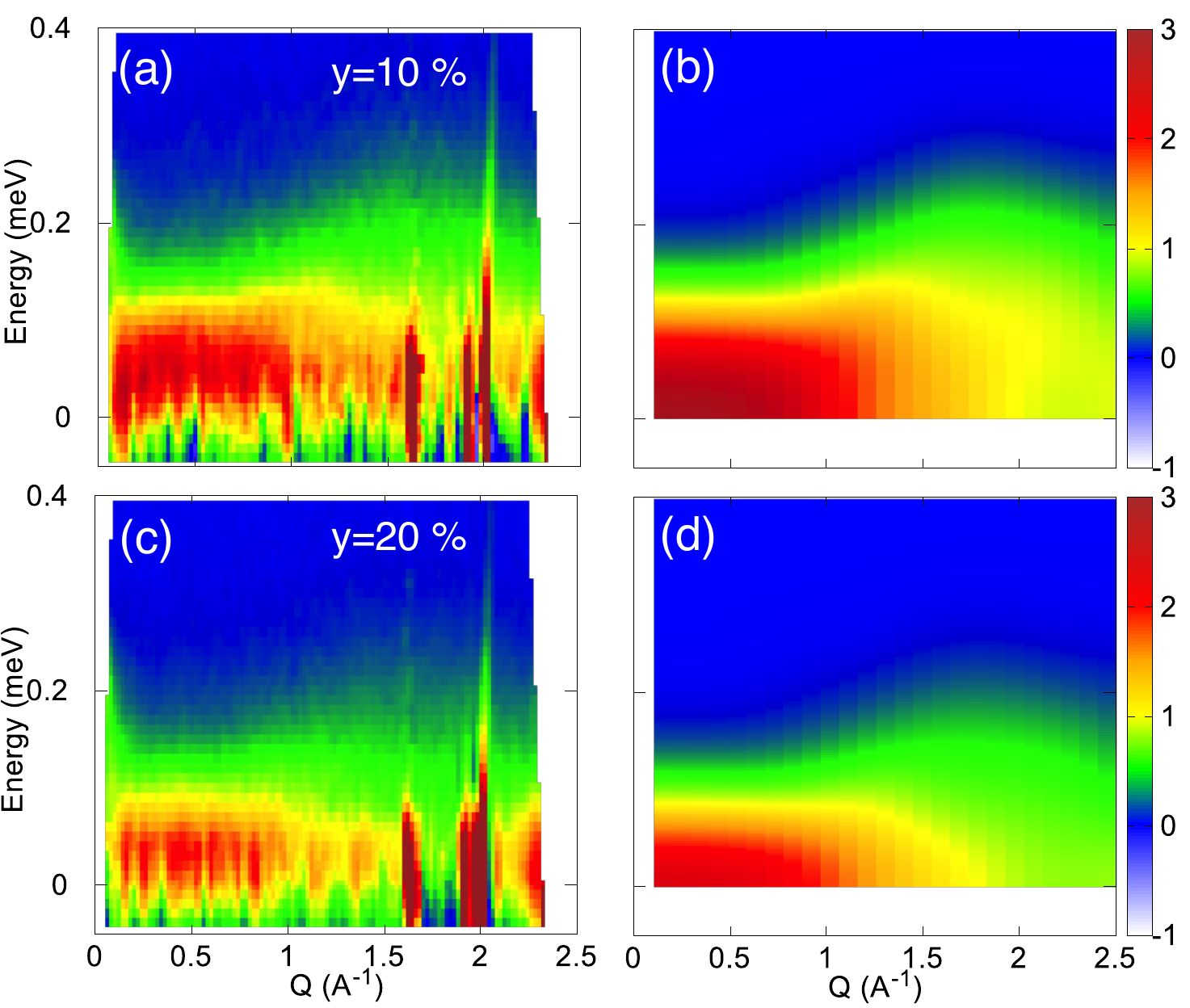}
\caption{\label{fig_IN6_NdLa} Low energy excitations of \ndzry\ measured at 50 mK on SHARP (a) for $y=10$~\% and (c) $y=20$~\%.The plotted data were obtained following the procedure explained in Section \ref{Experimentaltechniques}. (b)(d) Calculations performed with the parameters detailed in Table \ref{table_J}, and including a 60~$\mu$eV broadening (full width at half maximum) for each spin wave mode in order to mimic the experimental resolution.}
\end{figure}

\begin{figure}[h!]
\includegraphics[width=7.5cm]{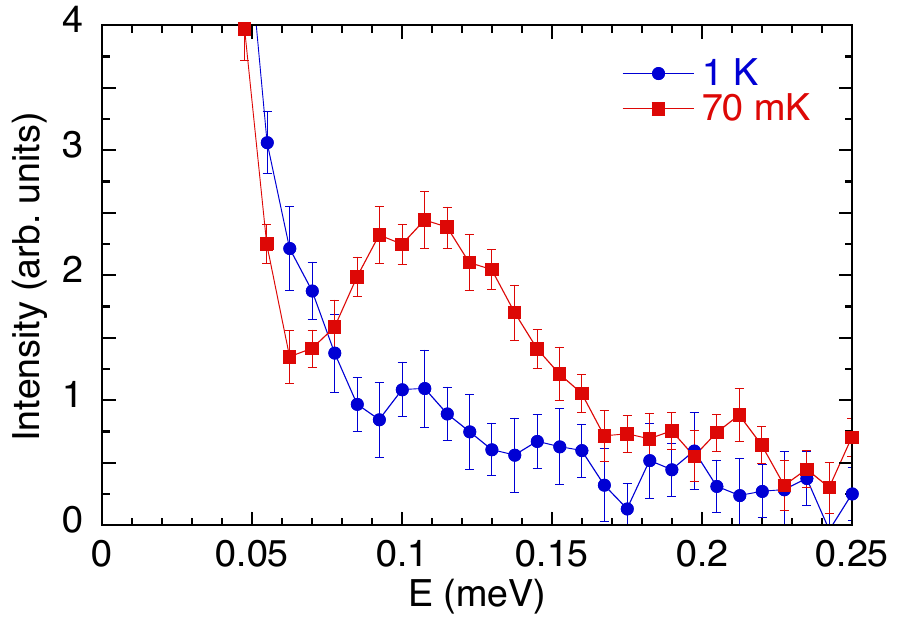}
\caption{\label{fig_FOCUS_NdMox} Energy scan in NdMox measured on FOCUS at 70~mK (red squares) and 1 K (blue dots), in which the signal has been integrated over the $Q$-range $[0.5; 1.6]$ \AA$^{-1}$.}
\end{figure}

The lack of single crystals for \ndzry\ and NdMox prevents the observation of pinch points in the excitations of these samples. Nevertheless, inelastic measurements show in all measured samples the persistence of a flat mode as can be seen in Figures \ref{fig_IN6_NdLa} and \ref{fig_FOCUS_NdMox}. The energy $\epsilon_0$ of the flat mode is decreased to about 50 and 25 $\mu$eV in \ndzry\ with $y=10$ and 20 \% respectively, while it is increased in the NdMox compound ($\epsilon_0 \approx 110\ \mu$eV). Dispersive excitations are hardly seen in the NdMox sample but are present in the \ndzry\ and qualitatively similar to the ones of the pure \ndzr\ compound; still, features are broadened, so that the gap looks almost suppressed in \ndzry\ with the present resolution (70 $\mu$eV).

These results confirm the conclusion reached in the previous section: the magnetic state in the substituted compounds is of the same nature as in the parent compound \ndzr. Disorder and / or dilution seem to renormalize the parameters (ordering temperature, ordered magnetic moment, gap) and to induce some distribution but without affecting strongly the ground state itself. 

\section{Determination of the Hamiltonian parameters}
To go further in the understanding of the role of disorder, it is necessary to quantitatively estimate the microscopic parameters which govern the magnetic state, namely the magnetic couplings. 

CEF measurements show that in spite of the loss of symmetry induced by local disorder, the magnetic moment of the \nd\ ions is well described by a dipolar-octupolar wavefunction. As a result, the XYZ Hamiltonian, which was shown to be relevant for this kind of magnetic ions \cite{Huang_2014} can be used to investigate the magnetic couplings in the system. This Hamiltonian is written: 
\begin{equation}
{\cal H} = \sum_{\langle i,j \rangle}\left[{\sf J}_x \tau^x_i \tau^x_j + {\sf J}_y \tau^y_i \tau^y_j + {\sf J}_z \tau^z_i \tau^z_j + {\sf J}_{xz} (\tau^x_i \tau^z_j+\tau^z_i \tau^x_j) \right]
\label{Hxyz}
\end{equation} 
where $\tau^{x,y,z}_i$ are the components of the pseudo-spin and ${\sf J}_a$ the coupling parameters ($a=x,y,z,xz$). 
$\tau^z$ is oriented along the local $\langle 111 \rangle$ axes of the tetrahedra, and corresponds to the magnetic component probed by magnetization or neutron scattering (at low $Q$). The dipolar-octupolar character of the pseudo-spin implies that $\tau^x$ and $\tau^z$ components transform like a pseudo-vector (or a dipole) but that the $\tau^y$ component transforms as an octupole. 

The  ${\sf J}_{xz}$ term can be removed from Equation \ref{Hxyz} by rotating the $x$ and $z$ axes in the $({\bf x},{\bf z})$ plane by an angle $\theta$ \cite{Huang_2014, Benton_2016}: 
\begin{equation}
\begin{aligned}
{\cal H}_{\rm XYZ} = & \sum_{\langle i,j \rangle}\left[\tilde{\sf J}_x \tilde \tau^{\tilde x}_i \tilde \tau^{\tilde x}_j + \tilde{\sf J}_y \tilde \tau^{\tilde y}_i \tilde \tau^{\tilde y}_j + \tilde{\sf J}_z \tilde \tau^{\tilde z}_i \tilde \tau^{\tilde z}_j \right] \\
& {\rm with} \quad \tan (2 \theta)=\frac{2 {\sf J}_{xz}}{{\sf J}_x - {\sf J}_z}
 \label{Hxyz_tilde}
\end{aligned}
\end{equation} 
where the $\tilde{\sf J}_a$ ($a=x,y,z$) are the interaction parameters in the rotated frame $({\bf \tilde{x}},{\bf \tilde{z}})$ (see Appendix \ref{appendix_XYZ}).

When an all-in--all-out state is stabilized as in the compounds studied here, the $\theta$ angle is simply related to the ordered magnetic moment $m_{\rm ord}$ along $\bf z$ and the saturated magnetic moment $m_{\rm sat}$ \cite{Benton_2016} by the equation: 
\begin{equation}
\cos(\theta)=\frac{m_{\rm ord}}{m_{\rm sat}}
\label{theta}
\end{equation}
Accounting for zero-point quantum fluctuations slightly affects this equation, leading to \cite{Benton_2016, Xu_2019}: 
\begin{equation}
\cos(\theta) \approx \frac{m_{\rm ord}}{0.9 m_{\rm sat}}
\label{theta2}
\end{equation}

The coupling parameters can be obtained from the measured magnetic scattering function $S(Q,\omega)$ using spin wave theory, as previously done for \ndzr\ and \ndtiA\ \cite{Leger_2021, Xu_2019}. In the absence of single crystals, the dispersing branches in $S(Q,\omega)$ are broadened by the powder average, making an accurate determination of the parameters difficult. Nevertheless, in the specific case studied here, distinctive features are present in the spectra, which enable an estimation of the parameters, even for powder samples: the gap to the flat mode $\epsilon_0$, the energy $\epsilon_1$ of the dispersion at the Brillouin zone boundary (${\bf Q}=(110),\ (112),\ (221)$) and the top of the dispersion at the Brillouin zone center $\epsilon_2$ (${\bf Q}=(220),\ (113)$). They are given by the following relations \cite{Benton_2016, Xu_2019}: 
\begin{align}
\epsilon_0 &= \sqrt{(3 |\tilde{\sf J}_z|-\tilde{\sf J}_x)(3 |\tilde{\sf J}_z|-\tilde{\sf J}_y)} \label{E_0} \\
\epsilon_1 &=\sqrt{(3 |\tilde{\sf J}_z|+\tilde{\sf J}_x)(3 |\tilde{\sf J}_z|+\tilde{\sf J}_y)} \label{E_1} \\
\epsilon_2 &=3\sqrt{( |\tilde{\sf J}_z|+\tilde{\sf J}_x)(|\tilde{\sf J}_z|+\tilde{\sf J}_y)} \label{E_2} 
\end{align}
$\epsilon_0$ can directly be observed in powder spectra but it is not as straightforward for $\epsilon_1$ and $\epsilon_2$. In \ndzr\ and \ndhf, $\tilde{\sf J}_y$ is small compared to the other parameters, we have thus set it to zero to better constrain the fit. By analyzing energy cuts at several $Q$, we could then estimate $\epsilon_2$ and deduce the parameters $\tilde{\sf J}_x$ and $\tilde{\sf J}_z$ of \ndlaA\ and \ndlaB, which are reported in Table \ref{table_J} (note that the sign of $\tilde{\sf J}_z$ is fixed by the all-in--all-out nature of the ground state).
The calculated powder average spectra corresponding to these parameters are shown in Figures~\ref{fig_IN6_NdLa}(b) and (d).

\begin{table}
\begin{tabular}{*6{c}}
\hline
\hline
Sample	& $\tilde{{\sf J}}_{x}$ & $\tilde{{\sf J}}_{y}$ & $\tilde{{\sf J}}_{z}$ & $\tilde{{\sf J}}_{x}/\tilde{{\sf J}}_{z}$& $\theta$ (rad) \\
\hline
\multicolumn{6}{c}{Single crystals} \\
\ndzr\   \cite{Leger_2021}  	& 1.18	& -0.03	& -0.53 	& -2.20 	& 1.23 \\
\ndzr\   \cite{Xu_2019}  		& 1.06	& 0.16	& -0.53 	& -1.97 	& 0.98 \\  
\ndtiA\ \cite{Leger_2021} 		& 0.97	& 0.21	& -0.53	& -1.83	& 1.08 \\
\ndtiB 					& 1.26	& -0.05	& -0.57	& -2.23	& 1.09 \\ 
 \ndhf\ \cite{Samartzis_2022}	& 1.23	& 0.09	& -0.66 	& -1.86	& 1.26 \\ 
\hline
\multicolumn{6}{c}{Polycrystalline samples} \\ 
\ndlaA   					& 1.19	& 0		& -0.48 	& -2.50	& 0.99 \\ 
\ndlaB     					& 0.935	& 0		& -0.34	& -2.76 	& 1.01 \\ 
{\it NdMox} 				& 0.25	& 0		& -0.47	& -0.53	& 1.02 \\
{\it \ndscnb}    				& -0.07  	& 0   		& -0.37 	& 0.19 	& 1.08 \\   
{\it \ndgasb}				& -0.44	& 0	   	& -0.91 	& 0.48	& 0.73 \\ 
\hline
\hline
\end{tabular}
\caption{\label{table_J} Estimated couplings (in K) in \nd\ pyrochlore magnets. ${\tilde {\sf J}}_{y}$ was fixed to zero in all samples, except \ndzr, \ndzrx\ and \ndhf\ for which single crystals are available. $\theta$ was obtained according to Equation \ref{theta2}. {\sf J} parameters were determined using the dispersion inferred from inelastic neutron scattering data only in \ndzr, \ndzrx, and \ndzry\ compounds (see Appendix {\ref{J_Tip10} for \ndtiB}). For the other ones (in italics), they were deduced from $\epsilon_0$, $\theta$ and $T_{\rm CW}$. \ndgasb\ and \ndscnb\ parameters are deduced from values given in Ref. \onlinecite{Gomez_2021} and \onlinecite{Scheie_2021} respectively, following the procedure detailed in the text. }
\end{table}

In systems where low energy inelastic neutron scattering measurements are not of high enough quality so that only $\epsilon_0$ can be determined, the Curie-Weiss temperature can also be used to estimate the coupling parameters. Indeed, it is given by the relation \cite{Benton_2016}:
\begin{equation}
T_{\rm CW}=\frac{{\sf J}_z}{2}=\frac{\tilde{\sf J}_z \cos^2(\theta) + \tilde{\sf J}_x \sin^2(\theta)}{2}
\label{thetaCW}
\end{equation}
where the coupling parameters are in K. Nevertheless, as previously mentioned, the determination of the Curie-Weiss temperatures $T_{\rm CW}$ from susceptibility are far from accurate given their low values, so that the resulting errors in the coupling parameters may be important. 
After determining $\theta$ from Equation \ref{theta2} and assuming $\tilde{\sf J}_y=0$, $\tilde{\sf J}_x$ and $\tilde{\sf J}_z$ can thus be determined using Equations \ref{E_0} and \ref{thetaCW}. We used this method to estimate the couplings in NdMox. To get a broader view, we also applied it to other Nd pyrochlore systems where $m_{\rm ord}$, $T_{\rm CW}$ and $\epsilon_0$ are known from the literature, namely \ndscnb\ and \ndgasb. 
It is, however, worth noting that when the interaction parameters could be determined independently, the computed $T_{\rm CW}$ is different from the experimental one (see Appendix \ref{J_Theta}), except in the study by Xu {\it et al.} \cite{Xu_2019}. In addition to the experimental uncertainty in $T_{\rm CW}$, the determination of $\theta$ may also be at the origin of these discrepancies; the ordered moment can indeed be underestimated due to thermalization issues. The coupling parameters obtained by this method may thus be considered cautiously.	

\section{Field induced behavior}
\begin{figure}[h!]
\centering{\includegraphics[width=7.5cm]{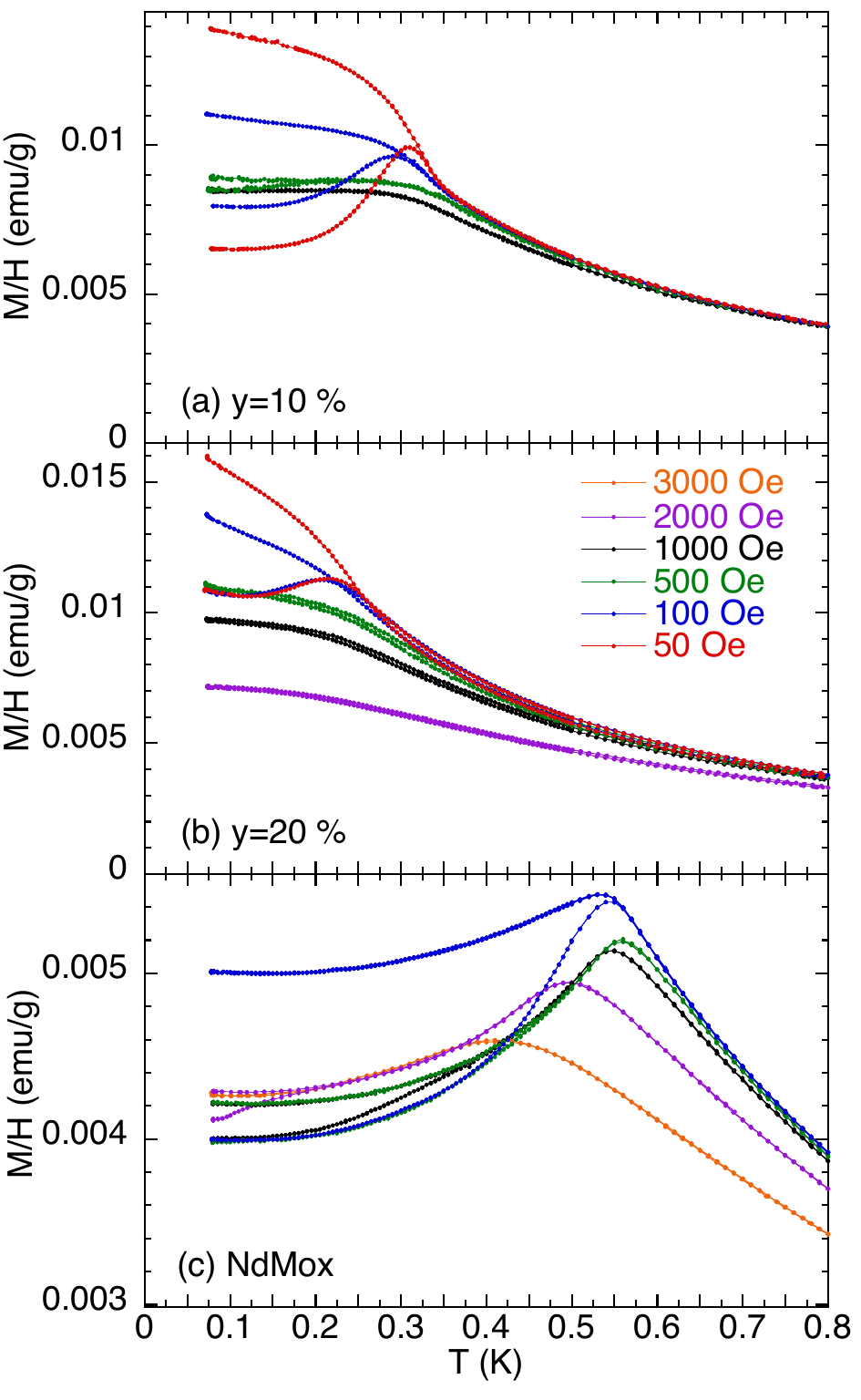}}
\caption{$M/H$ versus temperature measured at several applied magnetic fields $H$ in \ndzry\ with (a) $y=10$~\%, (b) $y=20$ \% and in (c) NdMox. All curves were measured using a ZFC-FC procedure. The derivatives of the FC curves as a function of temperature are shown in Appendix~\ref{comp_data}.}
\label{fig_ZFC-FC}
\end{figure}
The behavior of the substituted samples when a magnetic field is applied is qualitatively similar to the \ndzr\ sample (except, of course, for \ndlaC\ which does not order down to 75 mK), as detailed in Ref.~\onlinecite{Leger_2021b} for the \ndzrx\ compounds. 
The lack of single crystals for \ndzry\ and NdMox samples prevents from a detailed study under applied magnetic field. Several key features can nevertheless be observed. 
As shown in Figure \ref{fig_ZFC-FC}, a freezing is observed below $T_{\rm N}$ that manifests as a splitting in Zero Field Cooled - Field Cooled (ZFC-FC) magnetization measurements as a function of temperature. This effect has already been observed in several pyrochlore compounds, including Nd-based ones, and is probably due to the trapping of magnetic moments at the domain-walls. This irreversibility is suppressed when increasing the magnetic field: it is not visible anymore at 500~Oe in \ndlaB, 1000~Oe in \ndlaA\ and at 3000 Oe in NdMox. This hierarchy is consistent with the different ordering temperatures observed in the three compounds. 

In both \ndzry\ compounds, the magnetization at low field keeps increasing when decreasing the temperature, as previously observed in the ac susceptibility. This effect is more important in the $y=20$ \% sample and may be due to the existence of Nd spins which remain paramagnetic down to the lowest temperature due to the dilution. 

\begin{figure}[h!]
\centering{\includegraphics[width=7.5cm]{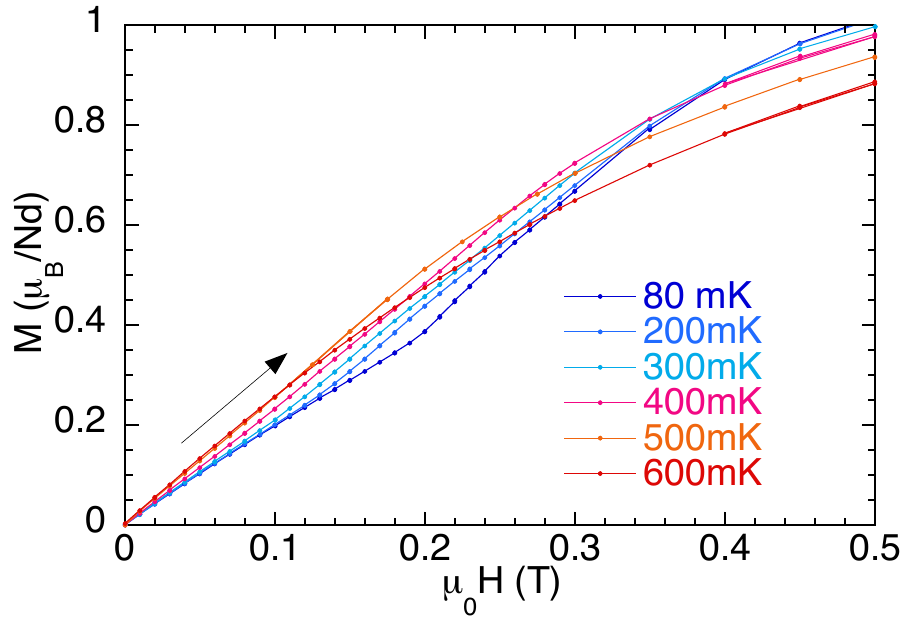}}
\caption{$M$ versus $H$ at several temperatures in NdMox. The sample was first ``saturated" in -1 T and the field was ramped up to 1 T.}
\label{fig_NdMox_MH}
\end{figure}

The field induced behavior was shown to be very similar to the pure compound in the \ndzrx\ samples \cite{Leger_2021b}. In \ndzry\ and NdMox samples, the field induced transitions could not be probed due to the polycrystalline character of the samples. Nevertheless, in NdMox a clear inflection point could be observed in the isothermal magnetization curve at low temperature (see Figure \ref{fig_NdMox_MH}), which may feature the field induced transition observed along $[111]$ in \ndzr\ \cite{Lhotel_2015, Lhotel_2018, Xu_2019}. The transition field is however one order of magnitude larger than in \ndzr. It reaches 0.225 T at low temperature, and progressively decreases when increasing the temperature, reaching zero at the N\'eel temperature. 

\section{Discussion}
The present study about the effects of disorder on Nd-based pyrochlores highlights the robustness of the all-in--all-out ordered state with a dipolar-otcupolar character stabilized in these compounds. 
Disorder introduced on the transition metal site is expected to disturb the local symmetry of the Nd$^{3+}$ ions, and thus to affect the single ion properties. Nevertheless, even if the $D_{3d}$ symmetry is suppressed on a significant proportion of sites in the most substituted \ndzrx\ sample or in NdMox, the crystal electric field properties are not strongly modified. Disorder manifests as a broadening of the crystal field excitations but the nature and the position of the levels remain similar. As a consequence the data can be reproduced by crystal field wavefunctions close to the ones obtained in pure samples. In particular, the obtained Nd$^{3+}$ ground doublet retains the dipolar-octupolar character which enables the exotic properties reported in the pure compound. The main effect seems to be on the value of the ground doublet moment, which increases slightly with disorder.

\subsection{Substitution on the non-magnetic site}

Upon substituting titanium on the zirconium site, the Curie-Weiss temperature is only slightly modified, hence a small change of the exchange parameters, which is not that surprising since structural parameters are very similar in the substituted and the pure samples (see Table~\ref{Table_charac}). Nevertheless, the N\'eel temperature increases with the substitution rate, as well as the all-in--all-out ordered moment along the $\langle 111 \rangle$ directions. This indicates that the ordering is reinforced and slightly pushed along the local $\langle 111 \rangle$ directions, which means a smaller $\theta$ angle. In terms of interactions, this would imply that the absolute value of the ${\tilde{\sf J}}_z$ parameter is increased compared to  ${\tilde{\sf J}}_x$ in these substituted polycrystalline samples. 

The Hamiltonian parameters determined from inelastic neutron scattering data analysis on single crystals (see Table \ref{table_J}) do not however provide clear trends. Strong variations are obtained even in the reported parameters for pure \ndzr\ samples, which may be due to distinct fit methods but also result from different crystal growth protocols. Indeed, these samples additionally have different $T_{\rm N}$ and ordered moments.  As shown in Ref. \onlinecite{Leger_2021b}, it is also worth mentioning that single crystals have smaller ordering temperature and ordered moment along ${\bf z}$ than polycrystalline samples. In spite of the robustness of global properties with respect to substitutions, the details of the microscopic parameters are thus sensitive to subtle variations in the composition.

\begin{figure}[h!]
\centering{\includegraphics[width=9cm]{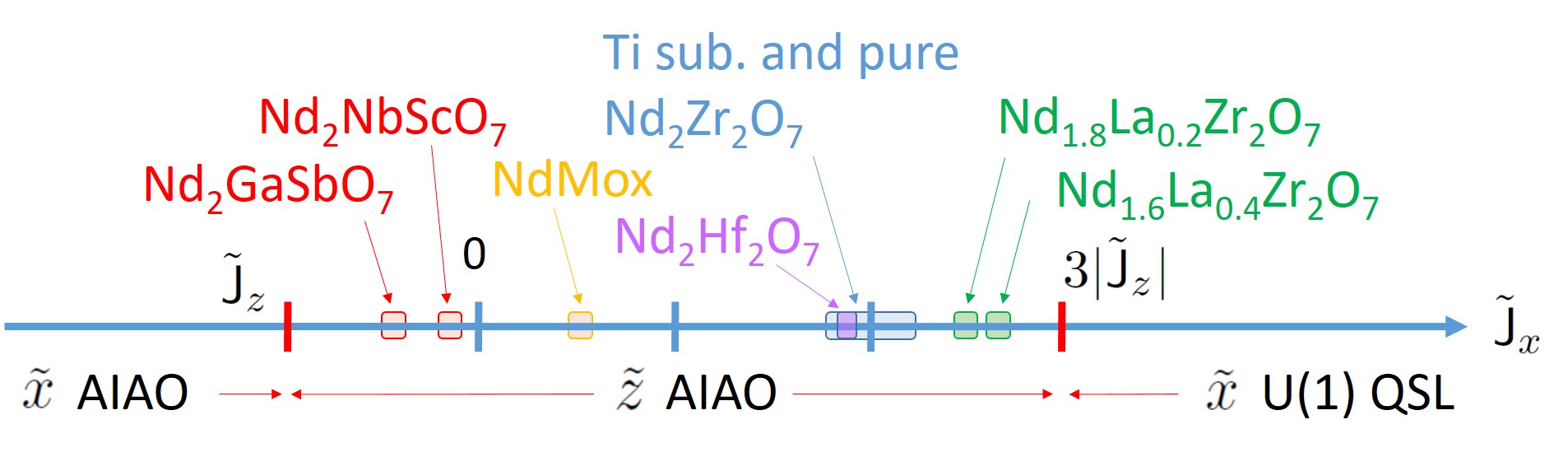}}
\caption{Sketch of the classical phase diagram for dipolar-octupolar pyrochlores assuming $\tilde{\sf J}_z <0$. 
The $\tilde{x}$ all-in--all-out phase is stabilized for $\tilde{\sf J}_x < \tilde{\sf J}_z <0$, while the $\tilde{x}$ spin ice is favored for $\tilde{\sf J}_x > 3|\tilde{\sf J}_z|$ (these two phases have their moments oriented along the $\tilde x$ direction). In quantum models, this $\tilde{x}$ spin ice phase is a $U(1)$ spin liquid phase, and the phase boundary is slightly shifted \cite{Benton_2020, Patri_2020}. In between arises the $\tilde{z}$ all-in--all-out (with the moments oriented along the $\tilde z$ direction) observed in pure and substituted \ndzr. The position of the different compounds listed in Table \ref{table_J} are indicated by colored rectangles.} 
\label{Fig_diagphase}
\end{figure}

In NdMox, where disorder is much stronger, the Curie-Weiss temperature is close to zero, while the ordering temperature and the flat mode energy significantly increase. This corresponds to a much smaller ${\tilde{\sf J}}_x$ value than in the pure compound, while ${\tilde{\sf J}}_z$ is preserved. The $\theta$ value is however little affected, since the ordered moment along $\langle 111 \rangle$ remains similar. This stability of $\theta$ is surprising since it cannot be simply related to a microscopic parameter of the system (see Equation \ref{Hxyz_tilde}). 
The main difference between NdMox and other measured compounds arises from the significant decrease of the lattice parameter (see Table \ref{Table_charac}). It is comparable to the one obtained in \ndsn\ \cite{Bertin_2015}, and \ndscnb\ \cite{Scheie_2021,Mauws_2021}.  \ndscnb\ has similar $T_{\rm CW}$ and ordered moment as NdMox but a smaller N\'eel temperature (370 mK). \ndsn\ has a much larger N\'eel temperature (910 mK) and $z$ ordered moment (1.7 $\mu_{\rm B}$). \ndgasb\  \cite{Gomez_2021}, which has the smallest reported lattice parameter (10.34 \AA) in the family, has properties similar to \ndsn. In all these compounds characterized by a reduced lattice parameter, $|{\tilde {\sf J}}_x/{\tilde{\sf J}}_z|$ is significantly decreased. In particular, ${\tilde {\sf J}}_x$ strongly decreases and possibly changes sign in \ndgasb\ and \ndscnb\ (see Table \ref{table_J}). As illustrated in the phase diagram of Figure \ref{Fig_diagphase}, the competition between the high temperature Coulomb phase and the all-in--all-out state is then much reduced, in favor of the all-in--all-out state. Starting from \ndzr\ characterized by $|{\tilde {\sf J}}_x/{\tilde{\sf J}}_z|\approx 2$, such samples are then located deeper in the $\tilde{z}$ all-in--all-out phase. Note that this also results in a higher transition temperature. Further studies are needed to enlighten the relation between these interaction parameters and microscopic properties.

\subsection{Substitution on the magnetic site}
Even more interesting is the case of non-magnetic substitution on the magnetic site. When diluting up to 20 \% the magnetic ions with non-magnetic ions, our measurements show only quantitative changes in the behavior but the physics remain the same: (i) the system reaches an ordered all-in--all-out ground state along the ${\tilde z}$ direction; (ii) the apparent reduced ordered moment along ${\bf z}$ remains in the same range as the pure sample (see Table \ref{Table_charac}); and (iii) the excitations show the same characteristics, a gapped flat mode associated to dispersive excitations (see Figure \ref{fig_IN6_NdLa}). The transition as well as Curie-Weiss temperatures nevertheless decrease gradually with dilution. $T_{\rm N}$ is below 75 mK at 40 \% dilution, despite a positive $T_{\rm CW}$ of 170 mK and a substitution rate quite far from the percolation threshold (61 \% of dilution) \cite{Stauffer}. The inelastic spectra for these dilute systems with $y=10$ and 20 \% can be reproduced with average interaction parameters (see Figure \ref{fig_IN6_NdLa} and Table \ref{table_J}). As expected, these parameters decrease with dilution, but interestingly the ratio $|{\tilde {\sf J}}_x/{\tilde{\sf J}_z}|$ increases. This is seen experimentally through the decrease of the spin ice flat mode towards the elastic line.  In Figure \ref{Fig_diagphase}, this corresponds to moving towards the boundary of the $U(1)$ spin liquid phase \cite{Benton_2016}. 
These results suggest that by weakening the all-in--all-out ordering, magnetic dilution favors the unconventional Coulomb phase observed in these compounds. To confirm this hypothesis, it would be interesting to measure the magnetic diffuse scattering in the 40 \% sample, and see whether or not it shows the spin ice like pattern down to very low temperature. 

\begin{figure}[h!]
\centering{\includegraphics[width=8cm]{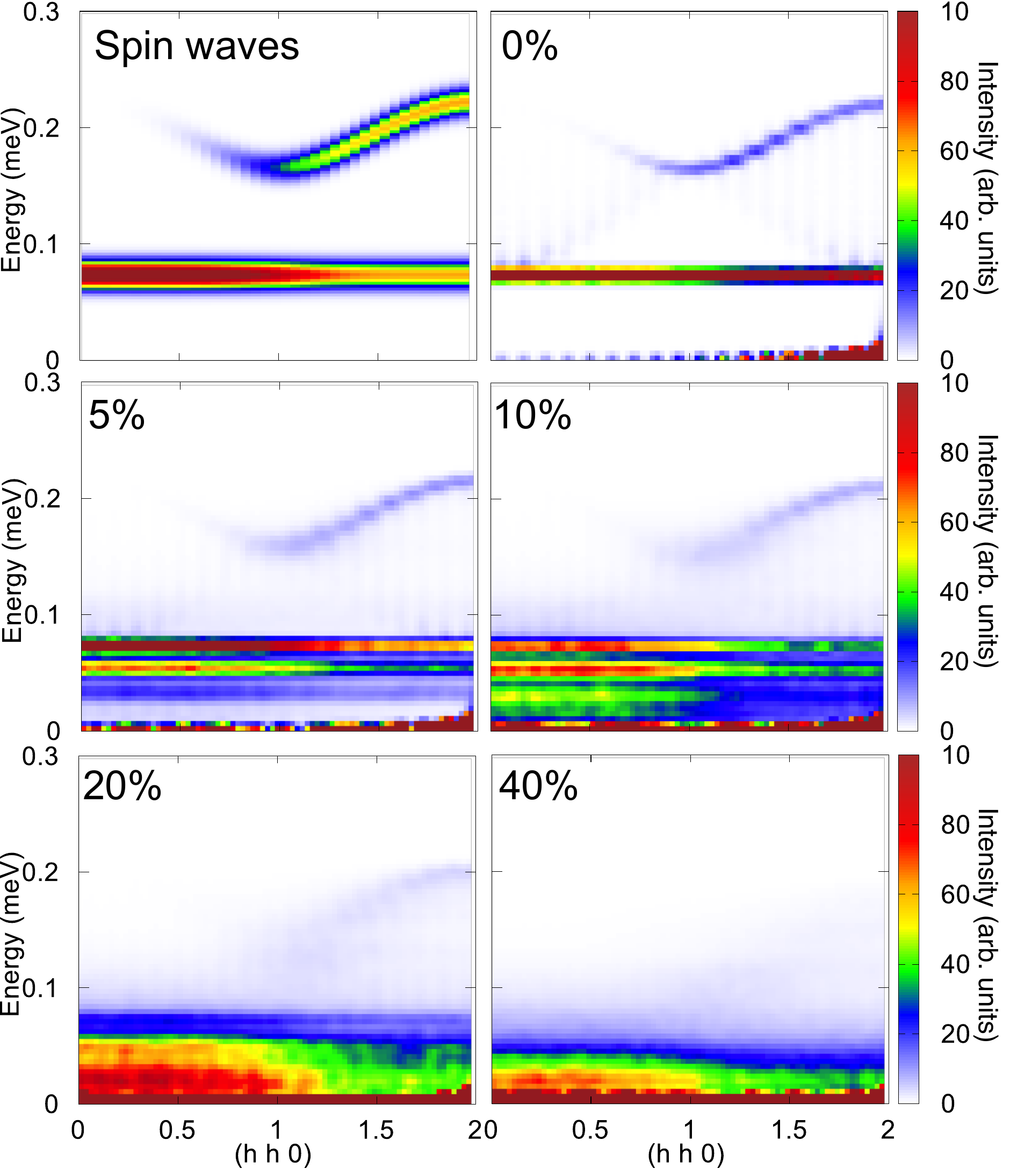}}
\caption{Calculated inelastic neutron scattering spectra along $(hh0)$ at zero temperature and for different substitution rates. The top left figure shows the spin wave calculations, to be compared with the 0 \% calculations.}
\label{disorder2}
\end{figure}

To go further in the understanding of magnetic dilution, we have performed numerical simulations of the spin dynamics in a dilute system (see Appendix \ref{simulations}), assuming interaction parameters close to the pure system ($\tilde{\sf J}_x=1,~\tilde{\sf J}_z=-0.5$ K). As shown in Figure \ref{disorder2}, when the dilution rate increases, the intensity of the dispersive mode decreases, while its width increases. The flat mode persists even at high substitution rates, but its position in energy decreases, approaching the elastic line. It condenses for a substitution rate of about 50 \%. Supernumerary flat levels below the gap to the main flat mode are clearly present in the systems with low dilution levels. We believe they reflect the presence of sites that no longer have six, but less neighbors due to substitution. These calculations are consistent with the dynamics observed in \ndzry\ samples, i.e., the persistence of the flat and dispersive modes, the broadening of the modes and the decrease of the gap. This suggests that the microscopic interaction parameters may not be strongly affected by the dilution on the Nd site. The apparent increase of $|{\tilde {\sf J}}_x/{\tilde{\sf J}_z}|$ determined based upon the average model is thus a consequence of the dilution and of the random fields it creates, which apparently tend to favor the disordered Coulomb phase with respect to the all-in--all-out ordered state. 

\section{Conclusions}
Our study of substitutions in \ndzr\ samples, on both the magnetic and non magnetic sites, demonstrates the remarkable stability of the dipolar-octupolar properties. The ground state remains this unusual all-in--all-out antiferromagnetic state, characterized by a partially ordered magnetic moment along the local $\langle 111 \rangle$ directions. The excitations, which exhibit both a flat mode and a dispersive one (the so-called dynamic fragmentation \cite{Benton_2016}), also persist in the samples where ordering could be observed. 

In all studied samples here and in the literature, substitution on the non-magnetic site, or replacement of zirconium by another element, tend to decrease the lattice parameter. Concomitantly, our study tends to show that the ratio $|{\tilde {\sf J}}_x/{\tilde{\sf J}}_z|$ decreases, which means that the system moves away from the $U(1)$ spin liquid state \cite{Benton_2016}. While no clear picture emerges up to now to explain the value of these couplings, this suggests that to reach this unconventional spin liquid state in Nd-based pyrochlore compounds, one should try to expand the lattice parameter. The dilution on the magnetic site may be an option, since \ndzry\ systems show an increase of $|{\tilde {\sf J}}_x/{\tilde{\sf J}}_z|$. The presence of specific diffuse scattering in the more diluted sample, that would indicate the persistence of the Coulomb phase, nevertheless remains to be investigated. \\

\begin{figure*}[t!]
\centering{\includegraphics[width=\textwidth]{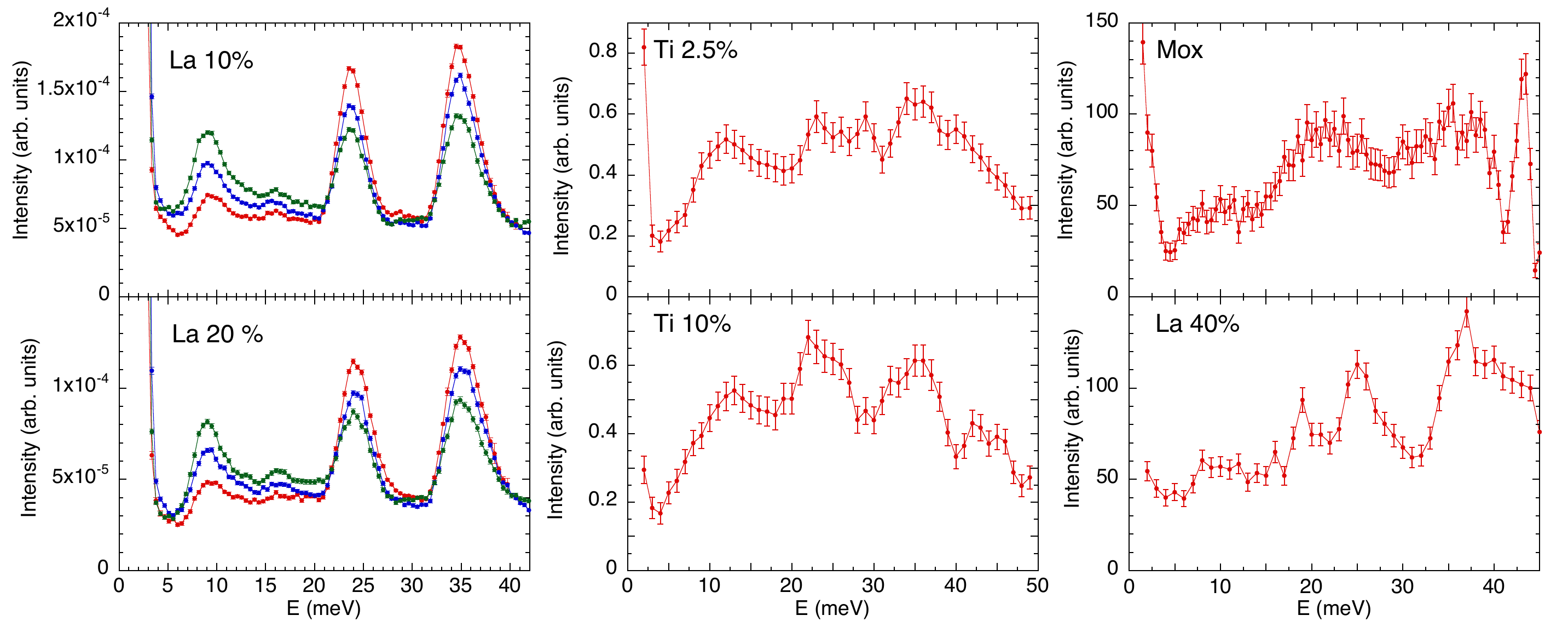}}
\caption{ Intensity vs energy obtained from constant $Q$-cuts in the inelastic neutron scattering maps shown in Figure \ref{fig_CEF}. For \ndzry\ with $y=10$ and 20 \%, $Q$-cuts are performed within $Q=2.5$ (red dots), 3.5 (blue dots), and 5 \AA\ (green dots) $\pm 0.4$ \AA. For \ndzrx\ with $x=2.5$ and 10 \%, they are performed at $Q=3.2\pm 0.2$ \AA. For \ndzry\ with $y=40$ \% and NdMox, they are performed at $Q=3 \pm 0.2$ \AA\ and $Q=3.5 \pm 0.2$ \AA\ respectively. All cuts show the presence of the CEF modes discussed in the text around 25 and 35 meV, as well as a phonon mode around 10 meV. }
\label{Qcuts}
\end{figure*}

\section{Acknowledgments}

We gratefully acknowledge E. Suard and C. Colin for additional diffraction experiments at D2B and D1B at ILL respectively, C. Paulsen for the use of his magnetometers and N. Dragoe. 

This work is based on experiments performed at the Swiss spallation neutron source SINQ, Paul Scherrer Institute, Villigen, Switzerland.

We acknowledge financial support from the Agence Nationale de la Recherche under Grant No. ANR-19-CE30-0040 and ANR 19-CE30-0030-01, as well as from the 2FDN. 
The work at the University of Warwick was supported by EPSRC through grant EP/T005963/1.

\appendix
\section{CEF}
\label{appendix_cef}
Figure \ref{Qcuts} shows constant $Q$-cuts performed on the maps of Figure \ref{fig_CEF} which highlight the CEF level positions as well as the 10 meV phonon present in all the samples. 

The $B_{k,m}$ values determined from the CEF energy levels of the different samples are summarized in Table \ref{cef-mplt}. The corresponding wavefunctions for the pure \ndzr\ sample are detailed in Table \ref{cef2wf}.
\FloatBarrier
\begin{table}[h!]
\begin{tabular}{*{7}{c}}
\hline
\hline
Sample 				& $B_{2,0}$	& $B_{4,0}$ 	& $B_{4,3}$	& $B_{6,0}$	& $B_{6,3}$	& $B_{6,6}$ \\ \hline 
\ndzr\ \cite{Xu_2015}  	&  49.2 		&  408.9 		&  121.6 		&  148.1 		&  -98.0 		&  139.1 \\ 
\ndtiA 				&  54.1 		&  408.9 		&  121.6 		& 140.7 		&  -107.8 		&  139.1 \\ 
\ndtiB 				&  52.9 		&  408.9 		&  121.6 		&  137.0 		&  -112.7 		&  139.1 \\
\ndlaA 				&  51.7 		&  429.3 		&  124.6 		&  144.4 		&  -98.0 		&  135.6 \\ 
\ndlaB  				&  54.1 		&  449.8 		&  133.8 		&  148.1 		&  -102.9 		&  132.1 \\ 
\ndlaC  				&  61.5 	     	&  460.0 	    	&  121.6	     	&  129.6    	&  -85.7     	& 139.0 \\ 
NdMox 				&  56.6    		&  408.9 	    	& 139.8 	     	&  137.0    	&  -112.7 	    	& 128.7 \\ 
\hline
\hline
\end{tabular}
\caption{\label{cef-mplt} Wybourne coefficients in meV obtained for the different samples. }
\end{table}

\begin{table*}[t!]
\begin{tabular}{*{11}{c}}
\hline
\hline
\multirow{2}{*}{$m_J$}	& $E_0$    & $E'_0$ 		& $E_1$   		& $E'_1$ 		& $E_2$ 		& $E'_2$ 		& $E_3$ 		& $E'_3$  		&  $E_4$ 		& $E'_4$\\
					&  \multicolumn{2}{c}{0}     & \multicolumn{2}{c}{23 meV}   & \multicolumn{2}{c}{34 meV}	&  \multicolumn{2}{c}{35 meV}  &  \multicolumn{2}{c}{104 meV} \\
 \hline
 \hline
 $J=9/2$ &&&&&&&&&& \\
 $-9/2$ 				& 0.894  	&          		&           		&         		& -0.346  	& 0.223 		&         		&           		&			& \\
 $-7/2$ 				&          	&          		&            	& 0.148 		&           		&          		& -0.510  	&           		& 0.833 	& \\
 $-5/2$ 				&           	&          		& 0.735   	&         		&           		&         		&            	& -0.486		&			& 0.442 \\
 $-3/2$ 				& 0.261 	& 0.336 		&            	&          		& 0.842		& -0.294 		&            	&            	& 			&  \\
 $-1/2$ 				&           	&          		&            	& 0.652   	&          		&          		& 0.681   	&            	& 0.306 	& \\
 $+1/2$ 				&          	&          		& 0.652    	&           		&          		&          		&            	& 0.681   	& 			& -0.306 \\
 $+3/2$ 				& -0.336 	& 0.261 		&            	&        		& -0.294 		& 0.842  		&           		&            	&			&\\
 $+5/2$ 				&          	&          		&            	& 0.735 		&          		&          		& -0.486 		&           		& -0.442 	&\\
 $+7/2$ 				&          	&          		& -0.148   	&          		&           		&          		&            	& 0.510		&			& 0.833 \\
 $+9/2$ 				&          	& 0.894 		&            	&           		& -0.223 		&-0.346 		&           		&           		&			& \\
 \hline
  $J=11/2$ &&&&&&&&&& \\
  $-11/2$ 			&         	&          		&           		&           		&           		&           		&           		&-0.082		&			& 0.075 \\
  $-9/2$ 				& 0.109  	&          		&           		&           		&           		& 0.145 		&           		&            	&			& \\
  $-7/2$ 				&           	&          		&           		& -0.055  	&           		&           		& 0.147  		&           		&			& \\
  $-5/2$ 				&           	&          		& 0.053 		&           		&           		&           		&           		&-0.079    	& 			& -0.065\\
  $-3/2$ 				&           	&          		&           		&            	&           		&-0.098 		&           		&            	& 			& \\
  $-1/2$ 				&           	&          		&           		&            	&           		&          		&           		&             	& 0.059	& \\
  $+1/2$ 				&          	&          		&           		&            	&           		&          		&           		&            	&			& 0.059 \\
  $+3/2$ 				&          	&          		&           		&            	& -0.098      	&          		&          		&            	&			& \\
  $+5/2$ 				&          	&          		&            	& 0.053  		& 	  			&          		& -0.079  	&            	& 0.065	&  \\
  $+7/2$ 				&          	&          		& 0.055   	&            	&           		&         		&            	&-0.147  		&			& \\
  $+9/2$ 				&          	& 0.109  		&           		&           		& -0.145   	&         		&            	&            	& 			& \\
  $+11/2$ 			&        	&           		&           		&           		&            	&         		& -0.082   	&            	& -0.075 	& \\
\hline
\hline
\end{tabular}
\caption{\label{cef2wf} Coefficients of the \nd\ wavefunction expansion in \ndzr.}
\end{table*}

\begin{figure*}[h!t]
\centering{\includegraphics[width=\textwidth]{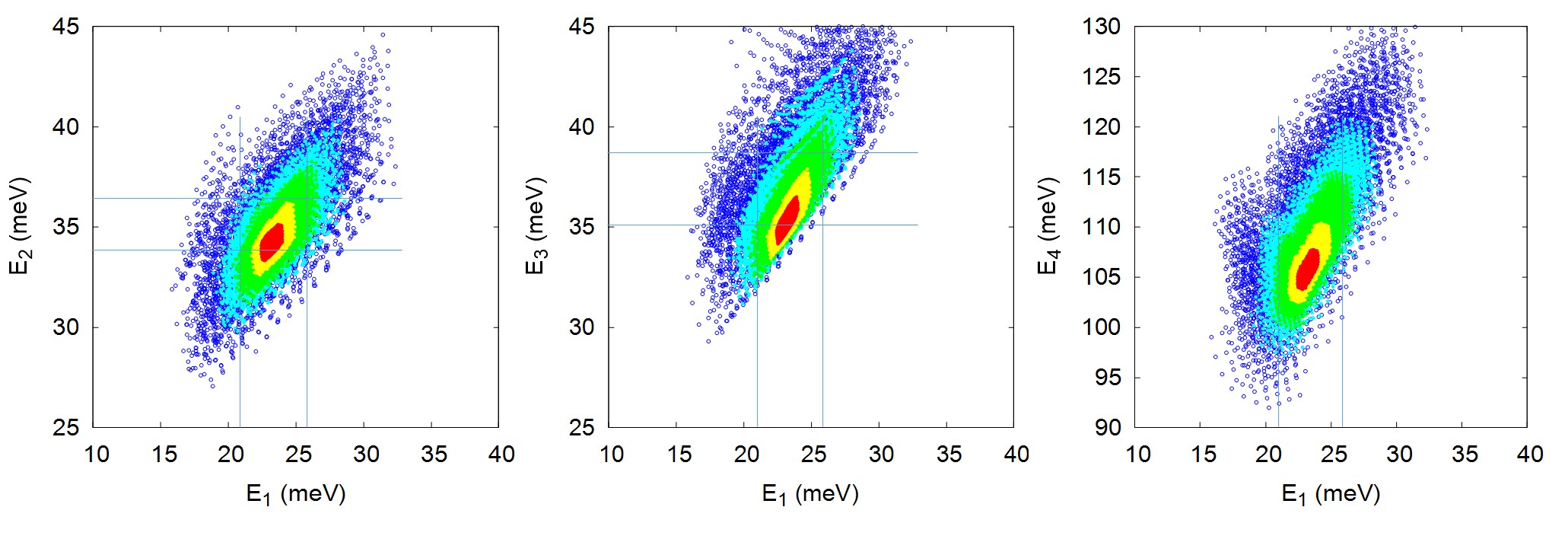}}
\caption{Energy of the crystalline field levels obtained by varying the $B_{k,m}$ coefficients in an interval of $\pm$ 5 (red), 10 (yellow), 20 (green), 30 (cyan) and 50 (blue) \% around the nominal values of Ref. \onlinecite{Xu_2015}. The vertical and horizontal lines depict the variation range observed in the various samples.} 
\label{cef-incertitude}
\end{figure*}

These parameters have been estimated from the average position of the energy levels only. To estimate the uncertainty on these parameters, the energy levels have been calculated in a systematic way, by varying the $B_{k,m}$ coefficients in an interval of 5, 10, 20, 30 and 50 \% around the values determined in Ref. \onlinecite{Xu_2015} for \ndzr\ (see line~1 of Table \ref{cef-mplt}). 

Figure \ref{cef-incertitude} presents the different CEF energy levels obtained, showing $E_{2,3,4}$ as a function of $E_1$, in the form of a scatter graph. On those graphs are also shown the energy range determined by INS experiments. These calculations thus allow us to estimate the uncertainty range on the $B_{k,m}$ values still compatible with measurements. They also indicate which local variation of the $B_{k,m}$ parameters could induce the broadening observed in certain experiments.

\FloatBarrier

\section{Complementary macroscopic measurements}
\label{comp_data}
\begin{figure}[h!]
\centering{\includegraphics[width=8cm]{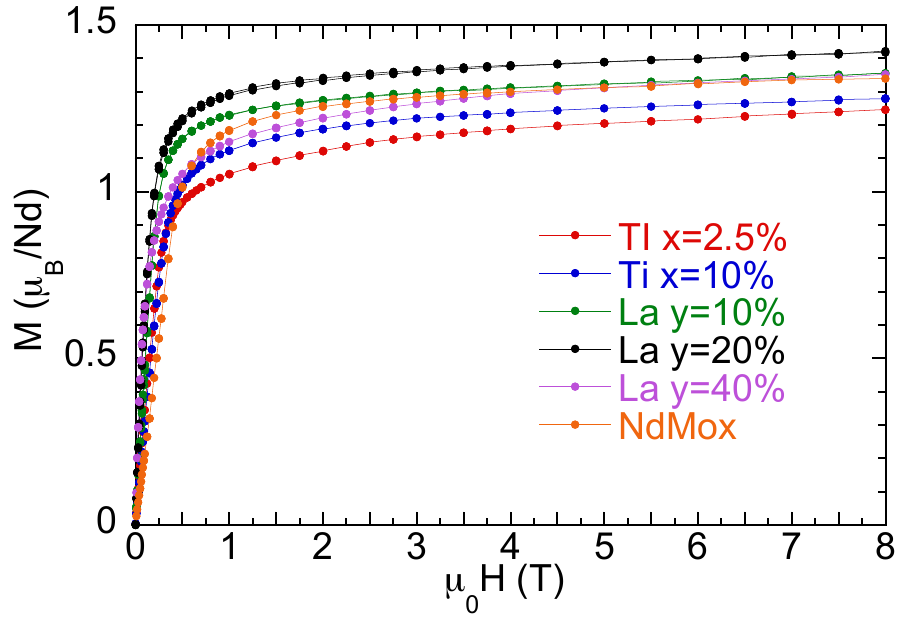}}
\caption{Magnetization $M$ vs magnetic field $H$ measured in all samples up to 8 T at about 100 mK. }
\label{fig_MH_8T}
\end{figure}

The magnetization curves measured at low temperature (typically around 100 mK) on all the studied samples are shown in Figure \ref{fig_MH_8T}, showing a finite slope at high field. 

\begin{figure}[h!]
\centering{\includegraphics[width=8cm]{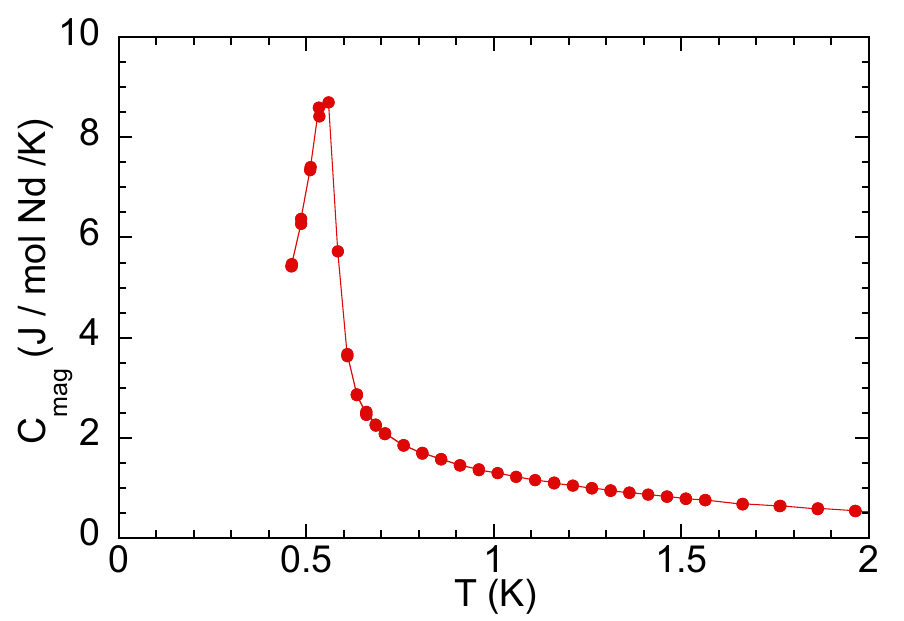}}
\caption{\label{NdMox_CvsT} Magnetic contribution to the specific heat $C$ vs temperature $T$ measured in NdMox. The YMox specific heat was subtracted from the raw data after a correction due to the different molecular weigths \cite{Hardy_2003}. }
\end{figure}

The temperature dependence of the NdMox specific heat is shown in Figure \ref{NdMox_CvsT}, showing the magnetic transition at 550 mK. 

\begin{figure}[h!]
\centering{\includegraphics[width=8cm]{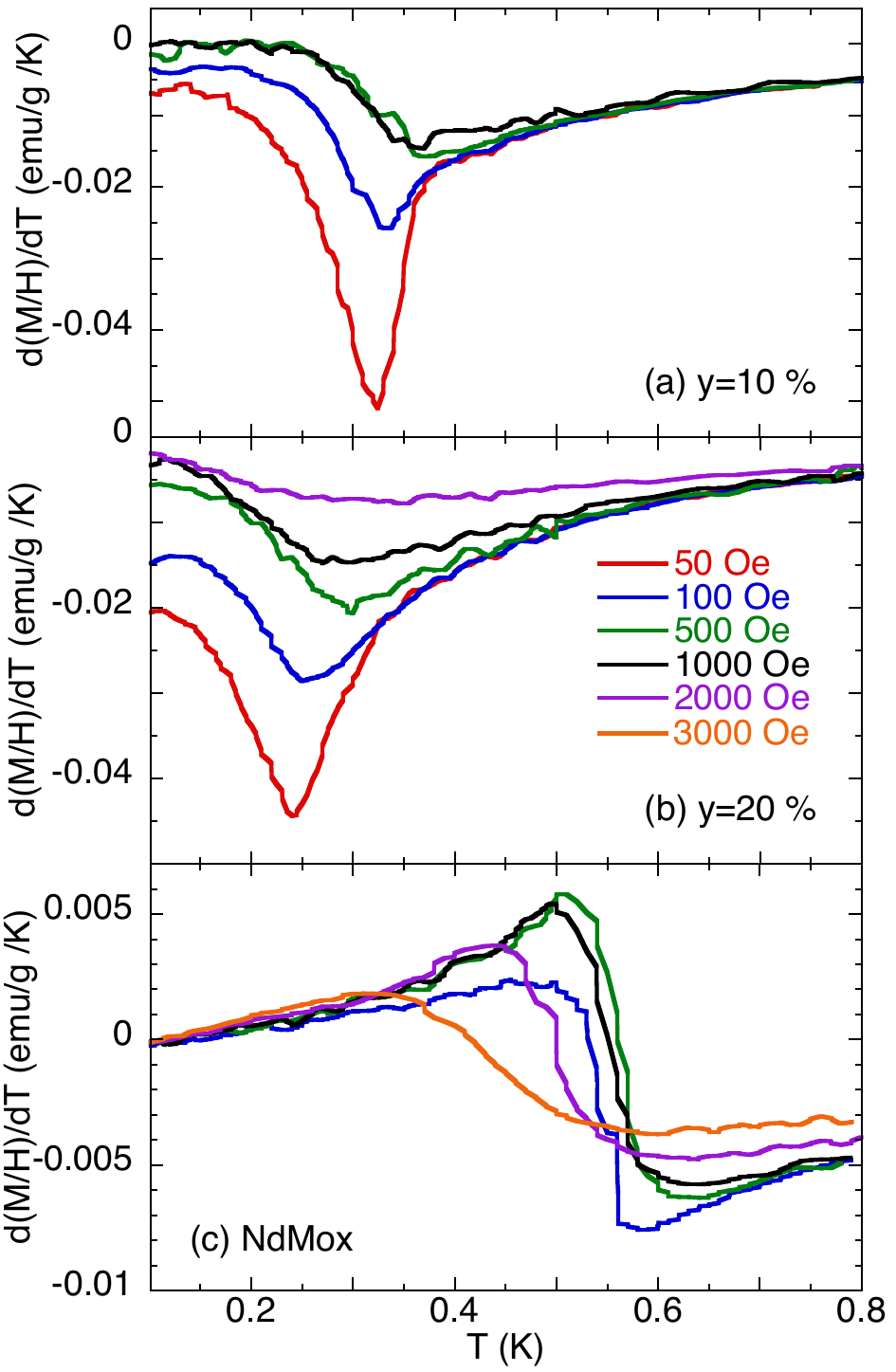}}
\caption{\label{fig_dMdT} Derivative of $M/H$ with respect to the temperature in \ndzry\ with (a) $y=10$~\%, (b) $y=20$ \% and in (c) NdMox for several applied fields between 50 and 3000~Oe. }
\end{figure}

The derivatives of the FC curves of Figure \ref{fig_ZFC-FC} are shown in Figure \ref{fig_dMdT}. In \ndzry\ ((a) and (b)), the inflection point in $M$ vs $T$, corresponding to the minimum in the derivative, smoothens and moves to higher temperatures when increasing the magnetic field. In NdMox, the minimum first moves to higher temperatures, and then shifts to lower temperatures for fields above 1000 Oe. 

\begin{figure*}[ht!]
\centering{\includegraphics[width=\textwidth]{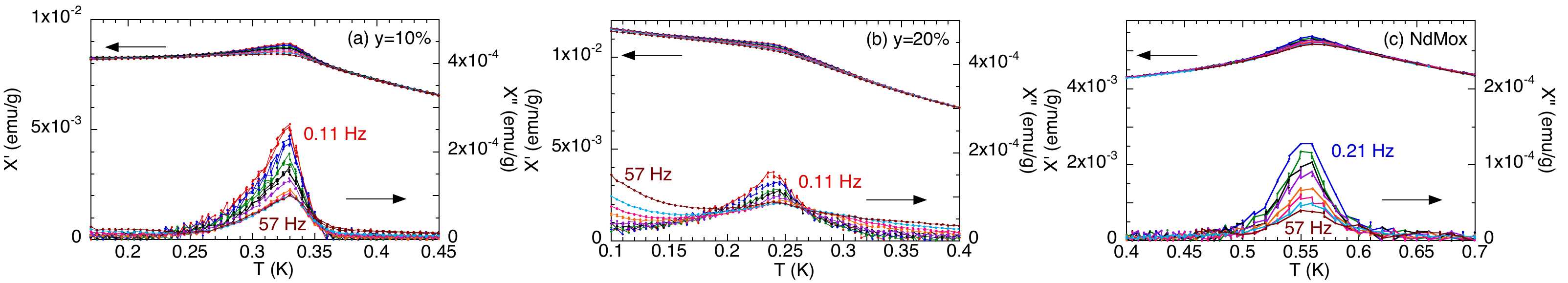}}
{\color{blue} \caption{\label{fig_Xac_multif} $\chi_{\rm ac}$ vs $T$ zoomed in around the transition temperature, for several frequencies in \ndzry\ with (a)~$y=10$~\%, (b) $y=20$ \% and in (c) NdMox.  }}
\end{figure*}

The frequency dependence of the susceptibility at the transition is shown on Figure \ref{fig_Xac_multif}. While the position of the maxima in $\chi'$ and $\chi''$ does not change with frequency in our frequency range (typically $f<200$ Hz), the amplitude of the peak strongly decreases, especially in $\chi''$ when the frequency increases. 

\section{XYZ Hamiltonian}
\label{appendix_XYZ}
To model the low energy properties of \nd\ pyrochlore magnets, we have introduced the XYZ Hamiltonian in Equations \ref{Hxyz} and \ref{Hxyz_tilde}. The ${\sf J}$ and $\tilde{{\sf J}}$ parameters in these equations are connected by the following relations: 
\begin{equation*}
\begin{cases}
\tilde{{\sf J}}_{x}=& {\sf J}_{x} \cos^2 \theta +{\sf J}_{z} \sin^2 \theta + {\sf J}_{xz} \sin 2 \theta \\
\tilde{{\sf J}}_{z}=& {\sf J}_{x} \sin^2 \theta +{\sf J}_{z} \cos^2 \theta - {\sf J}_{xz} \sin 2 \theta \\
\tilde{{\sf J}}_{y} =& {\sf J}_{y}
\end{cases}
\end{equation*}
or
\begin{equation*}
\begin{cases}
\tilde{{\sf J}}_{x}=& \frac{1}{2} \left( {\sf J}_{x}+{\sf J}_{z} + \sqrt{({\sf J}_{x}-{\sf J}_{z})^2+4{\sf J}_{xz}^2} \right)\\
\tilde{{\sf J}}_{z} =& \frac{1}{2} \left( {\sf J}_{x}+{\sf J}_{z} - \sqrt{({\sf J}_{x}-{\sf J}_{z})^2+4{\sf J}_{xz}^2} \right)\\ 
\tilde{{\sf J}}_{y} =& {\sf J}_{y}
\end{cases}
\end{equation*}
and conversely:
\begin{equation*}
\begin{cases}
{\sf J}_{xz} =& (\tilde{{\sf J}}_{x}-\tilde{{\sf J}}_{z}) \sin \theta \cos \theta \\
{\sf J}_{x} =& \tilde{{\sf J}}_{x} \cos^2 \theta + \tilde{{\sf J}}_{z} \sin^2 \theta \\
{\sf J}_{z} =& \tilde{{\sf J}}_{z} \cos^2 \theta + \tilde{{\sf J}}_{x} \sin^2 \theta \\
\end{cases}
\end{equation*}

\section{Interaction parameters for \ndtiB}
\label{J_Tip10}
\begin{figure}[!h]
\centering{\includegraphics[width=8.5cm]{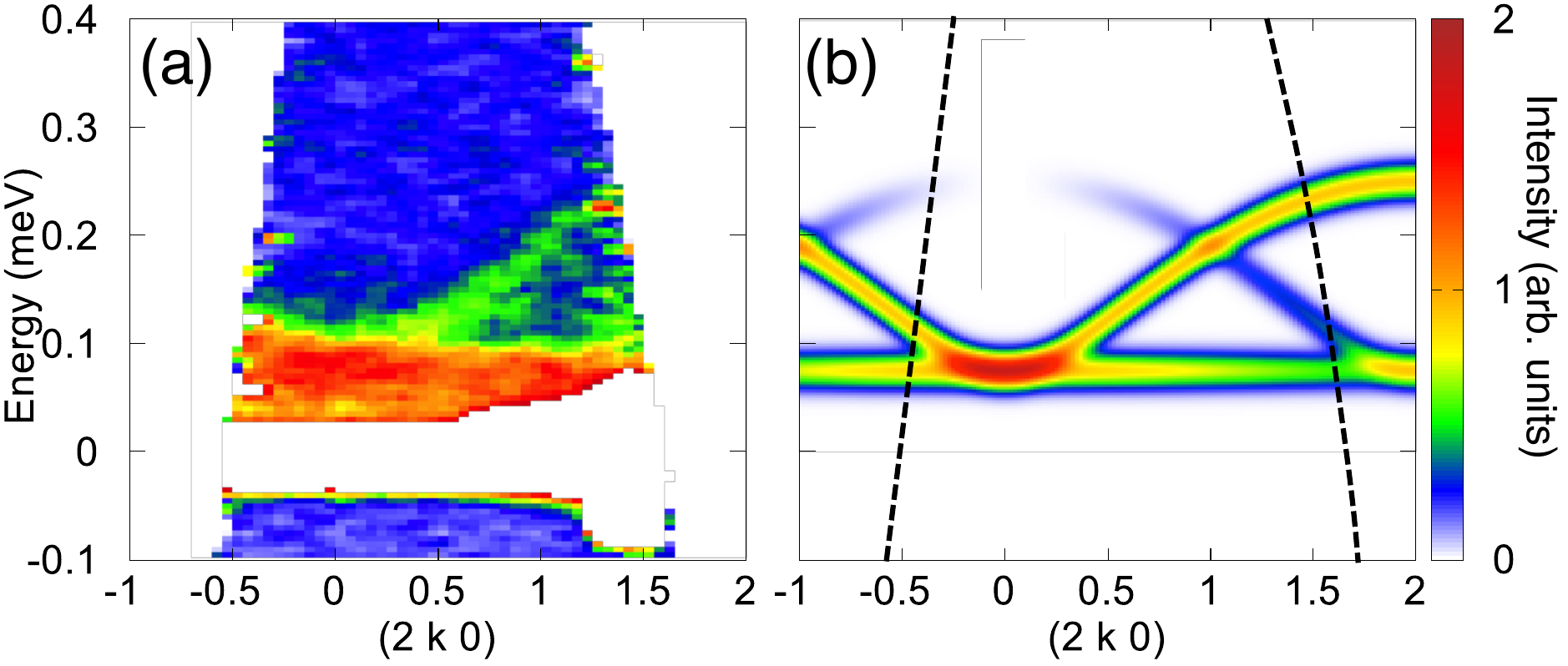}}
\caption{(a) Inelastic neutron scattering spectra measured in \ndtiB\ on IN5, and (b) calculated dispersion with the parameters of Table \ref{table_J}.}
\label{NdZrTip10_IN5}
\end{figure}

Inelastic neutron scattering on IN5 with the field applied along $[001]$
allowed us to determine the dispersion in the \ndtiB\ sample, and thus to access the parameters of the XYZ Hamiltonian using the Spinwave software \cite{Spinwave1,Spinwave2}. The obtained parameters are detailed in Table \ref{table_J}, and the measurements together with the calculations are shown in Figure \ref{NdZrTip10_IN5}.

In this case, since single crystal samples are available, neutron scattering experiments can be analyzed thoroughly and the coupling constants can be directly obtained from the fitting of the spectra. The uncertainty on those values is quite good, typically 10\%. 
In the present case, we could not directly use Eq. \ref{E_2} as the (220) wavevector is impossible to reach for this incident wavelength. Nevertheless, Eqs. \ref{E_0} and \ref{E_1} are already enough to put strong constraints on the parameters. 

\section{Comments about the Curie-Weiss temperatures}
\label{J_Theta}
The determination of the Curie-Weiss temperature in rare-earth based systems is often not as simple as it seems. This is especially true in Nd-based pyochlores where the Curie-Weiss temperature is less than 1 K, while fits of the susceptibility are performed in ranges of dozens of Kelvins. 
It is commonly believed that the determination is accurate when the Curie-Weiss fit is performed well below the energy of the first excited level ($\sim 25$ meV, i.e. 300 K in the case of \ndzr). This must nevertheless be taken with caution. In the Nd pyrochlores case, CEF calculations show that there is an important Van Vleck contribution so that the susceptibility is never perfectly linear. Associated to the fact that the Curie-Weiss value is very small, the obtained Curie-Weiss temperature thus varies depending on the chosen fit range. This is illustrated in Refs. \onlinecite{Xu_2015} and \onlinecite{Xu_2019} where two different fitting ranges were used, leading to two estimations of the Curie-Weiss temperature. 

In this paper, we have chosen the fitting range ($4-50$~K) based on our CEF calculations. Indeed, in this range the CEF susceptibility gives a $T_{\rm CW} \approx 0$, so that the fitted $T_{\rm CW}$ to the experimental data corresponds to the interactions contribution only. This is not the case for the fitting ranges in other articles \cite{Xu_2015, Xu_2019, Anand_2015, Gomez_2021, Scheie_2021}, which is thus expected to induce a shift of the obtained value. 

In addition, demagnetization corrections have to be taken into account because they shift the Curie-Weiss temperature towards larger values. The increase can reach around a hundred of milliKelvins in our case. 

In samples where the magnetic couplings could be determined from the inelastic neutron scattering spectra, the experimental value of $T_{\rm CW}$ can be compared to the one deduced from the couplings, which is given by \cite{Benton_2016}: 
$$T_{\rm CW}=\frac{1}{2k_{\rm B}}({\tilde {\sf J}}_z \cos^2(\theta) + {\tilde {\sf J}}_x \sin^2(\theta))$$
$\theta$ is deduced from neutron diffraction with a typical accuracy of about 10\%  as it is calculated using the ratio between the zero temperature ordered moment and the saturated moment. Note that due to the difficulty of thermalization of the samples, the ordered moment can be underestimated. Some uncertaintity also arises from the ${\sf J}$ values.  

The results are shown in Table \ref{table_TCW}. A discrepancy is observed between the values obtained from susceptibility ($T_{\rm CW\ meas}$) and the ones obtained from estimated couplings and $\theta$ ($T_{\rm CW\ calc}$). Nevertheless, there is not a systematic difference between the two values. This makes it difficult to identify the precise reasons for these discrepancies beyond the possible errors in the $T_{\rm CW\ meas}$ and $T_{\rm CW\ calc}$ determinations discussed above. 

\begin{table}
\begin{tabular}{*4{c}}
\hline
\hline
Sample	& $T_{\rm CW\ calc}$ (K) && $T_{\rm CW\ meas}$ (K) \\
\hline
\multicolumn{4}{c}{Single crystals} \\
\ndzr\    	& 0.280 && 0.330 \\
\ndtiA\ 	& 0.320 && 0.325 \\
\ndtiB 	& 0.435 && 0.300 \\ 
\hline
\multicolumn{4}{c}{Polycrystalline samples} \\ 
\ndlaA 	& 0.350 && 0.275 \\ 
\ndlaB  	& 0.285 && 0.220  \\ 
\hline
\hline
\end{tabular}
\caption{\label{table_TCW} Estimated Curie-Weiss temperatures from the coupling parameters and ordered moment ($T_{\rm CW\ calc}$) and the Curie-Weiss fits in the 4-50 K range on data corrected from demagnetization effects ($T_{\rm CW\ meas}$).  (Note that the Curie-Weiss temperatures for the pure and Ti substituted samples are different from Table \ref{Table_charac}, because the latter correspond to single crystal samples and not to powders.)  }
\end{table}

\section{Numerical simulations}
\label{simulations}
To better understand the remarkable stability of the all-in--all-out ground state in Nd pyrochlores, numerical simulations of spin dynamics have been carried out. To this end, a finite size system containing $N \times N\times N$ unit cells of the pyrochlore lattice was considered, among which $n$ sites were chosen at random, to simulate the substitution by lanthanum. These sites are therefore empty. As a first step, the mean-field equation based on Equation \ref{Hxyz} was solved until convergence. As a function of $n$, the pseudo-spin ordered magnetic moment decreases gradually. It does not reach exactly zero at the percolation threshold ($n/(16 N^3)\approx 0.61$), as expected due to the finite size of the system \cite{Henley_2001}.

Second, spin dynamics were calculated by solving the classical equations of motion. This includes integrating the differential equations governing the classical temporal evolution of a pseudo-spin at a given site. At the end of this stage, the correlations $\langle S_i S_j(t) \rangle$ are computed, which are Fourier transformed to obtain spectra as a function of energy transfer $\omega$. As \nd\ is a dipolar-octupolar Kramers ion, the actual magnetic moment is related to the pseudo-spin as $S_i =\eta_i \tau^z_i$, where $\eta_i$ denotes the local $\langle 111 \rangle$ direction. In addition, to benchmark the simulations, the $n=0$ calculations are compared to the spin-wave spectrum. Coupling parameters close to that of the pure sample $\tilde{\sf J}_x=1,~\tilde{\sf J}_z=-0.5$~K were used systematically. Results are shown in Figure \ref{disorder2}. \\

\bibliography{biblio}

\begin{thebibliography}{52}%
\makeatletter
\providecommand \@ifxundefined [1]{%
 \@ifx{#1\undefined}
}%
\providecommand \@ifnum [1]{%
 \ifnum #1\expandafter \@firstoftwo
 \else \expandafter \@secondoftwo
 \fi
}%
\providecommand \@ifx [1]{%
 \ifx #1\expandafter \@firstoftwo
 \else \expandafter \@secondoftwo
 \fi
}%
\providecommand \natexlab [1]{#1}%
\providecommand \enquote  [1]{``#1''}%
\providecommand \bibnamefont  [1]{#1}%
\providecommand \bibfnamefont [1]{#1}%
\providecommand \citenamefont [1]{#1}%
\providecommand \href@noop [0]{\@secondoftwo}%
\providecommand \href [0]{\begingroup \@sanitize@url \@href}%
\providecommand \@href[1]{\@@startlink{#1}\@@href}%
\providecommand \@@href[1]{\endgroup#1\@@endlink}%
\providecommand \@sanitize@url [0]{\catcode `\\12\catcode `\$12\catcode
  `\&12\catcode `\#12\catcode `\^12\catcode `\_12\catcode `\%12\relax}%
\providecommand \@@startlink[1]{}%
\providecommand \@@endlink[0]{}%
\providecommand \url  [0]{\begingroup\@sanitize@url \@url }%
\providecommand \@url [1]{\endgroup\@href {#1}{\urlprefix }}%
\providecommand \urlprefix  [0]{URL }%
\providecommand \Eprint [0]{\href }%
\providecommand \doibase [0]{https://doi.org/}%
\providecommand \selectlanguage [0]{\@gobble}%
\providecommand \bibinfo  [0]{\@secondoftwo}%
\providecommand \bibfield  [0]{\@secondoftwo}%
\providecommand \translation [1]{[#1]}%
\providecommand \BibitemOpen [0]{}%
\providecommand \bibitemStop [0]{}%
\providecommand \bibitemNoStop [0]{.\EOS\space}%
\providecommand \EOS [0]{\spacefactor3000\relax}%
\providecommand \BibitemShut  [1]{\csname bibitem#1\endcsname}%
\let\auto@bib@innerbib\@empty
\bibitem [{\citenamefont {Lacroix}(2011)}]{Lacroix_2011}%
  \BibitemOpen
  \bibfield  {author} {\bibinfo {author} {\bibfnamefont {C.}~\bibnamefont
  {Lacroix}},\ }\href {https://doi.org/10.1007/978-3-642-10589-0} {\emph
  {\bibinfo {title} {Introduction to Frustrated Magnetism}}},\ edited by\
  \bibinfo {editor} {\bibfnamefont {C.}~\bibnamefont {Lacroix}}, \bibinfo
  {editor} {\bibfnamefont {P.}~\bibnamefont {Mendels}},\ and\ \bibinfo {editor}
  {\bibfnamefont {F.}~\bibnamefont {Mila}}\ (\bibinfo  {publisher}
  {Springer-Verlag, Berlin},\ \bibinfo {year} {2011})\BibitemShut {NoStop}%
\bibitem [{\citenamefont {Moessner}\ and\ \citenamefont
  {Ramirez}(2006)}]{Moessner_2006}%
  \BibitemOpen
  \bibfield  {author} {\bibinfo {author} {\bibfnamefont {R.}~\bibnamefont
  {Moessner}}\ and\ \bibinfo {author} {\bibfnamefont {P.}~\bibnamefont
  {Ramirez}},\ }\bibfield  {title} {\bibinfo {title} {Geometrical
  frustration},\ }\href {https://doi.org/10.1063/1.2186278} {\bibfield
  {journal} {\bibinfo  {journal} {Physics Today}\ }\textbf {\bibinfo {volume}
  {59}},\ \bibinfo {pages} {24} (\bibinfo {year} {2006})}\BibitemShut {NoStop}%
\bibitem [{\citenamefont {Gardner}\ \emph {et~al.}(2010)\citenamefont
  {Gardner}, \citenamefont {Gingras},\ and\ \citenamefont
  {Greedan}}]{Gardner_2010}%
  \BibitemOpen
  \bibfield  {author} {\bibinfo {author} {\bibfnamefont {J.~S.}\ \bibnamefont
  {Gardner}}, \bibinfo {author} {\bibfnamefont {M.~J.~P.}\ \bibnamefont
  {Gingras}},\ and\ \bibinfo {author} {\bibfnamefont {J.~E.}\ \bibnamefont
  {Greedan}},\ }\bibfield  {title} {\bibinfo {title} {Magnetic pyrochlore
  oxides},\ }\href {https://doi.org/10.1103/RevModPhys.82.53} {\bibfield
  {journal} {\bibinfo  {journal} {Rev. Mod. Phys.}\ }\textbf {\bibinfo {volume}
  {82}},\ \bibinfo {pages} {53} (\bibinfo {year} {2010})}\BibitemShut {NoStop}%
\bibitem [{\citenamefont {Gingras}\ and\ \citenamefont
  {McClarty}(2014)}]{Gingras_2014}%
  \BibitemOpen
  \bibfield  {author} {\bibinfo {author} {\bibfnamefont {M.~J.~P.}\
  \bibnamefont {Gingras}}\ and\ \bibinfo {author} {\bibfnamefont {P.~A.}\
  \bibnamefont {McClarty}},\ }\bibfield  {title} {\bibinfo {title} {Quantum
  spin ice: a search for gapless quantum spin liquids in pyrochlore magnets},\
  }\href {https://doi.org/10.1088/0034-4885/77/5/056501} {\bibfield  {journal}
  {\bibinfo  {journal} {Rep. Prog. Phys.}\ }\textbf {\bibinfo {volume} {77}},\
  \bibinfo {pages} {056501} (\bibinfo {year} {2014})}\BibitemShut {NoStop}%
\bibitem [{\citenamefont {Harris}\ \emph {et~al.}(1997)\citenamefont {Harris},
  \citenamefont {Bramwell}, \citenamefont {McMorrow}, \citenamefont {Zeiske},\
  and\ \citenamefont {Godfrey}}]{Harris_1997}%
  \BibitemOpen
  \bibfield  {author} {\bibinfo {author} {\bibfnamefont {M.~J.}\ \bibnamefont
  {Harris}}, \bibinfo {author} {\bibfnamefont {S.~T.}\ \bibnamefont
  {Bramwell}}, \bibinfo {author} {\bibfnamefont {D.~F.}\ \bibnamefont
  {McMorrow}}, \bibinfo {author} {\bibfnamefont {T.}~\bibnamefont {Zeiske}},\
  and\ \bibinfo {author} {\bibfnamefont {K.~W.}\ \bibnamefont {Godfrey}},\
  }\bibfield  {title} {\bibinfo {title} {Geometrical frustration in the
  ferromagnetic pyrochlore {Ho}$_2${Ti}$_2${O}$_7$},\ }\href
  {https://doi.org/10.1103/PhysRevLett.79.2554} {\bibfield  {journal} {\bibinfo
   {journal} {Phys. Rev. Lett.}\ }\textbf {\bibinfo {volume} {79}},\ \bibinfo
  {pages} {2554} (\bibinfo {year} {1997})}\BibitemShut {NoStop}%
\bibitem [{\citenamefont {Ciomaga~Hatnean}\ \emph {et~al.}(2015)\citenamefont
  {Ciomaga~Hatnean}, \citenamefont {Lees}, \citenamefont {Petrenko},
  \citenamefont {Keeble}, \citenamefont {Balakrishnan}, \citenamefont
  {Gutmann}, \citenamefont {Klekovkina},\ and\ \citenamefont
  {Malkin}}]{Ciomaga_2015}%
  \BibitemOpen
  \bibfield  {author} {\bibinfo {author} {\bibfnamefont {M.}~\bibnamefont
  {Ciomaga~Hatnean}}, \bibinfo {author} {\bibfnamefont {M.~R.}\ \bibnamefont
  {Lees}}, \bibinfo {author} {\bibfnamefont {O.~A.}\ \bibnamefont {Petrenko}},
  \bibinfo {author} {\bibfnamefont {D.~S.}\ \bibnamefont {Keeble}}, \bibinfo
  {author} {\bibfnamefont {G.}~\bibnamefont {Balakrishnan}}, \bibinfo {author}
  {\bibfnamefont {M.~J.}\ \bibnamefont {Gutmann}}, \bibinfo {author}
  {\bibfnamefont {V.~V.}\ \bibnamefont {Klekovkina}},\ and\ \bibinfo {author}
  {\bibfnamefont {B.~Z.}\ \bibnamefont {Malkin}},\ }\bibfield  {title}
  {\bibinfo {title} {Structural and magnetic investigation of
  single-crystalline neodymium zirconate pyrochlore {Nd}$_2${Zr}$_2${O}$_7$},\
  }\href {https://doi.org/10.1103/PhysRevB.91.174416} {\bibfield  {journal}
  {\bibinfo  {journal} {Phys. Rev. B}\ }\textbf {\bibinfo {volume} {91}},\
  \bibinfo {pages} {174416} (\bibinfo {year} {2015})}\BibitemShut {NoStop}%
\bibitem [{\citenamefont {Lhotel}\ \emph {et~al.}(2015)\citenamefont {Lhotel},
  \citenamefont {Petit}, \citenamefont {Guitteny}, \citenamefont {Florea},
  \citenamefont {{Ciomaga Hatnean}}, \citenamefont {Colin}, \citenamefont
  {Ressouche}, \citenamefont {Lees},\ and\ \citenamefont
  {Balakrishnan}}]{Lhotel_2015}%
  \BibitemOpen
  \bibfield  {author} {\bibinfo {author} {\bibfnamefont {E.}~\bibnamefont
  {Lhotel}}, \bibinfo {author} {\bibfnamefont {S.}~\bibnamefont {Petit}},
  \bibinfo {author} {\bibfnamefont {S.}~\bibnamefont {Guitteny}}, \bibinfo
  {author} {\bibfnamefont {O.}~\bibnamefont {Florea}}, \bibinfo {author}
  {\bibfnamefont {M.}~\bibnamefont {{Ciomaga Hatnean}}}, \bibinfo {author}
  {\bibfnamefont {C.}~\bibnamefont {Colin}}, \bibinfo {author} {\bibfnamefont
  {E.}~\bibnamefont {Ressouche}}, \bibinfo {author} {\bibfnamefont {M.~R.}\
  \bibnamefont {Lees}},\ and\ \bibinfo {author} {\bibfnamefont
  {G.}~\bibnamefont {Balakrishnan}},\ }\bibfield  {title} {\bibinfo {title}
  {Fluctuations and all-in--all-out ordering in dipole-octupole
  {Nd}$_2${Zr}$_2${O}$_7$},\ }\href
  {https://doi.org/10.1103/PhysRevLett.115.197202} {\bibfield  {journal}
  {\bibinfo  {journal} {Phys. Rev. Lett.}\ }\textbf {\bibinfo {volume} {115}},\
  \bibinfo {pages} {197202} (\bibinfo {year} {2015})}\BibitemShut {NoStop}%
\bibitem [{\citenamefont {Xu}\ \emph {et~al.}(2015)\citenamefont {Xu},
  \citenamefont {Anand}, \citenamefont {Bera}, \citenamefont {Frontzek},
  \citenamefont {Abernathy}, \citenamefont {Casati}, \citenamefont
  {Siemensmeyer},\ and\ \citenamefont {Lake}}]{Xu_2015}%
  \BibitemOpen
  \bibfield  {author} {\bibinfo {author} {\bibfnamefont {J.}~\bibnamefont
  {Xu}}, \bibinfo {author} {\bibfnamefont {V.~K.}\ \bibnamefont {Anand}},
  \bibinfo {author} {\bibfnamefont {A.~K.}\ \bibnamefont {Bera}}, \bibinfo
  {author} {\bibfnamefont {M.}~\bibnamefont {Frontzek}}, \bibinfo {author}
  {\bibfnamefont {D.~L.}\ \bibnamefont {Abernathy}}, \bibinfo {author}
  {\bibfnamefont {N.}~\bibnamefont {Casati}}, \bibinfo {author} {\bibfnamefont
  {K.}~\bibnamefont {Siemensmeyer}},\ and\ \bibinfo {author} {\bibfnamefont
  {B.}~\bibnamefont {Lake}},\ }\bibfield  {title} {\bibinfo {title} {Magnetic
  structure and crystal field states of the pyrochlore antiferromagnet
  {Nd}$_2${Zr}$_2${O}$_7$},\ }\href
  {https://doi.org/10.1103/PhysRevB.92.224430} {\bibfield  {journal} {\bibinfo
  {journal} {Phys. Rev. B}\ }\textbf {\bibinfo {volume} {92}},\ \bibinfo
  {pages} {224430} (\bibinfo {year} {2015})}\BibitemShut {NoStop}%
\bibitem [{\citenamefont {Bertin}\ \emph {et~al.}(2015)\citenamefont {Bertin},
  \citenamefont {Dalmas~de R\'eotier}, \citenamefont {F\r{a}k}, \citenamefont
  {Marin}, \citenamefont {Yaouanc}, \citenamefont {Forget}, \citenamefont
  {Sheptyakov}, \citenamefont {Frick}, \citenamefont {Ritter}, \citenamefont
  {Amato}, \citenamefont {Baines},\ and\ \citenamefont {King}}]{Bertin_2015}%
  \BibitemOpen
  \bibfield  {author} {\bibinfo {author} {\bibfnamefont {A.}~\bibnamefont
  {Bertin}}, \bibinfo {author} {\bibfnamefont {P.}~\bibnamefont {Dalmas~de
  R\'eotier}}, \bibinfo {author} {\bibfnamefont {B.}~\bibnamefont {F\r{a}k}},
  \bibinfo {author} {\bibfnamefont {C.}~\bibnamefont {Marin}}, \bibinfo
  {author} {\bibfnamefont {A.}~\bibnamefont {Yaouanc}}, \bibinfo {author}
  {\bibfnamefont {A.}~\bibnamefont {Forget}}, \bibinfo {author} {\bibfnamefont
  {D.}~\bibnamefont {Sheptyakov}}, \bibinfo {author} {\bibfnamefont
  {B.}~\bibnamefont {Frick}}, \bibinfo {author} {\bibfnamefont
  {C.}~\bibnamefont {Ritter}}, \bibinfo {author} {\bibfnamefont
  {A.}~\bibnamefont {Amato}}, \bibinfo {author} {\bibfnamefont
  {C.}~\bibnamefont {Baines}},\ and\ \bibinfo {author} {\bibfnamefont
  {P.~J.~C.}\ \bibnamefont {King}},\ }\bibfield  {title} {\bibinfo {title}
  {{Nd}$_2${Sn}$_2${O}$_7$: an all-in/all-out pyrochlore with no divergence
  free field and anomalously slow paramagnetic spin dynamics},\ }\href
  {https://doi.org/10.1103/PhysRevB.92.144423} {\bibfield  {journal} {\bibinfo
  {journal} {Phys. Rev. B}\ }\textbf {\bibinfo {volume} {92}},\ \bibinfo
  {pages} {144423} (\bibinfo {year} {2015})}\BibitemShut {NoStop}%
\bibitem [{\citenamefont {Anand}\ \emph {et~al.}(2015)\citenamefont {Anand},
  \citenamefont {Bera}, \citenamefont {Xu}, \citenamefont
  {Herrmannsd{\"o}rfer}, \citenamefont {Ritter},\ and\ \citenamefont
  {Lake}}]{Anand_2015}%
  \BibitemOpen
  \bibfield  {author} {\bibinfo {author} {\bibfnamefont {V.~K.}\ \bibnamefont
  {Anand}}, \bibinfo {author} {\bibfnamefont {A.~K.}\ \bibnamefont {Bera}},
  \bibinfo {author} {\bibfnamefont {J.}~\bibnamefont {Xu}}, \bibinfo {author}
  {\bibfnamefont {T.}~\bibnamefont {Herrmannsd{\"o}rfer}}, \bibinfo {author}
  {\bibfnamefont {C.}~\bibnamefont {Ritter}},\ and\ \bibinfo {author}
  {\bibfnamefont {B.}~\bibnamefont {Lake}},\ }\bibfield  {title} {\bibinfo
  {title} {Observation of long-range magnetic ordering in pyrohafnate
  {Nd}$_2${Hf}$_2${O}$_7$: a neutron diffraction study},\ }\href
  {https://doi.org/10.1103/PhysRevB.92.184418} {\bibfield  {journal} {\bibinfo
  {journal} {Phys. Rev. B}\ }\textbf {\bibinfo {volume} {92}},\ \bibinfo
  {pages} {184418} (\bibinfo {year} {2015})}\BibitemShut {NoStop}%
\bibitem [{\citenamefont {Anand}\ \emph {et~al.}(2017)\citenamefont {Anand},
  \citenamefont {Abernathy}, \citenamefont {Adroja}, \citenamefont {Hillier},
  \citenamefont {Biswas},\ and\ \citenamefont {Lake}}]{Anand_2017}%
  \BibitemOpen
  \bibfield  {author} {\bibinfo {author} {\bibfnamefont {V.~K.}\ \bibnamefont
  {Anand}}, \bibinfo {author} {\bibfnamefont {D.~L.}\ \bibnamefont
  {Abernathy}}, \bibinfo {author} {\bibfnamefont {D.~T.}\ \bibnamefont
  {Adroja}}, \bibinfo {author} {\bibfnamefont {A.~D.}\ \bibnamefont {Hillier}},
  \bibinfo {author} {\bibfnamefont {P.~K.}\ \bibnamefont {Biswas}},\ and\
  \bibinfo {author} {\bibfnamefont {B.}~\bibnamefont {Lake}},\ }\bibfield
  {title} {\bibinfo {title} {Muon spin relaxation and inelastic neutron
  scattering investigations of the all-in/all-out antiferromagnet
  {Nd}$_2${Hf}$_2${O}$_7$},\ }\href
  {https://doi.org/10.1103/PhysRevB.95.224420} {\bibfield  {journal} {\bibinfo
  {journal} {Phys. Rev. B}\ }\textbf {\bibinfo {volume} {95}},\ \bibinfo
  {pages} {224420} (\bibinfo {year} {2017})}\BibitemShut {NoStop}%
\bibitem [{\citenamefont {Petit}\ \emph {et~al.}(2016)\citenamefont {Petit},
  \citenamefont {Lhotel}, \citenamefont {Canals}, \citenamefont {{Ciomaga
  Hatnean}}, \citenamefont {Ollivier}, \citenamefont {Mutka}, \citenamefont
  {Ressouche}, \citenamefont {Wildes}, \citenamefont {Lees},\ and\
  \citenamefont {Balakrishnan}}]{Petit_2016}%
  \BibitemOpen
  \bibfield  {author} {\bibinfo {author} {\bibfnamefont {S.}~\bibnamefont
  {Petit}}, \bibinfo {author} {\bibfnamefont {E.}~\bibnamefont {Lhotel}},
  \bibinfo {author} {\bibfnamefont {B.}~\bibnamefont {Canals}}, \bibinfo
  {author} {\bibfnamefont {M.}~\bibnamefont {{Ciomaga Hatnean}}}, \bibinfo
  {author} {\bibfnamefont {J.}~\bibnamefont {Ollivier}}, \bibinfo {author}
  {\bibfnamefont {H.}~\bibnamefont {Mutka}}, \bibinfo {author} {\bibfnamefont
  {E.}~\bibnamefont {Ressouche}}, \bibinfo {author} {\bibfnamefont {A.~R.}\
  \bibnamefont {Wildes}}, \bibinfo {author} {\bibfnamefont {M.~R.}\
  \bibnamefont {Lees}},\ and\ \bibinfo {author} {\bibfnamefont
  {G.}~\bibnamefont {Balakrishnan}},\ }\bibfield  {title} {\bibinfo {title}
  {Observation of magnetic fragmentation in spin ice},\ }\href
  {https://doi.org/10.1038/NPHYS3710} {\bibfield  {journal} {\bibinfo
  {journal} {Nat. Phys.}\ }\textbf {\bibinfo {volume} {12}},\ \bibinfo {pages}
  {746} (\bibinfo {year} {2016})}\BibitemShut {NoStop}%
\bibitem [{\citenamefont {Lhotel}\ \emph {et~al.}(2018)\citenamefont {Lhotel},
  \citenamefont {Petit}, \citenamefont {Ciomaga~Hatnean}, \citenamefont
  {Ollivier}, \citenamefont {Mutka}, \citenamefont {Ressouche}, \citenamefont
  {Lees},\ and\ \citenamefont {Balakrishnan}}]{Lhotel_2018}%
  \BibitemOpen
  \bibfield  {author} {\bibinfo {author} {\bibfnamefont {E.}~\bibnamefont
  {Lhotel}}, \bibinfo {author} {\bibfnamefont {S.}~\bibnamefont {Petit}},
  \bibinfo {author} {\bibfnamefont {M.}~\bibnamefont {Ciomaga~Hatnean}},
  \bibinfo {author} {\bibfnamefont {J.}~\bibnamefont {Ollivier}}, \bibinfo
  {author} {\bibfnamefont {H.}~\bibnamefont {Mutka}}, \bibinfo {author}
  {\bibfnamefont {E.}~\bibnamefont {Ressouche}}, \bibinfo {author}
  {\bibfnamefont {M.~R.}\ \bibnamefont {Lees}},\ and\ \bibinfo {author}
  {\bibfnamefont {G.}~\bibnamefont {Balakrishnan}},\ }\bibfield  {title}
  {\bibinfo {title} {Evidence for dynamic kagome ice},\ }\href
  {https://doi.org/10.1038/s41467-018-06212-2} {\bibfield  {journal} {\bibinfo
  {journal} {Nat. Commun.}\ }\textbf {\bibinfo {volume} {9}},\ \bibinfo {pages}
  {3786} (\bibinfo {year} {2018})}\BibitemShut {NoStop}%
\bibitem [{\citenamefont {Xu}\ \emph {et~al.}(2019)\citenamefont {Xu},
  \citenamefont {Benton}, \citenamefont {Anand}, \citenamefont {Islam},
  \citenamefont {Guidi}, \citenamefont {Ehlers}, \citenamefont {Feng},
  \citenamefont {Su}, \citenamefont {Sakai}, \citenamefont {Gegenwart},\ and\
  \citenamefont {Lake}}]{Xu_2019}%
  \BibitemOpen
  \bibfield  {author} {\bibinfo {author} {\bibfnamefont {J.}~\bibnamefont
  {Xu}}, \bibinfo {author} {\bibfnamefont {O.}~\bibnamefont {Benton}}, \bibinfo
  {author} {\bibfnamefont {V.~K.}\ \bibnamefont {Anand}}, \bibinfo {author}
  {\bibfnamefont {A.~T. M.~N.}\ \bibnamefont {Islam}}, \bibinfo {author}
  {\bibfnamefont {T.}~\bibnamefont {Guidi}}, \bibinfo {author} {\bibfnamefont
  {G.}~\bibnamefont {Ehlers}}, \bibinfo {author} {\bibfnamefont
  {E.}~\bibnamefont {Feng}}, \bibinfo {author} {\bibfnamefont {Y.}~\bibnamefont
  {Su}}, \bibinfo {author} {\bibfnamefont {A.}~\bibnamefont {Sakai}}, \bibinfo
  {author} {\bibfnamefont {P.}~\bibnamefont {Gegenwart}},\ and\ \bibinfo
  {author} {\bibfnamefont {B.}~\bibnamefont {Lake}},\ }\bibfield  {title}
  {\bibinfo {title} {Anisotropic exchange {H}amiltonian, magnetic phase diagram
  and domain inversion of {Nd}$_2${Zr}$_2${O}$_7$},\ }\href
  {https://doi.org/10.1103/PhysRevB.99.144420} {\bibfield  {journal} {\bibinfo
  {journal} {Phys. Rev. B}\ }\textbf {\bibinfo {volume} {99}},\ \bibinfo
  {pages} {144420} (\bibinfo {year} {2019})}\BibitemShut {NoStop}%
\bibitem [{\citenamefont {Benton}(2016)}]{Benton_2016}%
  \BibitemOpen
  \bibfield  {author} {\bibinfo {author} {\bibfnamefont {O.}~\bibnamefont
  {Benton}},\ }\bibfield  {title} {\bibinfo {title} {Quantum origins of moment
  fragmentation in {Nd}$_2${Zr}$_2${O}$_7$},\ }\href
  {https://doi.org/10.1103/PhysRevB.94.104430} {\bibfield  {journal} {\bibinfo
  {journal} {Phys. Rev. B}\ }\textbf {\bibinfo {volume} {94}},\ \bibinfo
  {pages} {104430} (\bibinfo {year} {2016})}\BibitemShut {NoStop}%
\bibitem [{\citenamefont {Huang}\ \emph {et~al.}(2014)\citenamefont {Huang},
  \citenamefont {Chen},\ and\ \citenamefont {Hermele}}]{Huang_2014}%
  \BibitemOpen
  \bibfield  {author} {\bibinfo {author} {\bibfnamefont {Y.-P.}\ \bibnamefont
  {Huang}}, \bibinfo {author} {\bibfnamefont {G.}~\bibnamefont {Chen}},\ and\
  \bibinfo {author} {\bibfnamefont {M.}~\bibnamefont {Hermele}},\ }\bibfield
  {title} {\bibinfo {title} {Quantum spin ice and topological phases from
  dipolar-octupolar doublets on the pyrochlore lattice},\ }\href
  {https://doi.org/10.1103/PhysRevLett.112.167203} {\bibfield  {journal}
  {\bibinfo  {journal} {Phys. Rev. Lett.}\ }\textbf {\bibinfo {volume} {112}},\
  \bibinfo {pages} {167203} (\bibinfo {year} {2014})}\BibitemShut {NoStop}%
\bibitem [{\citenamefont {Xu}\ \emph {et~al.}(2020)\citenamefont {Xu},
  \citenamefont {Benton}, \citenamefont {Islam}, \citenamefont {Guidi},
  \citenamefont {Ehlers},\ and\ \citenamefont {Lake}}]{Xu_2020}%
  \BibitemOpen
  \bibfield  {author} {\bibinfo {author} {\bibfnamefont {J.}~\bibnamefont
  {Xu}}, \bibinfo {author} {\bibfnamefont {O.}~\bibnamefont {Benton}}, \bibinfo
  {author} {\bibfnamefont {A.~T. M.~N.}\ \bibnamefont {Islam}}, \bibinfo
  {author} {\bibfnamefont {T.}~\bibnamefont {Guidi}}, \bibinfo {author}
  {\bibfnamefont {G.}~\bibnamefont {Ehlers}},\ and\ \bibinfo {author}
  {\bibfnamefont {B.}~\bibnamefont {Lake}},\ }\bibfield  {title} {\bibinfo
  {title} {Order out of a {C}oulomb phase and {H}iggs transition: Frustrated
  transverse interactions of {Nd}$_2${Zr}$_2${O}$_7$},\ }\href
  {https://doi.org/10.1103/PhysRevLett.124.097203} {\bibfield  {journal}
  {\bibinfo  {journal} {Phys. Rev. Lett.}\ }\textbf {\bibinfo {volume} {124}},\
  \bibinfo {pages} {097203} (\bibinfo {year} {2020})}\BibitemShut {NoStop}%
\bibitem [{\citenamefont {L\'eger}\ \emph
  {et~al.}(2021{\natexlab{a}})\citenamefont {L\'eger}, \citenamefont {Lhotel},
  \citenamefont {Ciomaga~Hatnean}, \citenamefont {Ollivier}, \citenamefont
  {Wildes}, \citenamefont {Raymond}, \citenamefont {Ressouche}, \citenamefont
  {Balakrishnan},\ and\ \citenamefont {Petit}}]{Leger_2021}%
  \BibitemOpen
  \bibfield  {author} {\bibinfo {author} {\bibfnamefont {M.}~\bibnamefont
  {L\'eger}}, \bibinfo {author} {\bibfnamefont {E.}~\bibnamefont {Lhotel}},
  \bibinfo {author} {\bibfnamefont {M.}~\bibnamefont {Ciomaga~Hatnean}},
  \bibinfo {author} {\bibfnamefont {J.}~\bibnamefont {Ollivier}}, \bibinfo
  {author} {\bibfnamefont {A.~R.}\ \bibnamefont {Wildes}}, \bibinfo {author}
  {\bibfnamefont {S.}~\bibnamefont {Raymond}}, \bibinfo {author} {\bibfnamefont
  {E.}~\bibnamefont {Ressouche}}, \bibinfo {author} {\bibfnamefont
  {G.}~\bibnamefont {Balakrishnan}},\ and\ \bibinfo {author} {\bibfnamefont
  {S.}~\bibnamefont {Petit}},\ }\bibfield  {title} {\bibinfo {title} {Spin
  dynamics and unconventional {C}oulomb phase in {Nd}$_2${Zr}$_2${O}$_7$},\
  }\href {https://doi.org/10.1103/PhysRevLett.126.247201} {\bibfield  {journal}
  {\bibinfo  {journal} {Phys. Rev. Lett.}\ }\textbf {\bibinfo {volume} {126}},\
  \bibinfo {pages} {247201} (\bibinfo {year} {2021}{\natexlab{a}})}\BibitemShut
  {NoStop}%
\bibitem [{\citenamefont {Benton}(2020)}]{Benton_2020}%
  \BibitemOpen
  \bibfield  {author} {\bibinfo {author} {\bibfnamefont {O.}~\bibnamefont
  {Benton}},\ }\bibfield  {title} {\bibinfo {title} {Ground-state phase diagram
  of dipolar-octupolar pyrochlores},\ }\href
  {https://doi.org/10.1103/PhysRevB.102.104408} {\bibfield  {journal} {\bibinfo
   {journal} {Phys. Rev. B}\ }\textbf {\bibinfo {volume} {102}},\ \bibinfo
  {pages} {104408} (\bibinfo {year} {2020})}\BibitemShut {NoStop}%
\bibitem [{\citenamefont {Patri}\ \emph {et~al.}(2020)\citenamefont {Patri},
  \citenamefont {Hosoi},\ and\ \citenamefont {Kim}}]{Patri_2020}%
  \BibitemOpen
  \bibfield  {author} {\bibinfo {author} {\bibfnamefont {A.~S.}\ \bibnamefont
  {Patri}}, \bibinfo {author} {\bibfnamefont {M.}~\bibnamefont {Hosoi}},\ and\
  \bibinfo {author} {\bibfnamefont {Y.~B.}\ \bibnamefont {Kim}},\ }\bibfield
  {title} {\bibinfo {title} {Distinguishing dipolar and octupolar quantum spin
  ices using contrasting magnetostriction signatures},\ }\href
  {https://doi.org/10.1103/PhysRevResearch.2.023253} {\bibfield  {journal}
  {\bibinfo  {journal} {Phys. Rev. Research}\ }\textbf {\bibinfo {volume}
  {2}},\ \bibinfo {pages} {023253} (\bibinfo {year} {2020})}\BibitemShut
  {NoStop}%
\bibitem [{\citenamefont {Taniguchi}\ \emph {et~al.}(2013)\citenamefont
  {Taniguchi}, \citenamefont {Kadowaki}, \citenamefont {Takatsu}, \citenamefont
  {F\r{a}k}, \citenamefont {Ollivier}, \citenamefont {Yamazaki}, \citenamefont
  {Sato}, \citenamefont {Yoshizawa}, \citenamefont {Shimura}, \citenamefont
  {Sakakibara}, \citenamefont {Hong}, \citenamefont {Goto}, \citenamefont
  {Yaraskavitch},\ and\ \citenamefont {Kycia}}]{Taniguchi_2013}%
  \BibitemOpen
  \bibfield  {author} {\bibinfo {author} {\bibfnamefont {T.}~\bibnamefont
  {Taniguchi}}, \bibinfo {author} {\bibfnamefont {H.}~\bibnamefont {Kadowaki}},
  \bibinfo {author} {\bibfnamefont {H.}~\bibnamefont {Takatsu}}, \bibinfo
  {author} {\bibfnamefont {B.}~\bibnamefont {F\r{a}k}}, \bibinfo {author}
  {\bibfnamefont {J.}~\bibnamefont {Ollivier}}, \bibinfo {author}
  {\bibfnamefont {T.}~\bibnamefont {Yamazaki}}, \bibinfo {author}
  {\bibfnamefont {T.~J.}\ \bibnamefont {Sato}}, \bibinfo {author}
  {\bibfnamefont {H.}~\bibnamefont {Yoshizawa}}, \bibinfo {author}
  {\bibfnamefont {Y.}~\bibnamefont {Shimura}}, \bibinfo {author} {\bibfnamefont
  {T.}~\bibnamefont {Sakakibara}}, \bibinfo {author} {\bibfnamefont
  {T.}~\bibnamefont {Hong}}, \bibinfo {author} {\bibfnamefont {K.}~\bibnamefont
  {Goto}}, \bibinfo {author} {\bibfnamefont {L.~R.}\ \bibnamefont
  {Yaraskavitch}},\ and\ \bibinfo {author} {\bibfnamefont {J.~B.}\ \bibnamefont
  {Kycia}},\ }\bibfield  {title} {\bibinfo {title} {Long-range order and
  spin-liquid states of polycrystalline {Tb$_{2+x}$Ti$_{2-x}$O$_{7+y}$}},\
  }\href {https://doi.org/10.1103/PhysRevB.87.060408} {\bibfield  {journal}
  {\bibinfo  {journal} {Phys. Rev. B}\ }\textbf {\bibinfo {volume} {87}},\
  \bibinfo {pages} {060408(R)} (\bibinfo {year} {2013})}\BibitemShut {NoStop}%
\bibitem [{\citenamefont {Shirai}\ \emph {et~al.}(2017)\citenamefont {Shirai},
  \citenamefont {Freitas}, \citenamefont {Lago}, \citenamefont {Bramwell},
  \citenamefont {Ritter},\ and\ \citenamefont {\v{Z}ivkovi\'c}}]{Shirai_2017}%
  \BibitemOpen
  \bibfield  {author} {\bibinfo {author} {\bibfnamefont {M.}~\bibnamefont
  {Shirai}}, \bibinfo {author} {\bibfnamefont {R.~S.}\ \bibnamefont {Freitas}},
  \bibinfo {author} {\bibfnamefont {J.}~\bibnamefont {Lago}}, \bibinfo {author}
  {\bibfnamefont {S.~T.}\ \bibnamefont {Bramwell}}, \bibinfo {author}
  {\bibfnamefont {C.}~\bibnamefont {Ritter}},\ and\ \bibinfo {author}
  {\bibfnamefont {I.}~\bibnamefont {\v{Z}ivkovi\'c}},\ }\bibfield  {title}
  {\bibinfo {title} {Doping-induced quantum crossover in
  er$_2$ti$_{2-x}$sn$_x$o$_7$},\ }\href
  {https://doi.org/10.1103/PhysRevB.96.180411} {\bibfield  {journal} {\bibinfo
  {journal} {Phys. Rev. B}\ }\textbf {\bibinfo {volume} {96}},\ \bibinfo
  {pages} {180411(R)} (\bibinfo {year} {2017})}\BibitemShut {NoStop}%
\bibitem [{\citenamefont {Arpino}\ \emph {et~al.}(2017)\citenamefont {Arpino},
  \citenamefont {Trump}, \citenamefont {Scheie}, \citenamefont {McQueen},\ and\
  \citenamefont {Koohpayeh}}]{Arpino_2017}%
  \BibitemOpen
  \bibfield  {author} {\bibinfo {author} {\bibfnamefont {K.~E.}\ \bibnamefont
  {Arpino}}, \bibinfo {author} {\bibfnamefont {B.~A.}\ \bibnamefont {Trump}},
  \bibinfo {author} {\bibfnamefont {A.~O.}\ \bibnamefont {Scheie}}, \bibinfo
  {author} {\bibfnamefont {T.~M.}\ \bibnamefont {McQueen}},\ and\ \bibinfo
  {author} {\bibfnamefont {S.~M.}\ \bibnamefont {Koohpayeh}},\ }\bibfield
  {title} {\bibinfo {title} {Impact of stoichiometry of {Yb$_2$Ti$_2$O$_7$} on
  its physical properties},\ }\href
  {https://doi.org/10.1103/PhysRevB.95.094407} {\bibfield  {journal} {\bibinfo
  {journal} {Phys. Rev. B}\ }\textbf {\bibinfo {volume} {95}},\ \bibinfo
  {pages} {094407} (\bibinfo {year} {2017})}\BibitemShut {NoStop}%
\bibitem [{\citenamefont {Robert}\ \emph {et~al.}(2015)\citenamefont {Robert},
  \citenamefont {Lhotel}, \citenamefont {Remenyi}, \citenamefont {Sahling},
  \citenamefont {Mirebeau}, \citenamefont {Decorse}, \citenamefont {Canals},\
  and\ \citenamefont {Petit}}]{Robert_2015}%
  \BibitemOpen
  \bibfield  {author} {\bibinfo {author} {\bibfnamefont {J.}~\bibnamefont
  {Robert}}, \bibinfo {author} {\bibfnamefont {E.}~\bibnamefont {Lhotel}},
  \bibinfo {author} {\bibfnamefont {G.}~\bibnamefont {Remenyi}}, \bibinfo
  {author} {\bibfnamefont {S.}~\bibnamefont {Sahling}}, \bibinfo {author}
  {\bibfnamefont {I.}~\bibnamefont {Mirebeau}}, \bibinfo {author}
  {\bibfnamefont {C.}~\bibnamefont {Decorse}}, \bibinfo {author} {\bibfnamefont
  {B.}~\bibnamefont {Canals}},\ and\ \bibinfo {author} {\bibfnamefont
  {S.}~\bibnamefont {Petit}},\ }\bibfield  {title} {\bibinfo {title} {Spin
  dynamics in the presence of competing ferromagnetic and antiferromagnetic
  correlations in yb$_2$ti$_2$o$_7$},\ }\href
  {https://doi.org/10.1103/PhysRevB.92.064425} {\bibfield  {journal} {\bibinfo
  {journal} {Phys. Rev. B}\ }\textbf {\bibinfo {volume} {92}},\ \bibinfo
  {pages} {064425} (\bibinfo {year} {2015})}\BibitemShut {NoStop}%
\bibitem [{\citenamefont {Jaubert}\ \emph {et~al.}(2015)\citenamefont
  {Jaubert}, \citenamefont {Benton}, \citenamefont {Rau}, \citenamefont
  {Oitmaa}, \citenamefont {Singh}, \citenamefont {Shannon}, ,\ and\
  \citenamefont {Gingras}}]{Jaubert_2015}%
  \BibitemOpen
  \bibfield  {author} {\bibinfo {author} {\bibfnamefont {L.~D.~C.}\
  \bibnamefont {Jaubert}}, \bibinfo {author} {\bibfnamefont {O.}~\bibnamefont
  {Benton}}, \bibinfo {author} {\bibfnamefont {J.~G.}\ \bibnamefont {Rau}},
  \bibinfo {author} {\bibfnamefont {J.}~\bibnamefont {Oitmaa}}, \bibinfo
  {author} {\bibfnamefont {R.~R.~P.}\ \bibnamefont {Singh}}, \bibinfo {author}
  {\bibfnamefont {N.}~\bibnamefont {Shannon}}, ,\ and\ \bibinfo {author}
  {\bibfnamefont {M.~J.~P.}\ \bibnamefont {Gingras}},\ }\bibfield  {title}
  {\bibinfo {title} {Are multiphase competition and order by disorder the keys
  to understanding yb$_2$ti$_2$o$_7$?},\ }\href
  {https://doi.org/10.1103/PhysRevLett.115.267208} {\bibfield  {journal}
  {\bibinfo  {journal} {Phys. Rev. Lett.}\ }\textbf {\bibinfo {volume} {115}},\
  \bibinfo {pages} {267208} (\bibinfo {year} {2015})}\BibitemShut {NoStop}%
\bibitem [{\citenamefont {Scheie}\ \emph {et~al.}(2020)\citenamefont {Scheie},
  \citenamefont {Kindervater}, \citenamefont {Zhang}, \citenamefont
  {Changlani}, \citenamefont {Sala}, \citenamefont {Ehlers}, \citenamefont
  {Heinemann}, \citenamefont {Tucker}, \citenamefont {Koohpayeh},\ and\
  \citenamefont {Broholm}}]{Scheie_2020}%
  \BibitemOpen
  \bibfield  {author} {\bibinfo {author} {\bibfnamefont {A.}~\bibnamefont
  {Scheie}}, \bibinfo {author} {\bibfnamefont {J.}~\bibnamefont {Kindervater}},
  \bibinfo {author} {\bibfnamefont {S.}~\bibnamefont {Zhang}}, \bibinfo
  {author} {\bibfnamefont {H.~J.}\ \bibnamefont {Changlani}}, \bibinfo {author}
  {\bibfnamefont {G.}~\bibnamefont {Sala}}, \bibinfo {author} {\bibfnamefont
  {G.}~\bibnamefont {Ehlers}}, \bibinfo {author} {\bibfnamefont
  {A.}~\bibnamefont {Heinemann}}, \bibinfo {author} {\bibfnamefont {G.~S.}\
  \bibnamefont {Tucker}}, \bibinfo {author} {\bibfnamefont {S.~M.}\
  \bibnamefont {Koohpayeh}},\ and\ \bibinfo {author} {\bibfnamefont
  {C.}~\bibnamefont {Broholm}},\ }\bibfield  {title} {\bibinfo {title}
  {Multiphase magnetism in {Yb$_2$Ti$_2$O$_7$}},\ }\bibfield  {journal}
  {\bibinfo  {journal} {Proc. Natl. Acad. Sci. USA}\ }\textbf {\bibinfo
  {volume} {117}},\ \href {https://doi.org/10.1073/pnas.2008791117}
  {10.1073/pnas.2008791117} (\bibinfo {year} {2020})\BibitemShut {NoStop}%
\bibitem [{\citenamefont {Zoghlin}\ \emph {et~al.}(2021)\citenamefont
  {Zoghlin}, \citenamefont {Schmehr}, \citenamefont {Holgate}, \citenamefont
  {Dally}, \citenamefont {Liu}, \citenamefont {Laurita},\ and\ \citenamefont
  {Wilson}}]{Zoghlin_2021}%
  \BibitemOpen
  \bibfield  {author} {\bibinfo {author} {\bibfnamefont {E.}~\bibnamefont
  {Zoghlin}}, \bibinfo {author} {\bibfnamefont {J.}~\bibnamefont {Schmehr}},
  \bibinfo {author} {\bibfnamefont {C.}~\bibnamefont {Holgate}}, \bibinfo
  {author} {\bibfnamefont {R.}~\bibnamefont {Dally}}, \bibinfo {author}
  {\bibfnamefont {Y.}~\bibnamefont {Liu}}, \bibinfo {author} {\bibfnamefont
  {G.}~\bibnamefont {Laurita}},\ and\ \bibinfo {author} {\bibfnamefont {S.~D.}\
  \bibnamefont {Wilson}},\ }\bibfield  {title} {\bibinfo {title} {Evaluating
  the effects of structural disorder on the magnetic properties of
  {Nd$_2$Zr$_2$O$_7$}},\ }\href
  {https://doi.org/10.1103/PhysRevMaterials.5.084403} {\bibfield  {journal}
  {\bibinfo  {journal} {Phys. Rev. Mater.}\ }\textbf {\bibinfo {volume} {5}},\
  \bibinfo {pages} {084403} (\bibinfo {year} {2021})}\BibitemShut {NoStop}%
\bibitem [{\citenamefont {L\'eger}\ \emph
  {et~al.}(2021{\natexlab{b}})\citenamefont {L\'eger}, \citenamefont {Lhotel},
  \citenamefont {Ressouche}, \citenamefont {Beauvois}, \citenamefont {Damay},
  \citenamefont {Paulsen}, \citenamefont {Al-Mawla}, \citenamefont {Suard},
  \citenamefont {Ciomaga~Hatnean}, \citenamefont {Balakrishnan},\ and\
  \citenamefont {Petit}}]{Leger_2021b}%
  \BibitemOpen
  \bibfield  {author} {\bibinfo {author} {\bibfnamefont {M.}~\bibnamefont
  {L\'eger}}, \bibinfo {author} {\bibfnamefont {E.}~\bibnamefont {Lhotel}},
  \bibinfo {author} {\bibfnamefont {E.}~\bibnamefont {Ressouche}}, \bibinfo
  {author} {\bibfnamefont {K.}~\bibnamefont {Beauvois}}, \bibinfo {author}
  {\bibfnamefont {F.}~\bibnamefont {Damay}}, \bibinfo {author} {\bibfnamefont
  {C.}~\bibnamefont {Paulsen}}, \bibinfo {author} {\bibfnamefont
  {A.}~\bibnamefont {Al-Mawla}}, \bibinfo {author} {\bibfnamefont
  {E.}~\bibnamefont {Suard}}, \bibinfo {author} {\bibfnamefont
  {M.}~\bibnamefont {Ciomaga~Hatnean}}, \bibinfo {author} {\bibfnamefont
  {G.}~\bibnamefont {Balakrishnan}},\ and\ \bibinfo {author} {\bibfnamefont
  {S.}~\bibnamefont {Petit}},\ }\bibfield  {title} {\bibinfo {title}
  {Field-temperature phase diagram of the enigmatic
  {Nd}$_2$({Zr}$_{1-x}${Ti}$_x$)$_2${O}$_7$ pyrochlore magnets},\ }\href
  {https://doi.org/10.1103/PhysRevB.103.214449} {\bibfield  {journal} {\bibinfo
   {journal} {Phys. Rev. B}\ }\textbf {\bibinfo {volume} {103}},\ \bibinfo
  {pages} {214449} (\bibinfo {year} {2021}{\natexlab{b}})}\BibitemShut
  {NoStop}%
\bibitem [{\citenamefont {{Ciomaga Hatnean}}\ \emph {et~al.}(2015)\citenamefont
  {{Ciomaga Hatnean}}, \citenamefont {Lees},\ and\ \citenamefont
  {Balakrishnan}}]{Ciomaga_2015b}%
  \BibitemOpen
  \bibfield  {author} {\bibinfo {author} {\bibfnamefont {M.}~\bibnamefont
  {{Ciomaga Hatnean}}}, \bibinfo {author} {\bibfnamefont {M.~R.}\ \bibnamefont
  {Lees}},\ and\ \bibinfo {author} {\bibfnamefont {G.}~\bibnamefont
  {Balakrishnan}},\ }\bibfield  {title} {\bibinfo {title} {Growth of
  single-crystals of rare-earth zirconate pyrochlores, ${L}n_2${Zr}$_2${O}$_7$
  (with ${L}n$ = {La}, {Nd}, {Sm}, and {Gd}) by the floating zone technique},\
  }\href {https://doi.org/10.1016/j.jcrysgro.2015.01.037} {\bibfield  {journal}
  {\bibinfo  {journal} {J. Cryst. Growth}\ }\textbf {\bibinfo {volume} {418}},\
  \bibinfo {pages} {1} (\bibinfo {year} {2015})}\BibitemShut {NoStop}%
\bibitem [{\citenamefont {Vayer}\ \emph {et~al.}(2022)\citenamefont {Vayer},
  \citenamefont {Decorse}, \citenamefont {B{\'e}rardan}, \citenamefont
  {Dragoe},\ and\ \citenamefont {Dragoe}}]{Vayer_2022}%
  \BibitemOpen
  \bibfield  {author} {\bibinfo {author} {\bibfnamefont {F.}~\bibnamefont
  {Vayer}}, \bibinfo {author} {\bibfnamefont {C.}~\bibnamefont {Decorse}},
  \bibinfo {author} {\bibfnamefont {D.}~\bibnamefont {B{\'e}rardan}}, \bibinfo
  {author} {\bibfnamefont {D.}~\bibnamefont {Dragoe}},\ and\ \bibinfo {author}
  {\bibfnamefont {N.}~\bibnamefont {Dragoe}},\ }\bibfield  {title} {\bibinfo
  {title} {Investigation of the chemical versatility in high entropy
  pyrochlores},\ }\href {https://doi.org/10.1111/jace.18922} {\bibfield
  {journal} {\bibinfo  {journal} {J. Am. Ceram. Soc.}\ }\textbf {\bibinfo
  {volume} {106}},\ \bibinfo {pages} {2601} (\bibinfo {year}
  {2022})}\BibitemShut {NoStop}%
\bibitem [{\citenamefont {Paulsen}(2001)}]{Paulsen01}%
  \BibitemOpen
  \bibfield  {author} {\bibinfo {author} {\bibfnamefont {C.}~\bibnamefont
  {Paulsen}},\ }\bibfield  {title} {\bibinfo {title} {Dc magnetic
  measurements},\ }in\ \href@noop {} {\emph {\bibinfo {booktitle} {Introduction
  to Physical Techniques in Molecular Magnetism: Structural and Macroscopic
  Techniques}}},\ \bibinfo {editor} {edited by\ \bibinfo {editor}
  {\bibfnamefont {F.}~\bibnamefont {Palacio}}, \bibinfo {editor} {\bibfnamefont
  {E.}~\bibnamefont {Ressouche}},\ and\ \bibinfo {editor} {\bibfnamefont
  {J.}~\bibnamefont {Schweizer}}}\ (\bibinfo  {publisher} {Servicio de
  Publicaciones de la Universidad de Zaragoza},\ \bibinfo {year}
  {2001})\BibitemShut {NoStop}%
\bibitem [{\citenamefont {Fischer}\ \emph {et~al.}(2000)\citenamefont
  {Fischer}, \citenamefont {Frey}, \citenamefont {Koch}, \citenamefont
  {K\"{o}nnecke}, \citenamefont {Pomjakushin}, \citenamefont {Schefer},
  \citenamefont {Thut}, \citenamefont {Schlumpf}, \citenamefont {B\"{u}rge},
  \citenamefont {Greuter}, \citenamefont {Bondt},\ and\ \citenamefont
  {Berruyer}}]{Fisher_2000}%
  \BibitemOpen
  \bibfield  {author} {\bibinfo {author} {\bibfnamefont {P.}~\bibnamefont
  {Fischer}}, \bibinfo {author} {\bibfnamefont {G.}~\bibnamefont {Frey}},
  \bibinfo {author} {\bibfnamefont {M.}~\bibnamefont {Koch}}, \bibinfo {author}
  {\bibfnamefont {M.}~\bibnamefont {K\"{o}nnecke}}, \bibinfo {author}
  {\bibfnamefont {V.}~\bibnamefont {Pomjakushin}}, \bibinfo {author}
  {\bibfnamefont {J.}~\bibnamefont {Schefer}}, \bibinfo {author} {\bibfnamefont
  {R.}~\bibnamefont {Thut}}, \bibinfo {author} {\bibfnamefont {N.}~\bibnamefont
  {Schlumpf}}, \bibinfo {author} {\bibfnamefont {R.}~\bibnamefont {B\"{u}rge}},
  \bibinfo {author} {\bibfnamefont {U.}~\bibnamefont {Greuter}}, \bibinfo
  {author} {\bibfnamefont {S.}~\bibnamefont {Bondt}},\ and\ \bibinfo {author}
  {\bibfnamefont {E.}~\bibnamefont {Berruyer}},\ }\bibfield  {title} {\bibinfo
  {title} {High-resolution powder diffractometer {HRPT} for thermal neutrons at
  {SINQ}},\ }\href {https://doi.org/10.1016/S0921-4526(99)01399-X} {\bibfield
  {journal} {\bibinfo  {journal} {Physica B}\ }\textbf {\bibinfo {volume}
  {276-278}},\ \bibinfo {pages} {146} (\bibinfo {year} {2000})}\BibitemShut
  {NoStop}%
\bibitem [{\citenamefont {F\r{a}k}\ \emph {et~al.}(2022)\citenamefont
  {F\r{a}k}, \citenamefont {Rols}, \citenamefont {Manzin},\ and\ \citenamefont
  {Meulien}}]{Fak_2022}%
  \BibitemOpen
  \bibfield  {author} {\bibinfo {author} {\bibfnamefont {B.}~\bibnamefont
  {F\r{a}k}}, \bibinfo {author} {\bibfnamefont {S.}~\bibnamefont {Rols}},
  \bibinfo {author} {\bibfnamefont {G.}~\bibnamefont {Manzin}},\ and\ \bibinfo
  {author} {\bibfnamefont {O.}~\bibnamefont {Meulien}},\ }\bibfield  {title}
  {\bibinfo {title} {Panther --- the new thermal neutron time-of-flight
  spectrometer at the ill},\ }\href
  {https://doi.org/10.1051/epjconf/202227202001} {\bibfield  {journal}
  {\bibinfo  {journal} {EPJ Web of Conferences}\ }\textbf {\bibinfo {volume}
  {272}},\ \bibinfo {pages} {02001} (\bibinfo {year} {2022})}\BibitemShut
  {NoStop}%
\bibitem [{\citenamefont {L\'eger}\ \emph
  {et~al.}(2021{\natexlab{c}})\citenamefont {L\'eger}, \citenamefont {F{\aa}k},
  \citenamefont {Lhotel}, \citenamefont {Petit},\ and\ \citenamefont
  {Zanotti}}]{doi_panther}%
  \BibitemOpen
  \bibfield  {author} {\bibinfo {author} {\bibfnamefont {M.}~\bibnamefont
  {L\'eger}}, \bibinfo {author} {\bibfnamefont {B.}~\bibnamefont {F{\aa}k}},
  \bibinfo {author} {\bibfnamefont {E.}~\bibnamefont {Lhotel}}, \bibinfo
  {author} {\bibfnamefont {S.}~\bibnamefont {Petit}},\ and\ \bibinfo {author}
  {\bibfnamefont {J.-M.}\ \bibnamefont {Zanotti}},\ }\bibfield  {title}
  {\bibinfo {title} {Magnetic moment fragmentation in {(NdLa)$_2$Zr$_2$O$_7$}
  pyrochlore magnets. {Institut Laue-Langevin (ILL)}},\ }\bibfield  {journal}
  {\bibinfo  {journal} {doi:}\ }\href
  {https://doi.org/10.5291/ILL-DATA.4-05-779} {10.5291/ILL-DATA.4-05-779}
  (\bibinfo {year} {2021}{\natexlab{c}})\BibitemShut {NoStop}%
\bibitem [{\citenamefont {Stuhr}\ \emph {et~al.}(2017)\citenamefont {Stuhr},
  \citenamefont {Roessli}, \citenamefont {Gvasaliya}, \citenamefont
  {R{\o}nnow}, \citenamefont {Filges}, \citenamefont {Graf}, \citenamefont
  {Bollhalder}, \citenamefont {Hohl}, \citenamefont {B\"{u}rge}, \citenamefont
  {Schild}, \citenamefont {Holitzner}, \citenamefont {Kaegi}, \citenamefont
  {Keller},\ and\ \citenamefont {M\"uhlebach}}]{Stuhr_2017}%
  \BibitemOpen
  \bibfield  {author} {\bibinfo {author} {\bibfnamefont {U.}~\bibnamefont
  {Stuhr}}, \bibinfo {author} {\bibfnamefont {B.}~\bibnamefont {Roessli}},
  \bibinfo {author} {\bibfnamefont {S.}~\bibnamefont {Gvasaliya}}, \bibinfo
  {author} {\bibfnamefont {H.~M.}\ \bibnamefont {R{\o}nnow}}, \bibinfo {author}
  {\bibfnamefont {U.}~\bibnamefont {Filges}}, \bibinfo {author} {\bibfnamefont
  {D.}~\bibnamefont {Graf}}, \bibinfo {author} {\bibfnamefont {A.}~\bibnamefont
  {Bollhalder}}, \bibinfo {author} {\bibfnamefont {D.}~\bibnamefont {Hohl}},
  \bibinfo {author} {\bibfnamefont {R.}~\bibnamefont {B\"{u}rge}}, \bibinfo
  {author} {\bibfnamefont {M.}~\bibnamefont {Schild}}, \bibinfo {author}
  {\bibfnamefont {L.}~\bibnamefont {Holitzner}}, \bibinfo {author}
  {\bibfnamefont {C.}~\bibnamefont {Kaegi}}, \bibinfo {author} {\bibfnamefont
  {P.}~\bibnamefont {Keller}},\ and\ \bibinfo {author} {\bibfnamefont
  {T.}~\bibnamefont {M\"uhlebach}},\ }\bibfield  {title} {\bibinfo {title} {The
  thermal triple-axis-spectrometer {EIGER} at the continuous spallation source
  {SINQ}},\ }\href {https://doi.org/10.1016/j.nima.2017.02.003} {\bibfield
  {journal} {\bibinfo  {journal} {Nucl. Instrum. Methods Phys. Res. Sect. A}\
  }\textbf {\bibinfo {volume} {853}},\ \bibinfo {pages} {16} (\bibinfo {year}
  {2017})}\BibitemShut {NoStop}%
\bibitem [{\citenamefont {Petit}\ \emph {et~al.}(2021)\citenamefont {Petit},
  \citenamefont {Berrod}, \citenamefont {L\'eger}, \citenamefont {Lhotel},\
  and\ \citenamefont {Zanotti}}]{doi_sharp}%
  \BibitemOpen
  \bibfield  {author} {\bibinfo {author} {\bibfnamefont {S.}~\bibnamefont
  {Petit}}, \bibinfo {author} {\bibfnamefont {Q.}~\bibnamefont {Berrod}},
  \bibinfo {author} {\bibfnamefont {M.}~\bibnamefont {L\'eger}}, \bibinfo
  {author} {\bibfnamefont {E.}~\bibnamefont {Lhotel}},\ and\ \bibinfo {author}
  {\bibfnamefont {J.-M.}\ \bibnamefont {Zanotti}},\ }\bibfield  {title}
  {\bibinfo {title} {Spin dynamics in {(NdLa)$_2$Zr$_2$O$_7$} pyrochlore
  magnets. {Institut Laue-Langevin (ILL)}},\ }\bibfield  {journal} {\bibinfo
  {journal} {doi:}\ }\href {https://doi.org/10.5291/ILL-DATA.CRG-2894}
  {10.5291/ILL-DATA.CRG-2894} (\bibinfo {year} {2021})\BibitemShut {NoStop}%
\bibitem [{\citenamefont {Petit}\ \emph {et~al.}(2018)\citenamefont {Petit},
  \citenamefont {Ciomaga~Hatnean}, \citenamefont {L\'eger}, \citenamefont
  {Lhotel},\ and\ \citenamefont {Ollivier}}]{doi_IN5_1}%
  \BibitemOpen
  \bibfield  {author} {\bibinfo {author} {\bibfnamefont {S.}~\bibnamefont
  {Petit}}, \bibinfo {author} {\bibfnamefont {M.}~\bibnamefont
  {Ciomaga~Hatnean}}, \bibinfo {author} {\bibfnamefont {M.}~\bibnamefont
  {L\'eger}}, \bibinfo {author} {\bibfnamefont {E.}~\bibnamefont {Lhotel}},\
  and\ \bibinfo {author} {\bibfnamefont {J.}~\bibnamefont {Ollivier}},\
  }\bibfield  {title} {\bibinfo {title} {{Nd$_2$Zr$_{2-x}$Ti$_x$O$_7$}: a new
  quantum spin ice candidate},\ }\bibfield  {journal} {\bibinfo  {journal}
  {doi:}\ }\href {https://doi.org/10.5291/ILL-DATA.4-05-715}
  {10.5291/ILL-DATA.4-05-715} (\bibinfo {year} {2018})\BibitemShut {NoStop}%
\bibitem [{\citenamefont {L\'eger}\ \emph {et~al.}(2019)\citenamefont
  {L\'eger}, \citenamefont {Ciomaga~Hatnean}, \citenamefont {Lhotel},
  \citenamefont {Ollivier},\ and\ \citenamefont {Petit}}]{doi_IN5_2}%
  \BibitemOpen
  \bibfield  {author} {\bibinfo {author} {\bibfnamefont {M.}~\bibnamefont
  {L\'eger}}, \bibinfo {author} {\bibfnamefont {M.}~\bibnamefont
  {Ciomaga~Hatnean}}, \bibinfo {author} {\bibfnamefont {E.}~\bibnamefont
  {Lhotel}}, \bibinfo {author} {\bibfnamefont {J.}~\bibnamefont {Ollivier}},\
  and\ \bibinfo {author} {\bibfnamefont {S.}~\bibnamefont {Petit}},\ }\bibfield
   {title} {\bibinfo {title} {{Nd$_2$Zr$_{2-x}$Ti$_x$O$_7$}: spin dynamics in a
  new quantum spin ice candidate. influence of the magnetic field along
  $[001]$},\ }\bibfield  {journal} {\bibinfo  {journal} {doi:}\ }\href
  {https://doi.org/10.5291/ILL-DATA.4-05-748} {10.5291/ILL-DATA.4-05-748}
  (\bibinfo {year} {2019})\BibitemShut {NoStop}%
\bibitem [{\citenamefont {Rodriguez-Carvajal}(1993)}]{Fullprof}%
  \BibitemOpen
  \bibfield  {author} {\bibinfo {author} {\bibfnamefont {J.}~\bibnamefont
  {Rodriguez-Carvajal}},\ }\bibfield  {title} {\bibinfo {title} {Recent
  advances in magnetic structure determination by neutron powder diffraction},\
  }\href {https://doi.org/10.1016/0921-4526(93)90108-I} {\bibfield  {journal}
  {\bibinfo  {journal} {Physica B}\ }\textbf {\bibinfo {volume} {192}},\
  \bibinfo {pages} {55} (\bibinfo {year} {1993})}\BibitemShut {NoStop}%
\bibitem [{\citenamefont {Harvey}\ \emph {et~al.}(2005)\citenamefont {Harvey},
  \citenamefont {Whittle}, \citenamefont {Lumpkin}, \citenamefont {Smith},\
  and\ \citenamefont {Redfern}}]{Harvey_2005}%
  \BibitemOpen
  \bibfield  {author} {\bibinfo {author} {\bibfnamefont {E.~J.}\ \bibnamefont
  {Harvey}}, \bibinfo {author} {\bibfnamefont {K.~R.}\ \bibnamefont {Whittle}},
  \bibinfo {author} {\bibfnamefont {G.~R.}\ \bibnamefont {Lumpkin}}, \bibinfo
  {author} {\bibfnamefont {R.~I.}\ \bibnamefont {Smith}},\ and\ \bibinfo
  {author} {\bibfnamefont {S.~A.~T.}\ \bibnamefont {Redfern}},\ }\bibfield
  {title} {\bibinfo {title} {Solid solubilities of
  {(LaNd)}$_2${(ZrTi)}$_2${O}$_7$ phases deduced by neutron diffraction},\
  }\href {https://doi.org/10.1016/j.jssc.2004.12.030} {\bibfield  {journal}
  {\bibinfo  {journal} {J. Solid. State Chem.}\ }\textbf {\bibinfo {volume}
  {178}},\ \bibinfo {pages} {800} (\bibinfo {year} {2005})}\BibitemShut
  {NoStop}%
\bibitem [{\citenamefont {Wybourne}(1965)}]{Wybourne_1965}%
  \BibitemOpen
  \bibfield  {author} {\bibinfo {author} {\bibfnamefont {B.~G.}\ \bibnamefont
  {Wybourne}},\ }\href@noop {} {\emph {\bibinfo {title} {Spectroscopic
  Properties of Rare Earths}}}\ (\bibinfo  {publisher} {Interscience, New
  York},\ \bibinfo {year} {1965})\BibitemShut {NoStop}%
\bibitem [{\citenamefont {Bramwell}\ \emph {et~al.}(2000)\citenamefont
  {Bramwell}, \citenamefont {Field}, \citenamefont {Harris},\ and\
  \citenamefont {Parkin}}]{Bramwell_2000}%
  \BibitemOpen
  \bibfield  {author} {\bibinfo {author} {\bibfnamefont {S.~T.}\ \bibnamefont
  {Bramwell}}, \bibinfo {author} {\bibfnamefont {M.~N.}\ \bibnamefont {Field}},
  \bibinfo {author} {\bibfnamefont {M.~J.}\ \bibnamefont {Harris}},\ and\
  \bibinfo {author} {\bibfnamefont {I.~P.}\ \bibnamefont {Parkin}},\ }\bibfield
   {title} {\bibinfo {title} {Bulk magnetization of the heavy rare earth
  titanate pyrochlores - a series of model frutrated magnets},\ }\href
  {https://doi.org/10.1088/0953-8984/12/4/308} {\bibfield  {journal} {\bibinfo
  {journal} {J. Phys. Condens. Matter}\ }\textbf {\bibinfo {volume} {12}},\
  \bibinfo {pages} {483} (\bibinfo {year} {2000})}\BibitemShut {NoStop}%
\bibitem [{\citenamefont {Petit}\ \emph {et~al.}(2017)\citenamefont {Petit},
  \citenamefont {Lhotel}, \citenamefont {Damay}, \citenamefont {Boutrouille},
  \citenamefont {Forget},\ and\ \citenamefont {Colson}}]{Petit_2017}%
  \BibitemOpen
  \bibfield  {author} {\bibinfo {author} {\bibfnamefont {S.}~\bibnamefont
  {Petit}}, \bibinfo {author} {\bibfnamefont {E.}~\bibnamefont {Lhotel}},
  \bibinfo {author} {\bibfnamefont {F.}~\bibnamefont {Damay}}, \bibinfo
  {author} {\bibfnamefont {P.}~\bibnamefont {Boutrouille}}, \bibinfo {author}
  {\bibfnamefont {A.}~\bibnamefont {Forget}},\ and\ \bibinfo {author}
  {\bibfnamefont {D.}~\bibnamefont {Colson}},\ }\bibfield  {title} {\bibinfo
  {title} {Long-range order in the dipolar $xy$ antiferromagnet
  {Er}$_2${Sn}$_2${O}$_7$},\ }\href
  {https://doi.org/10.1103/PhysRevLett.119.187202} {\bibfield  {journal}
  {\bibinfo  {journal} {Phys. Rev. Lett.}\ }\textbf {\bibinfo {volume} {119}},\
  \bibinfo {pages} {187202} (\bibinfo {year} {2017})}\BibitemShut {NoStop}%
\bibitem [{\citenamefont {Samartzis}\ \emph {et~al.}(2022)\citenamefont
  {Samartzis}, \citenamefont {Xu}, \citenamefont {Anand}, \citenamefont
  {Islam}, \citenamefont {Ollivier}, \citenamefont {Su},\ and\ \citenamefont
  {Lake}}]{Samartzis_2022}%
  \BibitemOpen
  \bibfield  {author} {\bibinfo {author} {\bibfnamefont {A.}~\bibnamefont
  {Samartzis}}, \bibinfo {author} {\bibfnamefont {J.}~\bibnamefont {Xu}},
  \bibinfo {author} {\bibfnamefont {V.~K.}\ \bibnamefont {Anand}}, \bibinfo
  {author} {\bibfnamefont {A.~T. M.~N.}\ \bibnamefont {Islam}}, \bibinfo
  {author} {\bibfnamefont {J.}~\bibnamefont {Ollivier}}, \bibinfo {author}
  {\bibfnamefont {Y.}~\bibnamefont {Su}},\ and\ \bibinfo {author}
  {\bibfnamefont {B.}~\bibnamefont {Lake}},\ }\bibfield  {title} {\bibinfo
  {title} {Pinch points and half-moons in dipolar-octupolar
  {Nd$_2$Hf$_2$O$_7$}},\ }\href {https://doi.org/10.1103/PhysRevB.106.L100401}
  {\bibfield  {journal} {\bibinfo  {journal} {Phys. Rev. B}\ }\textbf {\bibinfo
  {volume} {106}},\ \bibinfo {pages} {L100401} (\bibinfo {year}
  {2022})}\BibitemShut {NoStop}%
\bibitem [{\citenamefont {Gomez}\ \emph {et~al.}(2021)\citenamefont {Gomez},
  \citenamefont {Sarte}, \citenamefont {Zelensky}, \citenamefont {Hallas},
  \citenamefont {Gonzalez}, \citenamefont {Hong}, \citenamefont {Pace},
  \citenamefont {Calder}, \citenamefont {Stone}, \citenamefont {Su},
  \citenamefont {Feng}, \citenamefont {Le}, \citenamefont {Stock},
  \citenamefont {Attfield}, \citenamefont {Wilson}, \citenamefont {Wiebe},\
  and\ \citenamefont {Aczel}}]{Gomez_2021}%
  \BibitemOpen
  \bibfield  {author} {\bibinfo {author} {\bibfnamefont {S.~J.}\ \bibnamefont
  {Gomez}}, \bibinfo {author} {\bibfnamefont {P.~M.}\ \bibnamefont {Sarte}},
  \bibinfo {author} {\bibfnamefont {M.}~\bibnamefont {Zelensky}}, \bibinfo
  {author} {\bibfnamefont {A.~M.}\ \bibnamefont {Hallas}}, \bibinfo {author}
  {\bibfnamefont {B.~A.}\ \bibnamefont {Gonzalez}}, \bibinfo {author}
  {\bibfnamefont {K.~H.}\ \bibnamefont {Hong}}, \bibinfo {author}
  {\bibfnamefont {E.~J.}\ \bibnamefont {Pace}}, \bibinfo {author}
  {\bibfnamefont {S.}~\bibnamefont {Calder}}, \bibinfo {author} {\bibfnamefont
  {M.~B.}\ \bibnamefont {Stone}}, \bibinfo {author} {\bibfnamefont
  {Y.}~\bibnamefont {Su}}, \bibinfo {author} {\bibfnamefont {E.}~\bibnamefont
  {Feng}}, \bibinfo {author} {\bibfnamefont {M.~D.}\ \bibnamefont {Le}},
  \bibinfo {author} {\bibfnamefont {C.}~\bibnamefont {Stock}}, \bibinfo
  {author} {\bibfnamefont {J.~P.}\ \bibnamefont {Attfield}}, \bibinfo {author}
  {\bibfnamefont {S.~D.}\ \bibnamefont {Wilson}}, \bibinfo {author}
  {\bibfnamefont {C.~R.}\ \bibnamefont {Wiebe}},\ and\ \bibinfo {author}
  {\bibfnamefont {A.~A.}\ \bibnamefont {Aczel}},\ }\bibfield  {title} {\bibinfo
  {title} {Absence of moment fragmentation in the mixed ${B}$-site pyrochlore
  {Nd}$_2${GaSbO}$_7$},\ }\href {https://doi.org/10.1103/PhysRevB.103.214419}
  {\bibfield  {journal} {\bibinfo  {journal} {Phys. Rev. B}\ }\textbf {\bibinfo
  {volume} {103}},\ \bibinfo {pages} {214419} (\bibinfo {year}
  {2021})}\BibitemShut {NoStop}%
\bibitem [{\citenamefont {Scheie}\ \emph {et~al.}(2021)\citenamefont {Scheie},
  \citenamefont {Sanders}, \citenamefont {Gui}, \citenamefont {Qiu},
  \citenamefont {Prisk}, \citenamefont {Cava},\ and\ \citenamefont
  {Broholm}}]{Scheie_2021}%
  \BibitemOpen
  \bibfield  {author} {\bibinfo {author} {\bibfnamefont {A.}~\bibnamefont
  {Scheie}}, \bibinfo {author} {\bibfnamefont {M.}~\bibnamefont {Sanders}},
  \bibinfo {author} {\bibfnamefont {X.}~\bibnamefont {Gui}}, \bibinfo {author}
  {\bibfnamefont {Y.}~\bibnamefont {Qiu}}, \bibinfo {author} {\bibfnamefont
  {T.~R.}\ \bibnamefont {Prisk}}, \bibinfo {author} {\bibfnamefont {R.~J.}\
  \bibnamefont {Cava}},\ and\ \bibinfo {author} {\bibfnamefont
  {C.}~\bibnamefont {Broholm}},\ }\bibfield  {title} {\bibinfo {title} {Beyond
  magnons in {Nd}$_2${ScNbO}$_7$: An ising pyrochlore antiferromagnet with
  all-in -- all-out order and random fields},\ }\href
  {https://doi.org/10.1103/PhysRevB.104.134418} {\bibfield  {journal} {\bibinfo
   {journal} {Phys. Rev. B}\ }\textbf {\bibinfo {volume} {104}},\ \bibinfo
  {pages} {134418} (\bibinfo {year} {2021})}\BibitemShut {NoStop}%
\bibitem [{\citenamefont {Mauws}\ \emph {et~al.}(2021)\citenamefont {Mauws},
  \citenamefont {Hiebert}, \citenamefont {Rutherford}, \citenamefont {Zhou},
  \citenamefont {Huang}, \citenamefont {Stone}, \citenamefont {Butch},
  \citenamefont {Su}, \citenamefont {Choi}, \citenamefont {Yamani},\ and\
  \citenamefont {Wiebe}}]{Mauws_2021}%
  \BibitemOpen
  \bibfield  {author} {\bibinfo {author} {\bibfnamefont {C.}~\bibnamefont
  {Mauws}}, \bibinfo {author} {\bibfnamefont {N.}~\bibnamefont {Hiebert}},
  \bibinfo {author} {\bibfnamefont {M.~L.}\ \bibnamefont {Rutherford}},
  \bibinfo {author} {\bibfnamefont {H.~D.}\ \bibnamefont {Zhou}}, \bibinfo
  {author} {\bibfnamefont {Q.}~\bibnamefont {Huang}}, \bibinfo {author}
  {\bibfnamefont {M.~B.}\ \bibnamefont {Stone}}, \bibinfo {author}
  {\bibfnamefont {N.~P.}\ \bibnamefont {Butch}}, \bibinfo {author}
  {\bibfnamefont {Y.}~\bibnamefont {Su}}, \bibinfo {author} {\bibfnamefont
  {E.~S.}\ \bibnamefont {Choi}}, \bibinfo {author} {\bibfnamefont
  {Z.}~\bibnamefont {Yamani}},\ and\ \bibinfo {author} {\bibfnamefont {C.~R.}\
  \bibnamefont {Wiebe}},\ }\bibfield  {title} {\bibinfo {title} {Magnetic
  ordering in the ising antiferromagnetic pyrochlore {Nd}$_2${ScNbO}$_7$},\
  }\href {https://doi.org/10.1088/1361-648X/abf594} {\bibfield  {journal}
  {\bibinfo  {journal} {J. Phys.: Condens. Matter}\ }\textbf {\bibinfo {volume}
  {33}},\ \bibinfo {pages} {245802} (\bibinfo {year} {2021})}\BibitemShut
  {NoStop}%
\bibitem [{\citenamefont {Stauffer}(1985)}]{Stauffer}%
  \BibitemOpen
  \bibfield  {author} {\bibinfo {author} {\bibfnamefont {D.}~\bibnamefont
  {Stauffer}},\ }\href {https://doi.org/10.1201/9781315274386} {\emph {\bibinfo
  {title} {Introduction to percolation theory}}}\ (\bibinfo  {publisher}
  {Taylor and Francis},\ \bibinfo {year} {1985})\BibitemShut {NoStop}%
\bibitem [{\citenamefont {Hardy}\ \emph {et~al.}(2003)\citenamefont {Hardy},
  \citenamefont {Lambert}, \citenamefont {Lees},\ and\ \citenamefont
  {McK.~Paul}}]{Hardy_2003}%
  \BibitemOpen
  \bibfield  {author} {\bibinfo {author} {\bibfnamefont {V.}~\bibnamefont
  {Hardy}}, \bibinfo {author} {\bibfnamefont {S.}~\bibnamefont {Lambert}},
  \bibinfo {author} {\bibfnamefont {M.~R.}\ \bibnamefont {Lees}},\ and\
  \bibinfo {author} {\bibfnamefont {D.}~\bibnamefont {McK.~Paul}},\ }\bibfield
  {title} {\bibinfo {title} {Specific heat and magnetization study on single
  crystals of the frustrated quasi-one-dimensional oxide {Ca$_3$Co$_2$O$_6$}},\
  }\href {https://doi.org/10.1103/PhysRevB.68.014424} {\bibfield  {journal}
  {\bibinfo  {journal} {Phys. Rev. B}\ }\textbf {\bibinfo {volume} {68}},\
  \bibinfo {pages} {014424} (\bibinfo {year} {2003})}\BibitemShut {NoStop}%
\bibitem [{\citenamefont {Petit}(2011)}]{Spinwave1}%
  \BibitemOpen
  \bibfield  {author} {\bibinfo {author} {\bibfnamefont {S.}~\bibnamefont
  {Petit}},\ }\bibfield  {title} {\bibinfo {title} {Numerical simulations and
  magnetism},\ }\href {https://doi.org/10.1051/sfn/201112006} {\bibfield
  {journal} {\bibinfo  {journal} {JDN}\ }\textbf {\bibinfo {volume} {12}},\
  \bibinfo {pages} {105} (\bibinfo {year} {2011})}\BibitemShut {NoStop}%
\bibitem [{\citenamefont {Petit}(2010)}]{Spinwave2}%
  \BibitemOpen
  \bibfield  {author} {\bibinfo {author} {\bibfnamefont {S.}~\bibnamefont
  {Petit}},\ }\bibfield  {title} {\bibinfo {title} {Les ondes de spin},\ }\href
  {https://doi.org/10.1051/sfn/2010008} {\bibfield  {journal} {\bibinfo
  {journal} {JDN}\ }\textbf {\bibinfo {volume} {10}},\ \bibinfo {pages} {449}
  (\bibinfo {year} {2010})}\BibitemShut {NoStop}%
\bibitem [{\citenamefont {Henley}(2001)}]{Henley_2001}%
  \BibitemOpen
  \bibfield  {author} {\bibinfo {author} {\bibfnamefont {C.~L.}\ \bibnamefont
  {Henley}},\ }\bibfield  {title} {\bibinfo {title} {Effective {H}amiltonians
  and dilution effects in kagome and related anti-ferromagnets},\ }\href
  {https://doi.org/10.1139/cjp-79-11-12-1307} {\bibfield  {journal} {\bibinfo
  {journal} {Can. J. Phys.}\ }\textbf {\bibinfo {volume} {79}},\ \bibinfo
  {pages} {1307} (\bibinfo {year} {2001})}\BibitemShut {NoStop}%
\end{thebibliography}%

\end{document}